\documentclass[prd,superscriptaddress,twocolumn,tightenlines,nofootinbib, eqsecnum]{revtex4-2}

\usepackage{amsmath}
\usepackage{amsfonts}
\usepackage{amssymb}
\usepackage{physics}
\usepackage{bm}
\usepackage{mathrsfs}
\usepackage{graphicx}
\usepackage[colorlinks]{hyperref}
\usepackage[dvipsnames]{xcolor}
\usepackage{tikz}
\usepackage{mathtools}

\definecolor{darkgreen}{rgb}{0,0.5,0}
\usepackage{empheq}
\usepackage{ulem}
\usepackage{orcidlink}
\usepackage[capitalise]{cleveref}
\usepackage{ragged2e}
\usepackage{balance}

\hypersetup{
 unicode=false,          
 pdftoolbar=true,        
 pdfmenubar=true,        
 pdffitwindow=false,     
 pdfstartview={FitH},    
 pdftitle={My title},    
 pdfauthor={Author},     
 pdfsubject={Subject},   
 pdfcreator={Creator},   
 pdfproducer={Producer}, 
 pdfkeywords={keyword1} {key2} {key3}, 
 pdfnewwindow=true,      
 colorlinks=true,       
 linkcolor=Blue,          
 citecolor=cyan,        
 filecolor=magenta,      
 urlcolor=darkgreen           
}

\allowdisplaybreaks

\DeclareSymbolFontAlphabet{\mathrsfs}{rsfs}
\DeclareMathAlphabet{\mathcal}{OMS}{cmsy}{m}{n}

\newcommand{\be}{\begin{equation}}
\newcommand{\ee}{\end{equation}}
\newcommand{\bse}{\begin{subequations}}
\newcommand{\ese}{\end{subequations}}
\newcommand{\ba}{\begin{align}}
\newcommand{\ea}{\end{align}}
\newcommand{\nn}{\nonumber}
\newcommand{\p}{\partial}

\renewcommand{\dd}{\mathrm{d}}
\newcommand{\dE}{\mathrm{E}}
\newcommand{\dK}{\mathrm{K}}

\makeatletter
\g@addto@macro\bfseries{\boldmath}
\makeatother

\defcitealias{BBT22}{paper~I}
\defcitealias{T24_QK}{paper~II}

\begin{document}

\interfootnotelinepenalty=10000

\title{Constants of motion in gravitational self-force theory}

\author{David \textsc{Trestini}\,\orcidlink{0000-0002-4140-0591}}\email{david.trestini@southampton.ac.uk}

\author{Zachary \textsc{Nasipak}\,\orcidlink{0000-0002-5109-9704}}\email{z.nasipak@southampton.ac.uk}

\author{Adam \textsc{Pound}\,\orcidlink{0000-0001-9446-0638}}\email{a.pound@southampton.ac.uk}

\affiliation{School of Mathematical Sciences and STAG Research Centre, University of Southampton, Southampton SO17 1BJ, United Kingdom}

\date{\today}

\begin{abstract}

Synergies between self-force theory and other approaches to the gravitational two-body problem have traditionally relied on calculations of gauge-invariant observables as functions of orbital frequencies. However, in self-force theory, one can also define a complete set of constants of motion: energy, azimuthal angular momentum, and radial and polar actions. Here, we outline how directly utilizing these constants allows for more straightforward comparisons and hybridizations across the parameter space, as well as more streamlined waveform generation through flux-balance laws. Restricting to the case of eccentric, nonspinning binaries and first order in self-force, we compute the constants of motion and the corrections to fundamental frequencies numerically as well as analytically (to 9PN in a post-Newtonian expansion), establishing consistency with the highest-order (4PN) results  available from post-Newtonian theory. We also apply the results to identify the perturbed locations of special curves in the parameter space: circular orbits and the separatrix between bound and plunging orbits. 

\end{abstract}

\maketitle

\tableofcontents

\section{Introduction}
\label{sec:intro}

Much of the progress in gravitational-wave source modeling has relied on combining multiple methods~\cite{LeTiec:2014oez,LISAConsortiumWaveformWorkingGroup:2023arg}, each of which is accurate in a different regime of the two-body problem: post-Newtonian (PN) theory, accurate in the early inspiral~\cite{Blanchet:2013haa}; self-force (SF) theory, accurate for small mass ratios~\cite{Barack:2018yvs}; numerical relativity, best suited to comparable masses in the late inspiral, merger, and ringdown~\cite{Baumgarte:2010ndz}; black hole perturbation theory, applicable to the final ringdown~\cite{Berti:2025hly}; and most recently, post-Minkowskian (PM) theory, accurate for weak-field, unbound orbits~\cite{Damour:2016gwp}. Tools and results from these methods have been combined in a myriad of ways (e.g., Refs.~\cite{Barausse:2011dq, Antonelli:2019fmq, Pompili:2023tna, Ramos-Buades:2023ehm, Leather:2025nhu, Albanesi:2025txj, Buonanno:2024byg, Damour:2025uka, Damour:2007xr, Taracchini:2014zpa, Nagar:2022fep, Varma:2018mmi, Rifat:2019ltp, Ajith:2009bn, Santamaria:2010yb, Hannam:2013oca, London:2017bcn, Pratten:2020ceb, Damour:2009sm, Barack:2010ny, Bini:2013rfa, Bini:2013zaa, Kavanagh:2015lva, Bini:2019nra, Long:2024ltn, Huerta:2017kez,Paul:2024ujx,Wittek:2024pis, Iglesias:2025tnt,Honet:2025gge,Honet:2025lmk}), often through the framework of effective one-body (EOB) theory~\cite{Buonanno:1998gg,Buonanno:2000ef}.

This synergism is enabled by calculations of the same quantities in multiple formalisms in order to translate information between them. Frequently, the computed quantities are waveforms themselves, but in many cases the translation occurs at the level of orbital dynamics
---or at the level of waveforms as functions of orbital parameters~\cite{Warburton:2024xnr}. In PN and PM theory, the orbital dynamics is very often written in the language of Hamiltonian mechanics, with well-defined notions of conserved energy, angular momentum, and action variables~\cite{Damour:2016abl,Bernard:2017ktp,Bern:2022jvn,Driesse:2024xad}. In self-force theory, such conserved quantities have been essential for the understanding and exploitation of the leading-order~(0SF), geodesic motion of the secondary object (of mass $m_2$) around the primary black hole (of mass $m_1\geq m_2$)~\cite{Schmidt:2002qk,Hinderer:2008dm,Pound:2021qin}. They have also played a key role in modeling the effects of the secondary's spin~\cite{Witzany:2018ahb,Witzany:2019nml,Skoupy:2023lih,Ramond:2024ozy,Ramond:2024sfp,Piovano:2024yks,Grant:2024ivt,Witzany:2024ttz,Mathews:2025nyb}. But beyond the test-body limit, when the secondary object perturbs the primary's spacetime and experiences a self-force, such conserved quantities have been little explored. 

Notions of conserved energy, angular momentum, and radial action at first subleading order in the mass ratio $\varepsilon:=m_2/m_1$ (1SF) were first proposed in Refs.~\cite{LeTiec:2011ab,LeTiec:2015kgg}. There, they were effectively defined as the quantities satisfying the first law of binary mechanics~\cite{LeTiec:2011ab,Gralla:2012dm, Blanchet:2012at, LeTiec:2013iux, LeTiec:2015kgg,Fujita:2016igj, Blanchet:2017rcn,Ramond:2021mip,Ramond:2022ctc,Gonzo:2024xjk} for circular or eccentric orbits of the secondary around a Schwarzschild primary. Reference~\cite{Fujita:2016igj} expanded upon this approach and extended it to the case of generic bound orbits around a spinning, Kerr black hole, again defining the conserved quantities as the ones satisfying a first law of binary mechanics. However, this approach required an \textit{ad hoc} ``renormalization'' of the conserved quantities. More recently, a complete Hamiltonian analysis~\cite{Lewis:2025ydo} building on Refs.~\cite{Fujita:2016igj,Blanco:2022mgd} has shown that the ``renormalized'' variables precisely correspond to natural notions in Hamiltonian mechanics: a conserved energy that is the on-shell value of the Hamiltonian; and action variables---radial, polar, and azimuthal, the last of which represents azimuthal angular momentum---with the traditional meaning of invariant integrals over tori in the orbital phase space~\cite{Arnold}.

In this paper, we begin to explore the behavior and utility of these 1SF conserved quantities in more detail, specializing to the case of eccentric bound orbits around a nonspinning, Schwarzschild black hole. Access to these quantities brings several advantages: 
\begin{enumerate}
    \item It allows one to cast waveform generation in a streamlined form in terms of balance laws, in which the waveform's evolution is dictated by the slow evolution of the conserved quantities, and that slow evolution is itself determined from the gravitational-wave fluxes to infinity and down the primary's horizon. We expect this to be crucial at the first subleading order in $\varepsilon$---the so-called first postadiabatic (1PA) order~\cite{Pound:2021qin,Mathews:2025nyb}. Such 1PA models are required for modeling extreme-mass-ratio inspirals (EMRIs)~\cite{LISAConsortiumWaveformWorkingGroup:2023arg,Burke:2023lno}, key sources for space-based detectors such as LISA. Currently, the only extant 1PA waveform models~\cite{Wardell:2021fyy,Mathews:2025txc}, which have been restricted to quasicircular orbits, have relied on this balance-law approach. We expect their generalization to eccentric orbits to do so, as well.
    \item It facilitates communication between SF, PN, and PM theory. For example, formulations of waveform generation in terms of balance laws enable simple hybridizations at the level of conserved quantities and fluxes; see Refs.~\cite{Honet:2025gge,Honet:2025lmk}. As outlined in those references, hurdles in second-order self-force (2SF) calculations~\cite{PanossoMacedo:2024pox} mean that such a hybrid approach might be essential for the construction of 1PA waveform models across the binary parameter space in the near term. Direct applications to EOB models can also be envisioned.
    \item It provides a simple way of computing the corrections to the binary's fundamental frequencies, which have a central role in EMRI modeling~\cite{Pound:2021qin,Hughes:2021exa,Katz:2021yft,Wardell:2021fyy,Chapman-Bird:2025xtd,Mathews:2025nyb}. Traditionally, the translation of 1SF results into other formalisms has relied on computations of invariant observables as functions of these frequencies~\cite{LeTiec:2014oez,Akcay:2015pza,Barack:2018yvs}.
    \item Following the path laid out by Le Tiec~\cite{LeTiec:2015kgg}, it provides a clean characterization of the orbital phase space, helping to identify special surfaces such as circular orbits and the separatrix between bound and plunging orbits.
\end{enumerate}

To capitalize on these benefits, here we calculate the 1SF energy, angular momentum, radial action, and fundamental frequencies across the parameter space of bound orbits in Schwarzschild spacetime, using both numerical and analytical (post-Newtonian) methods\footnote{In the context of gravitational self-force, post-Newtonian expansions can be obtained to very high orders, but they are restricted to the leading or subleading order in the mass ratio; we refer to this as the ``analytical self force''~\cite{Sasaki:2003xr} or ``SF-PN.'' Conversely, ``traditional PN'' methods~\cite{Blanchet:2013haa} are those which can access all orders in the mass ratio; these are typically more involved, and current literature is restricted to lower orders.} from Refs.~\cite{Nasipak:2025tby,Munna:2022gio}. We verify that the results agree with those of traditional PN theory at all available PN orders (through 4PN~\cite{Trestini:2025yyc}), and we present several applications of them. We expect these results to be widely useful across the community, in the construction of 1PA SF models, in the validation of other models, and in the development of EOB and other hybrid models. For ease of use, we also delineate the region where high-order analytical results (carried to 9PN in the weak-field expansion) become more accurate than our numerical data.

The paper is organized as follows: In Sec.~\ref{sec:formalism}, we review the SF formalism and the 1SF conserved quantities. In Sec.~\ref{sec:geo}, we review the relevant equations at geodesic, 0SF order, including PN expansions of geodesic relationships. Section~\ref{sec:1SF} describes our calculation of the 1SF corrections, as well as the comparison between our numerical and analytical results. Section~\ref{sec:applications} applies our results to identify the location of special curves in the parameter space, as mentioned above, and outlines waveform generation schemes using our results. We conclude in Sec.~\ref{sec:conclusion} with a discussion of future applications. Some technical details and lengthy expressions are relegated to Appendices. We provide our data and PN expansions in the Supplemental Material~\cite{supp}.

We use geometric units with $G=c=1$ and the mostly positive signature $(-+++)$. Repeated indices are summed over even if they have the same vertical placement, as in $k_A\varphi_A := k_r\varphi_r+k_\phi\varphi_\phi$. 

\section{Formalism}
\label{sec:formalism}

Our review of conserved quantities in SF theory broadly follows Ref.~\cite{Lewis:2025ydo}, though with some minor changes in notation that we note along the way. Rather than specializing to conservative dynamics, we emphasize the role of the conserved quantities in the full, dynamical, radiating spacetime, situating them within the broader conceptual framework of multiscale expansions~\cite{Hinderer:2008dm,Miller:2020bft,Pound:2021qin,Miller:2023ers,Mathews:2025nyb,Wei:2025lva,Lewis:2025ydo} that underlies most modern SF calculations and enables rapid SF waveform generation~\cite{Chua:2020stf,Katz:2021yft,Chapman-Bird:2025xtd}.

\subsection{Multiscale formulation of self-force theory}

In gravitational SF theory, the spacetime metric is expanded in powers of the mass ratio $\varepsilon$,
\begin{equation}\label{eq:metric expansion}
    g_{\alpha\beta} = g^{(0)}_{\alpha\beta} + \varepsilon h^{(1)}_{\alpha\beta} + \varepsilon^2 h^{(2)}_{\alpha\beta} + {\cal O}(\varepsilon^3),
\end{equation}
and the secondary is reduced to a point particle orbiting in the spacetime of the primary. Here we restrict to the special case of a Schwarzschild background metric $g^{(0)}_{\alpha\beta}$ of mass $m_1$. Assuming the secondary is also nonspinning, its motion is geodesic not in $g_{\alpha\beta}$ (which is singular at the particle's position) or $g^{(0)}_{\alpha\beta}$ but in a certain effective metric~\cite{Detweiler:2002mi,Detweiler:2011tt,Pound:2012nt,Pound:2015tma,Pound:2017psq,Harte:2011ku,Harte:2025tmd}
\begin{equation}
    \tilde g_{\alpha\beta} = g^{(0)}_{\alpha\beta} + \varepsilon h^{{\rm R}(1)}_{\alpha\beta} + \varepsilon^2 h^{{\rm R}(2)}_{\alpha\beta} + {\cal O}(\varepsilon^3),
\end{equation}
which is regular and satisfies the vacuum Einstein equation even at the particle's position. The geodesic equation in $\tilde g_{\alpha\beta} $ can equivalently be written in terms of the background metric and perturbations as~\cite{Pound:2015fma}
\begin{equation}\label{eq:self-forced eom}
    \frac{D^2x^\alpha_p}{d\tau^2} = \varepsilon f^\alpha_{(1)} + \varepsilon^2 f^\alpha_{(2)} + {\cal O}(\varepsilon^3).
\end{equation}
Here $x^\alpha_p(\tau) = (t_p(\tau),r_p(\tau),\theta_p(\tau),\phi_p(\tau))$ is the particle's trajectory in Schwarzschild coordinates; $\tau$ is proper time measured in $g^{(0)}_{\alpha\beta}$; and $D/d\tau := u^\alpha \nabla_\alpha$ is the Schwarzschild covariant derivative along the particle's worldline, with $u^\alpha:=dx^\alpha_p/d\tau$. The first- and second-order self-forces per unit mass, $f^\alpha_{(1)}$ and $f^\alpha_{(2)}$, are constructed from $h^{{\rm R}(1)}_{\alpha\beta}$ and $h^{{\rm R}(2)}_{\alpha\beta}$. See Eqs.~(39) and (40) of Ref.~\cite{Pound:2015fma}, for example; we do not display these formulas here, because we advocate for the use of balance laws to \emph{avoid} explicit calculations of the self-forces.

The multiscale expansion of the orbital dynamics (and of the spacetime metric) is based on reformulating the self-forced motion in terms of fast and slow variables that cleanly separate oscillatory effects from slow evolution. These variables are best understood as coordinates on the orbital phase space. To exploit the stationarity of the background, and ultimately to obtain waveforms in terms of the time of an asymptotic observer, we use Schwarzschild time $t$ as the parameter along the particle's phase-space trajectory rather than proper time $\tau$. Since self-forced orbits in a Schwarzschild background are confined to a plane, we can also freely specify that they lie in the equatorial plane, $\theta=\pi/2$. With $t$ reduced to a parameter along the trajectory and with polar motion eliminated, the orbital phase space is reduced from eight dimensions to four, with phase-space coordinates given by the positions and momenta $(x^A_p,p_A)$, where $A=r,\phi$ and $p_A = m_2g_{AB}^{(0)}u^B$.\footnote{ Reference~\cite{Lewis:2025ydo} instead uses momenta $\tilde p_A = m_2 \tilde g_{A\beta}\frac{dx_p^\beta}{d\tilde\tau}$, which are canonically conjugate to $x^A_p$ at all SF orders in a certain \emph{pseudo}-Hamiltonian sense. $p_A$ in that reference denotes yet another momentum, which is conjugate to $x^A_p$ in the 1SF Hamiltonian we review below. The $p_A$ we use here is conjugate to $x^A_p$ only with respect to the background-geodesic Hamiltonian.} 

Appropriate fast variables $\varphi_A$ and slow variables $\pi_A$  are found via an $\varepsilon$-dependent transformation $(x^A_p,p_A)\mapsto (\varphi_A,\pi_A)$,\footnote{Our variables $(\varphi_A,\pi_A)$ are denoted $(\mathring\varphi^A,\mathring\pi_A)$ in Ref.~\cite{Lewis:2025ydo}, $(\varphi_A,p_\varphi^A)$ in Ref.~\cite{Pound:2021qin}, $(\mathring\psi^A,\mathring\pi_A)$ in Ref.~\cite{Mathews:2025nyb}, and $(\Phi_A,\alpha_A)$ in Refs.~\cite{Katz:2021yft,Chapman-Bird:2025xtd}. Different alphabets for the indices are also used.} which is chosen to enforce the requirements that (i)~the forces $f^\alpha_{(n)}(x^A_p,u_A)$ be $2\pi$-periodic in $\varphi_A$, and (ii)~the evolution equations for $\varphi_A$ and $\pi_A$, obtained from  the equation of motion~\eqref{eq:self-forced eom}, take the forms
\begin{align}
    \frac{d\varphi_A}{dt} & :=\Omega_A(\pi_B,\varepsilon) =  \Omega_A^{(0)}(\pi_B) + \varepsilon\Omega_A^{(1)}(\pi_B) + \cdots,\label{eq:phidot}\\
     \frac{d\pi_A}{dt} &\hphantom{:}= 
     \varepsilon\!\left[F_A^{(0)}(\pi_B) + \varepsilon F_A^{(1)}(\pi_B) + \cdots\right]\!.\label{eq:pidot}
\end{align}
The motion is approximately biperiodic, with independent  radial and azimuthal frequencies $\Omega_A=(\Omega_r,\Omega_\phi)$ that slowly evolve due to dissipation. The fast variables $\varphi_A=(\varphi_r,\varphi_\phi)$ describe the phases of this biperiodic motion; if dissipation were artificially ignored, the phase $\varphi_A$ would evolve by exactly $2\pi$ on each radial or azimuthal period $2\pi/\Omega_A ={\cal O}(m_1)$. The slow variables $\pi_A$ evolve due to dissipation on the long timescale of order $m_1/\varepsilon$ (and the frequencies hence evolve along with them). As we discuss in the body of this paper, there is considerable freedom in the choice of variables $\pi_A$, but they are typically chosen to be quasi-Keplerian parameters: a relativistic semilatus rectum and eccentricity, $\pi_A = (p, e)$, which provide the simplest parametrization of the original coordinates, $x^A_p(\varphi_B,\pi_B,\varepsilon)$ and $p_A(\varphi_B,\pi_B,\varepsilon)$. Explicit expressions for $\Omega^{(0)}_A(p,e)$ are given in Eqs.~\eqref{eqn:periods} and \eqref{eqn:OmegasofPE} below.

The transformation $(x^A_p,p_A)\mapsto (\varphi_A,\pi_A)$ is given in Refs.~\cite{Pound:2021qin,Mathews:2025nyb}, along with expressions for $\Omega^{(1)}_A$ and $F^{(n)}_A$ in terms of the forces $f^\alpha_{(n)}$. In practice, such transformations are typically found starting from coordinates $(\varphi^{(0)}_A,\pi^{(0)}_A)$ adapted to the zeroth-order, background-geodesic case, satisfying
\begin{equation}
\frac{d\varphi_A^{(0)}}{dt}=\Omega^{(0)}_A(\pi^{(0)}_B)\quad \text{and}\quad \frac{d\pi^{(0)}_A}{dt}=0
\end{equation}
when $\varepsilon=0$. Here, $\pi^{(0)}_A$ are constants of motion for a Schwarzschild geodesic, and $\varphi_A^{(0)}$ grows exactly linearly in time in that case. At finite $\varepsilon$, $d\varphi^{(0)}_A/dt$ and $d\pi^{(0)}_A/dt$ have oscillatory dependence on $\varphi^{(0)}_A$. The variables $(\varphi_A,\pi_A)$ are then found by performing a near-identity averaging transformation~\cite{VanDeMeent:2018cgn,Pound:2021qin,Lynch:2021ogr,Drummond:2023wqc,Mathews:2025nyb,Lewis:2025ydo},
\begin{align}
\varphi_A &= \varphi_A^{(0)} + \varepsilon \varphi_A^{(1)}(\varphi_B^{(0)},\pi^{(0)}_B) + {\cal O}(\varepsilon^2),\\
\pi_A &= \pi_A^{(0)} + \varepsilon \pi_A^{(1)}(\varphi_B^{(0)},\pi^{(0)}_B)  + {\cal O}(\varepsilon^2),
\end{align}
designed to ensure there are no oscillations in $d\varphi_A/dt$ or $d\pi_A/dt$.

The critical step in the multiscale expansion is to utilize these phase-space coordinates directly within the Einstein equations. We first adopt a time coordinate $s=t-\kappa(r)$ that reduces to advanced time $v=t+r^*$ at the horizon, Schwarzschild time $t$ at the particle, and retarded time $u=t-r^*$ at infinity. We then extend $(\varphi_A,\pi_A)$ off the worldline by taking them to be constant on slices of constant $s$, and we write the small-mass-ratio expansion~\eqref{eq:metric expansion} as
\begin{multline}\label{eq:metric multiscale expansion}
    g_{\alpha\beta} = g^{(0)}_{\alpha\beta}(x^i)+\varepsilon h^{(1)}_{\alpha\beta}(\varphi_A,\pi_A,x^i)\\
    +\varepsilon^2 h^{(2)}_{\alpha\beta}(\varphi_A,\pi_A,x^i)+{\cal O}(\varepsilon^3),
\end{multline}
where $x^i=(r,\theta,\phi)$ are the spatial coordinates on each $s=\text{constant}$ slice. Here, all time dependence in the metric is encoded in its dependence on $(\varphi_A,\pi_A)$, making the metric a function on phase space (crossed with space); this form of the metric is derived from first principles in Ref.~\cite{Lewis:2025ydo}. Taking the $r\to\infty$ limit of Eq.~\eqref{eq:metric multiscale expansion} then yields the waveform at infinity as a function of $(\varphi_A,\pi_A)$ and sky angles $(\theta,\phi)$. Typically, the functions in Eq.~\eqref{eq:metric multiscale expansion} and in the corresponding waveform are computed and stored as coefficients in a discrete Fourier series, 
\begin{equation}\label{eq:Fourier series}
h^{(n)}_{\alpha\beta} = \sum_{k_A\in\mathbb{Z}^2}h^{(n,k_A)}_{\alpha\beta}(\pi_B,x^i)e^{-ik_A\varphi_A}. 
\end{equation}
The waveform as a function of time is then generated by solving Eqs.~\eqref{eq:phidot} and \eqref{eq:pidot} and summing the Fourier modes~\cite{Katz:2021yft};\footnote{In addition to Eqs.~\eqref{eq:phidot} and \eqref{eq:pidot}, 1PA accuracy also requires evolution equations for the primary's mass and spin, but these will not be important for the present paper. For an initially nonspinning primary, the spin parameter remains small (of order $m_2$) throughout the inspiral~\cite{Miller:2020bft}, and it can be accounted for analytically as in Ref.~\cite{Mathews:2025txc}.} this is a rapid process by virtue of the fact that the right-hand sides of  Eqs.~\eqref{eq:phidot} and \eqref{eq:pidot} change slowly, allowing large time steps, as no oscillations need to be numerically resolved~\cite{VanDeMeent:2018cgn}. 

In this section, we have followed the traditional order counting in SF theory. An $n$th postadiabatic ($n$PA) term contributes to the phase evolution at order $\varepsilon^n$ relative to the leading, adiabatic (0PA) evolution; the numeric labels in Eqs.~\eqref{eq:phidot} and \eqref{eq:pidot} represent this postadiabatic counting. On the other hand, an $n$th-order self-force term ($n$SF) is computed from metric perturbations  that are of \emph{absolute} order~$\varepsilon^n$; the numeric labels in Eqs.~\eqref{eq:metric expansion}--\eqref{eq:self-forced eom} follow this counting. The two orderings differ because of the secular effect of dissipative forces: $\Omega^{(n)}_A$ is both $n$PA and $n$SF, while the dissipative forcing functions $F^{(n)}_A$ are $n$PA but $(n+1)$SF.

\subsection{Hamiltonian framework and conserved quantities}

The existence of the variables $(\varphi_A,\pi_A)$ is dependent on the fact that the zeroth-order, background-geodesic motion is an integrable Hamiltonian system~\cite{Schmidt:2002qk,Hinderer:2008dm}, with two independent constants of motion. A Hamiltonian analysis shows how those constants of motion naturally extend to 1SF order.

For a geodesic in Schwarzschild spacetime, the spacetime's Killing symmetries imply that the energy 
\begin{equation}\label{eq:E0 def}
E_{(0)}(p_A)=-p_t 
\end{equation}
and angular momentum 
\begin{equation}\label{eq:L0 def}
L_{(0)}(p_A) = p_\phi 
\end{equation}
are constants. 

These constants of motion, $P^{(0)}_A = (E_{(0)},L_{(0)})$, are in one-to-one correspondence with the quasi-Keplerian parameters $\pi_A=(p,e)$; the explicit geodesic relationships $P^{(0)}_A(\pi_B)$ are given in Eq.~\eqref{eqn:ELofPE} below. While these constants follow from spacetime symmetries, we can also obtain them from Hamiltonian methods: the Schwarzschild geodesic equation can be written as Hamilton's equations on the 4D phase space~\cite{Lewis:2025ydo},
\begin{align}
\frac{dx^A_p}{dt} &= \frac{\partial H^{(0)}_{\rm 4D}}{\partial p_A},\\
\frac{dp_A}{dt} &= -\frac{\partial H^{(0)}_{\rm 4D}}{\partial x^A_p},
\end{align}
with the Hamiltonian
\begin{equation}
H^{(0)}_{\rm 4D}(x^A_p,p_A) = \sqrt{-\frac{g^{AB}_{(0)}(r_p)p_A p_B+(m_2)^2}{g^{tt}_{(0)}(r_p)}}.
\end{equation}
Since the Hamiltonian is independent of $t$, its on-shell value is constant, equal to the energy~\eqref{eq:E0 def}, as one can see by rearranging $g^{\alpha\beta}_{(0)}p_\alpha p_\beta = -(m_2)^2$. Since the  Hamiltonian is independent of $\phi_p$, the angular momentum~\eqref{eq:L0 def} is also constant: $dp_\phi/dt = - \partial H^{(0)}_{\rm 4D}/\partial \phi_p=0$.

To reach the variables $\varphi^{(0)}_A$, we note that for bound, biperiodic geodesic orbits, surfaces of constant $P^{(0)}_A$ define 2D tori in phase space, and the biperiodic orbits wrap around the surfaces of these tori. We can define action variables $J^{(0)}_A=(J_{r(0)},J_{\phi(0)})$ as integrals over the tori,
\begin{equation}\label{eq:J0 def}
J^{(0)}_A  = \frac{1}{2\pi}\oint_{{\cal C}^{(0)}_A} p_B\, dx^B_p,    
\end{equation}
where ${\cal C}^{(0)}_A$ denotes one of the two topologically distinct loops on the torus. The action variables $J^{(0)}_A$ are  canonically conjugate (at 0SF) to the geodesic action angles $\varphi^{(0)}_A$, meaning that the geodesic frequencies can be written as $d\varphi_A^{(0)}/dt:=\Omega^{(0)}_A = \partial H^{(0)}_{\rm 4D}/\partial J^{(0)}_A$. Since $p_\phi=L^{(0)}$ is constant on each torus, we also have $J^{(0)}_\phi = L^{(0)}$. Note that $J_{r(0)}$ can also be written as a function of $P^{(0)}_A$; see Eq.~\eqref{eq:J0r} below. One can then work with whichever set of constants is convenient: $\pi^{(0)}_A$, $P^{(0)}_A$, or $J^{(0)}_A$.

Reference~\cite{Lewis:2025ydo} extended this analysis to the full 1PA dynamics, including dissipation. Specifically, Ref.~\cite{Lewis:2025ydo} showed that the evolution equations~\eqref{eq:phidot} and \eqref{eq:pidot} are equivalent (at 1PA order) to
\begin{align}
    \frac{d\varphi_A}{dt} &= \frac{\partial H_{\rm 4D}}{\partial J_A},\\
    \frac{dJ_A}{dt} &= \varepsilon\left[G^{(0)}_A(J_B) + \varepsilon G^{(1)}_A(J_B)\right],
\end{align}
where
\begin{equation}
    H_{\rm 4D}(J_A,\varepsilon) = E_{(0)}(J_A) - \frac{1}{4}\varepsilon m_2\left\langle\frac{d\tau}{dt} h_{uu}^{\mathrm{R}(1)}\right\rangle (J_A)
\end{equation}
is a Hamiltonian that governs the 1SF conservative sector, and $G^{(n)}_A$ are dissipative forcing functions that can be written in terms of a \emph{pseudo}-Hamiltonian (a certain two-point function on phase space that will not be needed in this paper). Here we have introduced the shorthand $h_{uu}^{\mathrm{R}(1)}:=h_{\alpha\beta}^{\mathrm{R}(1)}u^\alpha u^\beta$ as well as angular brackets to denote a torus average: $\langle \cdot \rangle := \frac{1}{(2\pi)^2}\oint \cdot\, d^2\varphi$. Due to the axial symmetry of the background, all functions we average will be independent of $\varphi_\phi$ when evaluated on the worldline, reducing the average to an average over one radial cycle, $\langle \cdot \rangle = \frac{1}{2\pi}\oint \cdot d\varphi_r$. We can equivalently write this as a time average over a radial period $T_r$: for any function $f(\varphi_r,J_A)$, we have
\begin{equation}
    \langle f \rangle(J_A) = \frac{1}{T_r}\int_{0}^{T_r} f(\Omega_r t,J_A) dt.
\end{equation}
The action variables can equivalently be replaced by quasi-Keplerian parameters $\pi_A$ in this equality.

The action variables $J_A$ are the unique variables canonically conjugate to $\varphi^A$ with respect to the Hamiltonian $H_{\rm 4D}$. In the conservative sector, they are constants, implying that conservative 1SF motion is restricted to tori of constant $J_A$, with $\varphi_A$ as coordinates on each torus. $J_A$ can trivially be written as integrals over these tori: 
\begin{equation}
J_A = \frac{1}{2\pi}\oint_{{\cal C}_A}J_B d\varphi_B = \frac{1}{2\pi}\oint_{{\cal C}_A}p^{\rm can}_B dx^B_p,
\end{equation}
where ${\cal C}_A$ are the two independent loops on the torus, and $p^{\rm can}_B$ (which we will not require) is canonically conjugate to $x^A_p$ with respect to $H_{\rm 4D}$. In the dissipative system, the system slowly evolves from one to torus to the next.

This Hamiltonian construction provides natural definitions of energy and angular momentum. As at 0SF order, the 1SF energy $E$ is the on-shell value of $H_{\rm 4D}$ (which, we note, is simply equal to $H_{\rm 4D}$'s value as a function of the action variables, since the Hamiltonian is independent of the angle variables). Similarly, the angular momentum is the azimuthal action variable, $L=J_\phi$. We can again work with any choice of variables: $\pi_A$, $P_A=(E,L)$, or $J_A$.

The conservative dynamics is encoded in the torus-averaged Detweiler-Barack-Sago redshift\footnote{\label{footnote:zSchw}We avoid labeling the terms here with $(0)$ and $(1)$ because they do not correspond to an expansion in powers of $\varepsilon$ at fixed phase-space coordinates. $z$ and $z_{\rm Schw}$ represent the two redshifts $d\tilde\tau/dt$ and $d\tau/dt$ along the \emph{same} accelerated orbit.} $\langle z\rangle=\langle d\tilde\tau/dt\rangle = \langle z_{\rm Schw} + \varepsilon z_{\rm 1SF}\rangle$~\cite{Detweiler:2008ft,Barack:2011ed}, where $\tilde\tau$ is proper time in the effective metric $\tilde g_{\alpha\beta}$, $z_{\rm Schw}=d\tau/dt= 1/u^t$, and 
\begin{equation}
z_{\rm 1SF}=-\frac{1}{2} z_{\rm Schw} h_{uu}^{\mathrm{R}(1)}.
\end{equation}
We can use the averaged redshift as a generating function for all of our quantities of interest. For this purpose, it is useful to expand $\langle z \rangle$ in powers of $\varepsilon$ at fixed values of the frequencies, where we use $(\Omega_A,\varphi_A)$ as phase-space coordinates and write $x^A_p=x^A_{(0)}(\Omega_B,\varphi_B) + \varepsilon x^A_{(1)}(\Omega_B,\varphi_B)$, leading to
\begin{align}
    \langle {z} \rangle({\Omega}_A,\varepsilon) = \langle {z}_{(0)} \rangle({\Omega}_A) + \varepsilon \langle {z}_{(1)} \rangle({\Omega}_A).
\end{align}
Here $x^A_{(0)}$ is the same function of $(\Omega_A,\varphi_A)$ as for a geodesic with those frequencies, $\langle {z}_{(0)} \rangle$ is $\langle z_{\rm Schw}\rangle$ evaluated at $x^A_p=x^A_{(0)}$, and
\begin{align}\label{eq:<z1>}
    \langle z_{(1)} \rangle = -\frac{1}{2} \langle z_{\rm Schw} h_{uu}^{\mathrm{R}(1)} \rangle.
\end{align}
The principal advantage of this expansion at fixed frequencies is that there are no additional terms at linear order in $\varepsilon$, while expansions at fixed values of other phase-space coordinates will generically contain additional ${\cal O}(\varepsilon)$ terms from the expansion of $z_{\rm Schw}$; see Appendix~D of Ref.~\cite{Lewis:2025ydo}.

It will be useful to work with dimensionless functions of dimensionless arguments rather than leaving functional dependence on $m_1$ implicit, as we have done up to this point. We define reduced, dimensionless counterparts to energy $E$, $L$, and $J_r$ via
\begin{align}
    E &= m_2 \hat{E}, 
    & 
    L &= m_1 m_2 \hat{L}, 
    &
    J_r &= m_1 m_2 \hat{J}_r,
\end{align}
along with dimensionless frequencies  \mbox{$\hat\Omega_A=m_1\Omega_A$.} 
Expanding $\hat P_A$ and $\hat J_A$ in powers of $\varepsilon$ at fixed $\hat\Omega_A$ yields
\begin{subequations} \label{eqn:ELJrSF}
\begin{align} \label{eqn:ESF}
    \hat{E}(\hat{\Omega}_A,\varepsilon) &= \hat{E}_{(0)}(\hat{\Omega}_A) + \varepsilon \hat{E}_{(1)}(\hat{\Omega}_A) , 
    \\
    \hat{L}(\hat{\Omega}_A,\varepsilon) &= \hat{L}_{(0)}(\hat{\Omega}_A) + \varepsilon  \hat{L}_{(1)}(\hat{\Omega}_A) , 
    \\ \label{eqn:JrSF}
    \hat{J}_r(\hat{\Omega}_A,\varepsilon) &= \hat{J}_{r(0)}(\hat{\Omega}_A)  + \varepsilon  \hat{J}_{r(1)}(\hat{\Omega}_A),
\end{align}
\end{subequations}
where there is now no hidden dependence on $m_1$. 
The 1SF terms in these expansions, which will be the focus of this paper, are given in terms of the redshift by~\cite{Lewis:2025ydo}
\begin{subequations}\label{eqn:E1J1Jr1SF}
\begin{align} \label{eqn:E1SF}
    \hat{E}_{(1)}(\hat{\Omega}_A) &= \frac{1}{2}\left(\langle z_{(1)} \rangle - \hat{\Omega}_A \,\frac{\p \langle z_{(1)} \rangle}{\p \hat{\Omega}_A}\right), 
    \\
    \hat{L}_{(1)}(\hat{\Omega}_A) &=  - \frac{1}{2} \frac{\p \langle z_{(1)} \rangle}{\p \hat{\Omega}_\phi}, 
    \\ \label{eqn:Jr1SF}
    \hat{J}_{r(1)}(\hat{\Omega}_A) &=  -\frac{1}{2} \frac{\p \langle z_{(1)} \rangle}{\p \hat{\Omega}_r},
\end{align}
\end{subequations}
where $\langle z_{(1)}\rangle=\langle z_{(1)}\rangle(\hat\Omega_A)$.

\section{Geodesic order}

\label{sec:geo}

In this section, we review the case of geodesic orbits in more detail, particularly the quasi-Keplerian parametrization and the relationships between the various constants of motion, largely following Ref.~\cite{Barack:2010tm}. We also carry out PN expansions of the geodesic relationships, defining alternative frequency variables that are useful for that purpose. 

\subsection{Exact expressions}
\label{sec:geoExact}

Rather than writing the geodesic motion directly in terms of the phase-space coordinates $(\varphi_A,\pi_A)$ used in waveform generation, we can more straightforwardly write it in terms of the (Darwin) relativistic anomaly $\chi$. In terms of this radial phase, Darwin's quasi-Keplerian parametrization of the geodesic radial motion is~\cite{Darwin:1961}
\begin{align}
    r_{(0)}(\chi,\pi_A) &= \frac{p m_1}{1+e \cos\chi} \,.
\end{align}
The orbital parameters $(p,e)$ are expressible in terms of the (geodesic) radii of periastron $r_{\rm min}$ and apastron $r_{\rm max}$ as
\begin{align}\label{eq:p_e_def_geo}
    p &= \frac{2 r_{\rm min} r_{\rm max}}{m_1(r_{\rm max}+r_{\rm min})} \,, &
    e &= \frac{r_{\rm max} - r_{\rm min}}{r_{\rm max}+ r_{\rm min}} \,.
\end{align}
The constants of motion can then be expressed in terms of $p$ and $e$. Explicitly, the reduced energy and angular momentum are
\begin{subequations}\label{eqn:ELofPE}\begin{align}
    \hat{E}_{(0)}(\pi_A) &= \sqrt{\frac{(p-2+2e)(p-2-2e)}{p(p-3-e^2)}} \,, \\
    \hat{L}_{(0)}(\pi_A) &= \frac{p}{\sqrt{p-3-e^2}} \,,
\end{align}\end{subequations}
and we define the semilatus rectum at the separatrix as
\begin{align}
    p_{\star} = 6+2e\,.
\end{align}
In the region of parameter space where $p>p_{\star}$ and \mbox{$0<e<1$,} the determinant of the Jacobian matrix between $(\hat{E}_{(0)},\hat{L}_{(0)})$ and $(p,e)$ is regular, so the map between these two pairs of variables is one-to-one.
The reduced radial action $\hat{J}_{r(0)}(\pi_A)$ is given by Eq.~\eqref{eq:J0 def}:
\begin{equation}\label{eq:J0r}
    \hat J_{r(0)} = \frac{1}{\pi}\int_{\hat{r}_{\rm min}}^{\hat{r}_{\rm max}}  d\hat r \,\frac{\sqrt{\hat r^4\hat E^2-(\hat r^2-2\hat r)(\hat L^2+\hat r^2)}}{\hat r^2-2\hat r}  \,,
\end{equation}
where we define the dimensionless radius $\hat r=r_{(0)}/m_1$ and use \mbox{$g^{rr}_{(0)} u_r^2=-1-g^{tt}_{(0)}u_t^2-g^{\phi\phi}_{(0)}u_\phi^2$} to rewrite the integrand.

The evolution of the coordinate time, azimuthal phase, and proper time along the orbit can be written as
\begin{subequations}\begin{align}
    \frac{\dd t_{(0)}}{\dd \chi} &=  \frac{ m_1\, p^2}{(p-2-2e \cos \chi)(1+e \cos\chi)^2}  \nonumber\\*
    &\qquad\times\sqrt{\frac{(p-2-2e)(p-2+2e)}{p-6 -2e \cos \chi}} \,,\\
    \frac{\dd \phi_{(0)}}{\dd \chi} &= \sqrt{\frac{p}{p-6-2e \cos\chi}} \,,\\
    \frac{\dd \tau_{(0)}}{\dd \chi} &= \frac{ m_1\, p^{3/2}}{(1+e\cos\chi)^2}\sqrt{\frac{p-3-e^2}{p-6-2e\cos\chi}}\,.
\end{align}\end{subequations}
Thus, the cumulative change of these times and phases over one radial period are
\begin{subequations}\label{eq:T Phi Tau}\begin{align}
    T_{r(0)} &= \!\int_0^{2\pi} \!\!\dd \chi \,\frac{\dd t_{(0)}}{\dd \chi}\,, \\
    \Phi_{(0)} &= \!\int_0^{2\pi} \!\! \dd \chi \,\frac{\dd \phi_{(0)}}{\dd \chi} \,,\\
    \mathcal{T}_{r(0)} &= \! \int_0^{2\pi} \!\! \dd \chi \,\frac{\dd \tau_{(0)}}{\dd \chi} \,.
\end{align}   
\end{subequations}
These definite integrals can be expressed in terms of elliptic integrals~\cite{Fujita:2009bp,VanDeMeent:2018cgn}.
Introducing the reduced periods $\hat{T}_r = T_r/m_1$ and $\hat{\mathcal{T}}_r = \mathcal{T}_r/m_1$ and restricting to the case of bound stable orbits $p >p_{\star}$, we find that at geodesic order they are given by 
\begin{subequations}\label{eqn:periods}
\begin{align}
    \hat T_{r(0)} &= \frac{2  p \, a_+  \,  a_-  \,  b_+}{(1-e^2)(p-4)} \dE(m_r) - \frac{2 p a_+ a_-}{ (1-e^2) b_+ }  \dK(m_r) \nonumber\\*
    & + \frac{4 a_+  \, a_- \! \left[p(p-1-3e^2) \! - \! 8(1-e^2)\right] }{(1-e)(1-e^2)(p-4)b_+}  \Pi \!\left(\! - \frac{2e}{1-e}\Big| m_r \!\!\right) \nonumber\\*
    & + \frac{16 a_-}{a_+ b_+} \Pi\left(\frac{4e}{p-2+2e}\Big| m_r\right),\\
    \Phi_{(0)} &=   \frac{4\sqrt{p}}{b_+} \dK(m_r), \\
    \hat{\mathcal{T}}_{r(0)} &=   \frac{2 p^{3/2} b_+ c_-}{(1-e^2)(p-4)} \dE\left(m_r\right)  - \frac{2 p^{3/2} c_-}{(1-e^2) b_+} \dK\left(m_r\right)  \nonumber\\*
    & + \frac{4 p^{3/2} c_-^3}{(1-e^2)(1-e)(p-4) b_+}  \Pi\left(-\frac{2e}{1-e} \Big| m_r \right), 
\end{align}\end{subequations}
where we have introduced the shorthands \mbox{$a_{\pm} = \sqrt{p-2 \pm 2e}$}, \mbox{$b_{\pm} = \sqrt{p-6 \pm 2e}$}, \mbox{$c_{\pm}=\sqrt{p-3\pm e^2}$}, and \mbox{$m_r = 4e/b_+^2$}. The elliptic integrals $\dE(\cdot)$, $\dK(\cdot)$, and $\Pi(\cdot|\cdot)$ are defined in \cref{app:elliptic} for completeness. 
To our knowledge, the above expression for $\mathcal{T}_r$ has never appeared in the literature, but the expressions for $T_r$ and $\Phi$ were known~\cite{VanDeMeent:2018cgn}. 

From the coordinate-time and proper-time periods and elapsed azimuthal angle, we can also define the orbital frequencies and orbit-averaged redshift
\begin{align} \label{eqn:OmegasofPE}
    \hat{\Omega}_r &= \frac{2\pi}{\hat{T}_r},
    &
    \hat{\Omega}_\phi &= \frac{\Phi}{\hat{T}_r},
    &
    \langle z_{\rm Schw} \rangle &= \frac{\hat{\mathcal{T}}_r}{\hat{T}_r}.
\end{align}
At geodesic order these are expressed in terms of $(p,e)$ using the formulas for $\hat T_{r(0)}$, $\Phi_{(0)}$, and $\hat{\cal T}_{r(0)}$ above:
\begin{align} \label{eqn:OmegasofPE_0}
    \hat{\Omega}_r^{(0)} &= \frac{2\pi}{\hat{T}_r^{(0)}},
    &
    \hat{\Omega}_\phi^{(0)} &= \frac{\Phi^{(0)}}{\hat{T}_r^{(0)}},
    &
    \langle z_{(0)} \rangle &= \frac{\hat{\mathcal{T}}_r^{(0)}}{\hat{T}_r^{(0)}}.
\end{align}
Note that alternative formulas for the frequencies exist in the literature (e.g., in the \texttt{Black Hole Perturbation Toolkit}~\cite{BHPToolkit}); they can be reconciled with our expressions using the relations~\eqref{eq:Pi_rule} and \eqref{eq:Pi_rule_applications} between elliptic integrals. 

The reduced radial action~\eqref{eq:J0r} can also be written directly in terms of elliptic integrals, as in Ref.~\cite{Witzany:2024ttz}. Equivalently, we can write it in terms of quantities given above by using the normalization condition $\hat\omega^{(0)}_\alpha \hat J^{(0)}_\alpha=-1$~\cite{Fujita:2016igj}, where $\hat \omega^{(0)}_\alpha := \langle z_{\rm (0)}\rangle^{-1}\hat\Omega^{(0)}_\alpha$ are the reduced orbital frequencies with respect to proper time, $\hat J_{t(0)} := - \hat E_{(0)}$, and $\hat\Omega^{(0)}_t:=1$. %
Rearranging the normalization condition yields
\begin{align} \label{eqn:Jrgeo}
    \hat{J}_{r(0)} = \frac{1}{2\pi}\left(\hat{T}_{r(0)}\ \hat{E}_{(0)} - \Phi_{(0)} \hat{L}_{(0)}- \hat{\mathcal{T}}_{r(0)}\right)\,.
\end{align}
We have checked that this expression is in agreement with Eq.~(34)  of Ref.~\cite{Witzany:2024ttz} in the appropriate limit.

The parameters $(p,e)$ as defined in Eq.~\eqref{eq:p_e_def_geo} provide globally valid phase-space coordinates everywhere outside the separatrix between bound and plunging orbits, which lies on the curve $p=p_{\star}$. For given initial values of $\phi_{(0)}$ and $\chi_{(0)}$, $p$ and $e$ uniquely specify a geodesic. However, even when restricting to stable bound orbits, for given initial $\phi_{(0)}$ and $\chi_{(0)}$, there exist \text{two} physically inequivalent orbits associated with the pair of frequencies $(\Omega_r, \Omega_\phi)$; this is the so-called \textit{isofrequency} pairing~\cite{Warburton:2013yj}. The isofrequency degeneracy curve, which separates the parameter space into two domains of such inequivalent orbits, is given by the condition 
\begin{align}
    \mathrm{det}\, \bm{J}  = 0 \,,
\end{align}
where the Jacobian matrix is defined by
\begin{align}
     \bm{J} = \begin{pmatrix}
        \frac{\partial \hat{\Omega}_r}{\partial  p} & \frac{\partial \hat{\Omega}_r}{\partial  e} \\
        \frac{\partial \hat{\Omega}_\phi}{\partial  p} & \frac{\partial \hat{\Omega}_\phi}{\partial  e}
    \end{pmatrix} \,.
\end{align}
This singular curve in parameter space will be highly relevant in our calculations of 1SF quantities. We will denote the relation between $p$ and $e$ on the isofrequency curve by $p_{\rm iso}(e)$.

\subsection{Post-Newtonian expansion}
\label{sec:geoPN}

The PN expansion is an expansion in small velocity, or equivalently, large separation. Here, it will translate to an expansion for $p \gg 1$ at fixed $e$ in the range $0\le e <1$. Alternatively, we can introduce another set of parameters that are more directly  related to the frequencies:
\begin{align}\label{eq:def_y_lambda}
    y =  \hat\Omega_\phi^{2/3} &&\text{and}&& \lambda = \frac{3 y}{\hat{\Omega}_\phi/\hat{\Omega}_r-1} \,, 
\end{align}
such that the PN expansion now corresponds to $y \ll 1$ with $0 < \lambda \le 1$.

Starting from Eqs.~\eqref{eqn:ELofPE} and \eqref{eqn:OmegasofPE}, we perform a PN expansion of the fundamental frequencies, the redshift, and the constants of motion. For this, we use the expansions of the elliptic integrals provided in Eq.~\eqref{eq:PNseries_E_K_Pi}.   The first few PN orders read
\begin{subequations}\label{seq:Omega_r_Omega_phi_PN_pe}\begin{align}\label{seq:Omega_r_PN_pe}
    \hat{\Omega}^{(0)}_r &= \left(\frac{1-e^2}{p}\right)^{3/2}\left[1- \frac{3(1-e^2)}{p}  + \mathcal{O}\!\left(\frac{1}{p^2}\right)\right], \\
    \label{seq:Omega_phi_PN_pe}
    \hat{\Omega}^{(0)}_\phi &= \left(\frac{1-e^2}{p}\right)^{3/2}\left[1+ \frac{3 e^2}{p}  + \mathcal{O}\!\left(\frac{1}{p^2}\right)\right],   \\
    \langle z_{(0)}\rangle &=  1 - \frac{3(1-e^2)}{2p}  + \mathcal{O}\!\left(\frac{1}{p^2}\right) \,,\\
    \hat{E}_{(0)} &=  - \frac{1-e^2}{2p} + \frac{3(1-e^2)^2}{8p^2} + \mathcal{O}\!\left(\frac{1}{p^3}\right) \,,\\
    \hat{L}_{(0)} &= \sqrt{p} + \frac{3+e^2}{2\sqrt{p}}  + \mathcal{O}\!\left(\frac{1}{p^{3/2}}\right) \,,\\
    \hat{J}_{r(0)} &=  \sqrt{p}\left[\frac{1}{\sqrt{1-e^2}}-1\right] + \mathcal{O}\!\left(\frac{1}{\sqrt{p}}\right) \,.
\end{align}\end{subequations}
The expansions are provided in the ancillary file \texttt{PN\_expressions.csv}~\cite{supp} at relative 12PN for $\hat{\Omega}_r^{(0)}$ and $\hat{\Omega}_\phi^{(0)}$; 10PN for $\langle z_{(0)}\rangle$; and 9PN for $\hat{E}_{(0)}$, $\hat{L}_{(0)}$, and $\hat{J}_{r(0)}$. This translates to (geodesic-order) expansions for the reduced frequency variables which read
\begin{subequations} \label{eqn:PNmap}
\begin{align}
    y_{(0)} &= \frac{1-e^2}{p}\left[1+ \frac{2 e^2}{p}  + \mathcal{O}\!\left(\frac{1}{p^2}\right)\right], \\
    \lambda_{(0)} &= (1-e^2)\left[1+ \frac{1}{p}\left(- \frac{9}{2} + \frac{7}{4}e^2 \right)  + \mathcal{O}\!\left(\frac{1}{p^2}\right)\right].
\end{align} 
\end{subequations}
The expansions for $y$ and $\lambda$ are provided, respectively, at 12PN and 11PN order in the ancillary file \texttt{PN\_expressions.wl}~\cite{supp}. Indeed, there is a loss of one PN order in $\lambda$ due to the degeneracy between radial and azimuthal frequencies at Newtonian order; see Ref.~\cite{Trestini:2025yyc} for a discussion.
Inverting the latter expansion, one finds that 
\begin{subequations}\begin{align}
    p &= \frac{\lambda_{(0)}}{y_{(0)}}\left[1 + y_{(0)}\left(- \frac{1}{4} + \frac{19}{4 \,\lambda_{(0)}}\right) + \mathcal{O}\bigl(y^2_{(0)}\bigr)\right] \,,\\
    e &= \sqrt{1-\lambda_{(0)}}\Biggl[1- \frac{y_{(0)}}{1-\lambda_{(0)}}\left( \frac{11}{8} + \frac{7}{8 \, \lambda_{(0)}}\right) + \mathcal{O}\bigl(y^2_{(0)}\bigr)\Biggr],
\end{align}\end{subequations}
where these expansions for $p$ and $e$ are provided, respectively, at 12PN and 11PN order in the  ancillary file \texttt{PN\_expressions.wl}~\cite{supp}.

Our 1SF results will focus on parametrizations in terms of $(p,e)$, and our comparisons with numerical results will indicate that the PN expansion is significantly more accurate when written as a large-$p$ expansion at fixed $e$ rather than a small-$y$ expansion at fixed $\lambda$. However, comparisons with results in strict PN theory are far simpler in terms of $(y,\lambda)$. For this reason, we also provide high-order PN expansions in terms of $(y,\lambda)$ in the ancillary file \texttt{PN\_expressions.wl}~\cite{supp}.

\section{First order}
\label{sec:1SF}

In this section, we numerically and analytically compute the 1SF corrections to the conserved quantities $(\hat{E}, \hat{L},\hat{J}_r)$ as functions of the frequencies $\hat{\Omega}_A$ across much of the bound-orbit parameter space, starting from the formulas~\eqref{eqn:E1J1Jr1SF} that express these functions in terms of the 1SF redshift correction $\langle z_{(1)}\rangle$. We obtain our numerical results on a dense grid up to $e=0.6$ and $p=p_{\star}+200$, suitable for use in waveform-generation codes, and we perform a detailed comparison with the analytical, 9PN results across that range. Our 9PN results also establish consistency with the 4PN results obtained independently within PN theory in Ref.~\cite{Trestini:2025yyc}.

\subsection{Quasi-Keplerian variables at 1SF order}

We are ultimately interested in the expressions (or numerical values) for the constants of motion as functions of the orbital frequencies, which represent gauge-invariant relationships between gauge-invariant quantities.   
But it is still useful to make use of the intermediate orbital parameters $\pi_A=(p,e)$. These quantities provide an intuitive geometric description of the leading-order dynamics, and  therefore  are a natural choice for tiling the parameter space and for parametrizing 1SF calculations. (See Sec.~\ref{sec:redshift-invariant}.) They are also the parameters used to store offline data for practical waveform generation, as reviewed in Sec.~\ref{sec:formalism}. Finally, as we will see in Sec.~\ref{sec:redshift-invariant}, the PN expansion of the redshift is  numerically more accurate when using $(p,e)$ rather than other parameters such as~$(y,\lambda)$.

However, there is considerable freedom in specifying the meaning of $(p,e)$ at 1SF order. Specifically, the multiscale expansion and waveform-generation framework are invariant under a transformation $\pi_A \to \pi_A + \varepsilon \delta \pi_A(\hat J_B)$ (i.e., a near-identity transformation that is independent of the angles $\varphi_A$)~\cite{Mathews:2025nyb}. This freedom was explored in the context of multiscale expansions of the full, dissipative system in Refs.~\cite{VanDeMeent:2018cgn,Pound:2021qin,Mathews:2025nyb,Lewis:2025ydo}, for example. In Refs.~\cite{Piovano:2024yks,Mathews:2025nyb}, two choices were highlighted that are particularly natural for us:\footnote{Another choice is to define $(p,e)$ in terms of the radial turning points of the perturbed dynamics. This is a natural extension of their geodesic definitions, but it depends on the choice of background coordinates, is only well defined in the absence of dissipation, and ties $(p,e)$ to the gauge of the metric perturbation $h^{(1)}_{\mu\nu}$, making it difficult for cross-gauge comparisons or determining their relations to $(\hat{E}, \hat{L}, \hat{J}_r)$ or $(\hat{\Omega}_r, \hat{\Omega}_\phi)$.}
\begin{enumerate}
    \item The \emph{fixed-constants} gauge, in which $(p,e)$ are related to $(\hat{E}, \hat{L})$ by the Schwarzschild-geodesic relationships~\eqref{eqn:ELofPE},
    \item The \emph{fixed-frequencies} gauge, in which $(p,e)$ are related to $(\hat{\Omega}_r, \hat{\Omega}_\phi)$ by the Schwarzschild-geodesic relationships~\eqref{eqn:OmegasofPE}.
\end{enumerate}
In other words, in the fixed-constants gauge, we have $\hat P_A(\pi_B,\varepsilon)=\hat P^{(0)}_A(\pi_B)$, independent of $\varepsilon$; in the fixed-frequencies gauge, we have $\hat \Omega_A(\pi_B,\varepsilon)=\hat \Omega^{(0)}_A(\pi_B)$.

In this work, we choose the latter, making it straightforward to parametrize $(\hat{E}, \hat{L}, \hat{J}_r)$ with $(p,e)$ in place of $(\hat{\Omega}_r, \hat{\Omega}_\phi)$ in Eq.~\eqref{eqn:ELJrSF}; this definition was previously adopted, e.g., in Eq.~(71) of Ref.~\cite{Barack:2011ed}. The fixed-constants gauge has the analogous advantage when calculating $\hat\Omega_A$ as a function of $\hat P_A$, rather than the converse:
\begin{equation}\label{eq:Omega(E,L)}
    \hat\Omega_A(\hat P_B,\varepsilon) =  \hat \Omega^{(0)}_A(\hat P_B) + \varepsilon \,\hat \Omega^{(1)}_A(\hat P_B).
\end{equation}
Here, we can use $(p,e)$ in place of $\hat P_A$ if we adopt a fixed-constant gauge. 

We focus on calculations of the corrections to the constants of motion at fixed frequencies, $\hat P^{(1)}_A(\hat\Omega_B)$; but for waveform generation, the converse, $\hat \Omega^{(1)}_A(\hat P_B)$, might be more useful; see Sec.~\ref{subsec:waveform_gen} below. Relating the two is straightforward. Substituting Eq.~\eqref{eq:Omega(E,L)} into Eq.~\eqref{eqn:ELJrSF} and expanding, we obtain
\begin{equation}\label{eq:Omega1}
    \Omega^{(1)}_A(\hat P_B) = -\frac{\partial \hat\Omega^{(0)}_A}{\partial\hat P_B}\hat P^{(1)}_B.
\end{equation}

\subsection{Redshift invariant}
\label{sec:redshift-invariant}

We compute $h_{\mu\nu}^{\mathrm{R}(1)}$, and subsequently $\langle z_{(1)} \rangle$, in a Fourier decomposition and an expansion in spherical harmonics, using the Python library \texttt{pybhpt}\footnote{\url{https://pybhpt.readthedocs.io/en/latest/}} \cite{pybhpt-0.9.8} and the methods outlined in Ref.~\cite{Nasipak:2025tby}. Crucially, with this approach, $\langle z_{(1)} \rangle$ reduces to a formally infinite sum over the spherical-harmonic $l$-modes,
\begin{align}
    \langle z_{(1)} \rangle = \sum_{\ell = 0}^\infty \langle z^\ell_{(1)} \rangle.
\end{align}
The convergence of the sum is impacted by the regularization procedure that is employed to obtain  $h_{uu}^\mathrm{R(1),\ell}$ and subsequently $\langle z^\ell_{(1)} \rangle$. For the mode sum regularization used in Ref.~\cite{Nasipak:2025tby}, $\langle z^\ell_{(1)} \rangle \sim \ell^{-2}$ as $\ell \rightarrow \infty$. Thus, truncating the mode sum at some finite $\ell_\mathrm{max}$ introduces an error $\sim \ell_\mathrm{max}^{-1}$. To accelerate the convergence of the mode-sum and improve the truncation error, we employ the same large-$\ell$ fitting procedure as described in Ref.~\cite{Nasipak:2025tby}, which produces an estimated value and error for $\langle z_{(1)} \rangle$. This fitting error is the dominant source of error in $\langle z_{(1)} \rangle$.

Consequently, we store the $\ell$-modes $\langle z^\ell_{(1)} \rangle$, the fitted value of $\langle z_{(1)} \rangle$, and its estimated error on a two-dimensional grid in $(p,e)$. Defining $\delta \equiv p - 6-2e$, the bounds of the domain are set by
\begin{align*}
    \delta_\mathrm{min} &= 10^{-3},
    &
    \delta_\mathrm{max} &= 2\times 10^{2},
    &
    e_\mathrm{max}=0.6,
\end{align*}
so that $\delta \in [\delta_\mathrm{min}, \delta_\mathrm{max}]$ and $e \in [0, e_\mathrm{max}]$.
Rather than sampling directly in $(p,e)$, we make use of intermediate coordinates $(u,w)$ defined by
\begin{subequations} \label{eqn:peToUW}
    \begin{align}
    u(p,e) &= \frac{\log \delta - \log \delta_\mathrm{min}}{\log \delta_\mathrm{max} - \log \delta _\mathrm{min}},
    \\
    w(p,e) &= \left[ \frac{e}{e_\mathrm{max}} \right]^{1/2},
\end{align}
\end{subequations}
to better capture the behavior near the separatrix.

The averaged redshift invariant is also known analytically as a PN expansion. It is provided in terms of $(p,e)$ at 10PN order in Ref.~\cite{Munna:2022gio}; see Sec.~V.B. for the 8.5PN result and  the \texttt{Black Hole Perturbation Toolkit} \cite{BHPToolkit} for the full 10PN expression (we reproduce the full result in our ancillary file \texttt{PN\_expressions.wl}~\cite{supp}). The conventions in that reference are the same as here: fixed frequency, with $(p,e)$ geodesically related to $(\Omega_r,\Omega_\phi)$. The redshift is exact in eccentricity up to 3.5PN; at 4PN, it is expressed in terms of an enhancement function $\Lambda_0(e)$, which is known exactly as an infinite sum over Bessel modes, and whose resummation is rigorously established and very accurate for all values of the eccentricity \cite{Munna:2022gio,Trestini:2025yyc}; beyond 4.5PN, some elements are controlled exactly and others are empirically resummed. The final resummed redshift is expressed in terms of regular functions that can be safely replaced by their small-$e$ expansion without spoiling the large-$e$ behavior of the redshift itself: for instance, the \texttt{chi4PNConv} function
corresponds to $-\frac{3}{2}\lambda_0(e)$, as defined in Ref.~\cite{Trestini:2025yyc}. In this work, we have replaced all of these regular functions with their expansions up to $e^{20}$.
The PN-expanded redshift finally reads 
\begin{align}\label{eq:z1_PN_pe}
    \langle z_{(1)} \rangle =  \frac{1-e^2}{p}\left\{1 - \frac{1-e^2}{p} + \mathcal{O}\left(\frac{1}{p^2}\right) \right\},
\end{align}
and we provide the full 10PN expression in the ancillary file \texttt{PN\_expressions.wl}~\cite{supp}. Using the PN map provided in \cref{eqn:PNmap}, we reexpress the redshift invariant in terms of the reduced frequencies $(y,\lambda)$. It reads 
\begin{align}
    \langle z_{(1)} \rangle =  y \left\{ 1 + y\left(1- \frac{2}{\lambda}\right) + \mathcal{O}(y^2) \right\},
\end{align}
and we provide the full 10PN  expression in the ancillary file \texttt{PN\_expressions.wl}~\cite{supp}. Note that this result agrees perfectly (at linear order in the mass ratio) with the 4PN redshift 
recently computed by one of us using traditional PN theory \cite{Trestini:2025yyc}. We additionally provide the partial derivatives of the redshift---namely $\partial_p \langle z \rangle$ and $\partial_e \langle z \rangle$---in the ancillary file \texttt{PN\_expressions.wl}~\cite{supp}.

Since the 1SF corrections to the constants of motion involve derivatives of the redshift, we also need to compute the latter numerically from our data. To compute numerical derivatives with respect to $p$ and $e$, we first subtract the 10PN expressions and normalize by the leading-order behavior, leading to
\begin{align}
    Z_{(1)} =  \frac{p}{(1-e^2)}\left[\langle z_{(1)}\rangle - \langle z_{(1)}\rangle^\mathrm{10PN} \right].
\end{align}
We then compute $\partial_u Z_{(1)}$ and $\partial_w Z_{(1)}$ using finite-difference stencils applied to the values of $Z_{(1)}$ on the $(u,w)$ grid. Two sources of uncertainty are considered:
\begin{enumerate}
\item The propagation of fitting uncertainties in $\langle z_{(1)}\rangle$.
\item Finite-difference truncation errors arising from the stencil order and step size.
\end{enumerate}
The fitting uncertainties are propagated in quadrature through the finite-difference stencils. Finite-difference errors are estimated by computing derivatives using four stencils: fourth-order stencils with step sizes $(\Delta u, \Delta w)$ and $(2\Delta u, 2\Delta w)$, and sixth-order stencils with the same step sizes. If all four results are consistent within their uncertainties, Richardson extrapolation is applied to the two fourth-order estimates to obtain the final derivatives, with uncertainties propagated in quadrature. If the four results are not consistent, the finite-difference uncertainty is estimated from the variance among the four estimates, and Richardson extrapolation of the two sixth-order results is used to determine the final derivative values at each grid point.

After computing $\partial_u Z_{(1)}$ and $\partial_w Z_{(1)}$, it is straightforward to obtain
\begin{subequations}\begin{align}
    \partial_p Z_{(1)} &= \frac{\partial u}{\partial p}\partial_u Z_{(1)},
    \\* \label{eqn:dedz1}
    \partial_e Z_{(1)} &= \frac{\partial u}{\partial e} \partial_u Z_{(1)} + \frac{\partial w}{\partial e} \partial_w Z_{(1)},
\end{align}\end{subequations}
and likewise,
\begin{subequations}\begin{align}
    \partial_p \langle z_{(1)} \rangle &= \frac{(1-e^2)}{p} \partial_p Z_{(1)} - \frac{(1-e^2)}{p^2} Z_{(1)} + \partial_p \langle z_{(1)} \rangle^\mathrm{PN},
    \\* 
    \partial_e \langle z_{(1)} \rangle &= \frac{(1-e^2)}{p} \partial_e Z_{(1)} - \frac{2e}{p} Z_{(1)} + \partial_e \langle z_{(1)} \rangle^\mathrm{PN}.
\end{align}\end{subequations}
Once again, we propagate errors by quadrature to estimate the numerical uncertainty in $\partial_p \langle z_{(1)} \rangle$ and $\partial_e \langle z_{(1)} \rangle$.

We find that derivatives with respect to $e$ require separate treatment at $e=0$. Because $\partial w/\partial e$ is singular at this point, we instead compute numerical derivatives with respect to $w^{2}$ using the same procedure as above, employing the highest-order finite-difference stencil for the final derivative values rather than Richardson extrapolation. We find $\partial_{e}\langle z_{(1)}\rangle(p,e=0)=0$ to be  within numerical error for all sampled values of $p$, as expected.

Because $\partial_{e}\langle z_{(1)}\rangle$ vanishes at $e=0$, the corrections $\hat{E}_{(1)}$, $\hat{L}_{(1)}$, and $\hat{J}_{r(1)}$ instead depend on $\partial_{e^{2}}\langle z_{(1)}\rangle = 2\,\partial^{2}_{e}\langle z_{(1)}\rangle$ in this limit; see Appendix~\ref{app:circular-numerics} for further discussion. However, we find that the finite-difference scheme on the original grid becomes numerically unstable near $e=0$, in part due to a delicate cancellation of the two terms on the right-hand side of \cref{eqn:dedz1} as $e\to0$. This leads to estimated uncertainties that are too large to extract a numerically significant value for $\partial^2_{e}\langle z_{(1)}\rangle(p,e=0)$. We therefore construct a new grid near $e=0$ with uniform spacing in $e^{2}$ at fixed $p$, and we recompute $\partial_{e^{2}}\langle z_{(1)}\rangle$ at $e=0$ using finite differencing. 
 
\begin{figure}[t!]
        \centering
        \includegraphics[width=\columnwidth,trim = {20pt 0pt 30pt 0pt},clip]{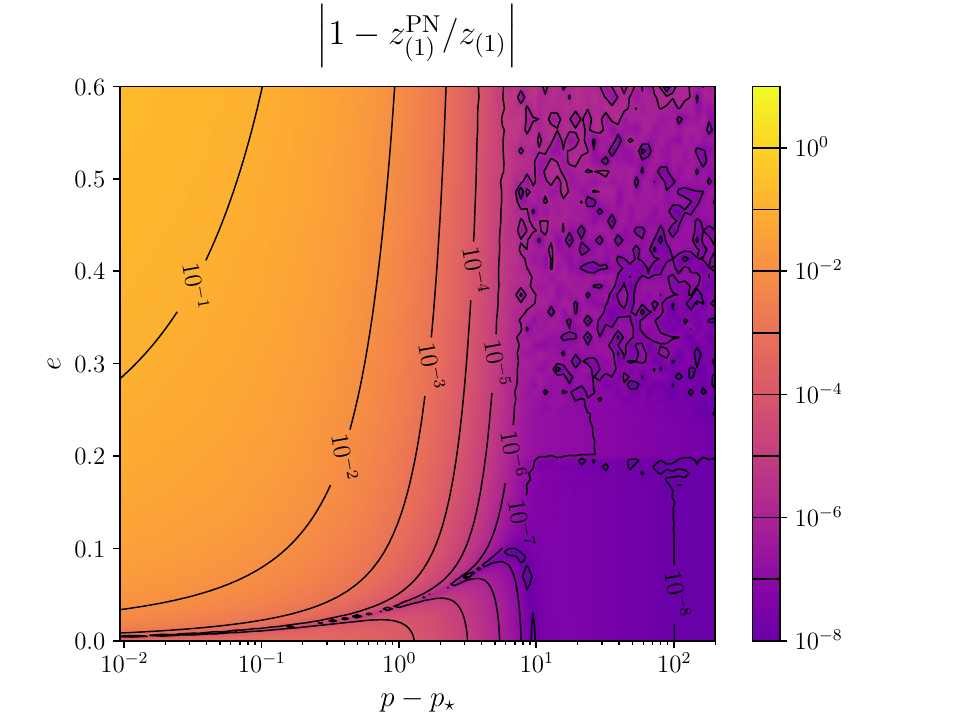}
        \caption{Comparison of the analytical 10PN expression for~$\langle z_{1} \rangle$, expanded in terms of $p$ and $e$, against our numerical results. On the horizontal axis, we use the semilatus rectum distance from the separatrix at $p_{\star}=6+2e$. The irregularities on the right side of the plot are due to numerical errors; in the large-$p$ region, we expect the PN expansion to be more accurate than the numerical results.}
        \label{fig:comparison_z1}
   
\end{figure}

We have compared the 10PN redshift $\langle z_{(1)}\rangle^\mathrm{10PN}$ in terms of $(p,e)$, as well as its partial derivatives with respect to $p$ and $e$, 
with the corresponding numerical data, 
and we have plotted the relative difference in Fig.~\ref{fig:comparison_z1}. Notably, we find that the 10PN expansion agrees with the numerical results to within at least 3 digits for all $p\gtrsim p_{\star}+2$. We observe only a very mild loss of accuracy with increasing $e$ (though our comparison is limited to $e\le0.6$). At the top right of the plot, the disagreement is dominated by error in our numerical results rather than inaccuracy of the 10PN expansion. We have checked that analogous comparisons in terms of $(y,\lambda)$ lead to significantly worse agreement; we therefore omit the comparison.

\subsection{Constants of motion}
\label{sec:1SFcorrections}

We now extract the constants of motion from the redshift data, using Eq.~\eqref{eqn:E1J1Jr1SF}. Since the derivatives of the redshift data are computed in terms of $(p,e)$, we use the chain rule to write 
\begin{align}
\label{eq:chain_rule}
    \frac{\partial \langle z_{(1)} \rangle}{\partial \hat{\Omega}_\phi} &= \frac{\partial \langle z_{(1)} \rangle}{\partial p} \frac{\partial p}{\partial \hat{\Omega}_\phi} + \frac{\partial \langle z_{(1)} \rangle}{\partial e} \frac{\partial e}{\partial \hat{\Omega}_\phi}\,,
\end{align}
and similarly for $\Omega_r$. Since the matrix elements of $(\partial \hat\Omega_A/\partial\pi_B)^{-1}$ are known exactly, we  have everything we need to compute the constants of motion numerically. 

However, we highlight that this approach suffers from \textit{large cancellations}, causing a significant loss of accuracy in our numerical results for the constants of motion relative to the underlying redshift data. To illustrate this, let us perform a numerical application of Eq.~\eqref{eq:chain_rule} for \mbox{$p=115.76626305141966$} and $e=0.56454$ (i.e., $u = 115/120$ and $w=98/100$); we find that
\begin{align}
    0.29 &= \underbrace{(-5.0243\times10^{-4}) \!\times\! (-9.2742\times10^6)}_{=456.96} 
    \\ \notag
    &\qquad \qquad + \underbrace{(-9.6365\times 10^{-2}) \!\times\! (4.8323\times10^{5})}_{=-465.67} \,.
\end{align}
Thus, the angular momentum loses $3$ digits of accuracy with respect to the numerical derivatives of the redshift. 
This feature is ubiquitous throughout the parameter space, particularly for higher eccentricities $e \gtrsim 0.4$ and near the separatrix $\delta \lesssim 10^{-2}$ and at large separations $p \gtrsim 10^2$. Consequently, to compute first-order corrections $\hat{E}_{(1)}$, $\hat{L}_{(1)}$, and $\hat{J}_{r(1)}$ to $\,\sim 2$ digits of precision, we must accurately resolve derivatives with respect to $p$ and $e$ to $\,\sim 5$ digits.

Moreover, in the special case where $e=0$, we find that $\p_e \langle z_{(1)}\rangle = 0$ whereas $\p e/\partial\hat\Omega_\phi$ and $\p e/\partial\hat\Omega_r$ diverge as $1/e$ when $e \rightarrow 0$. Thus, to compute the 1SF corrections to the constants of motion for $e=0$ using this method, we need to control the second derivative $\p_e^2 \langle z_{(1)} \rangle(p,e=0)$, as discussed in the previous section. Further details of our numerical approach in the circular limit are provided in Appendices \ref{app:circular} and \ref{app:circular-numerics}. For this work, we accept that these cancellations and singular limits are a limitation of our $(p,e)$ parametrization. Future work will investigate alternative parametrizations that mitigate these issues.

The ancillary file \texttt{data\_generic.csv}~\cite{supp} provides, for each data point, the values of the following quantities: $p$, $e$, $\hat{\Omega}_r$, $\hat{\Omega}_\phi$, $y$, $\lambda$, $\langle z_{(0)} \rangle$, $\hat{E}_{(0)}$, $\hat{L}_{(0)}$, $\hat{J}_{r(0)}$, $\langle z_{(1)} \rangle$, $\hat{E}_{(1)}$, $\hat{L}_{(1)}$, and $\hat{J}_{r(1)}$. Absolute error estimates are also provided. 

Following the same procedure, we also compute the constants of motion as PN expansions, starting from the 10PN expression for the 1SF redshift $\langle z_{(1)} \rangle$. 
This can be done analytically, and without any eccentricity reexpansion, thanks to the PN expansion of the Jacobian $(\partial \hat\Omega_A/\partial\pi_B)$. 
Note that at leading PN order, one has
\begin{subequations}\begin{align}
    \frac{\partial \langle z_{(1)} \rangle}{\partial p} \frac{\partial p}{\partial \hat{\Omega}_r} &= - \frac{\partial \langle z_{(1)} \rangle}{\partial e} \frac{\partial e}{\partial \hat{\Omega}_r} =   - \frac{p^{3/2}}{3\sqrt{1-e^2}} + \mathcal{O}(\sqrt{p}) \,,
\end{align}
such that ${\partial \langle z_{(1)} \rangle}/{\partial \hat{\Omega}_r} = \mathcal{O}(\sqrt{p})$  instead of the $\mathcal{O}(p^{3/2})$ scaling that one would naively expect. Even more remarkably, 
\begin{align}
    \frac{\partial \langle z_{(1)} \rangle}{\partial p} \frac{\partial p}{\partial \hat{\Omega}_\phi} &= - \frac{\partial \langle z_{(1)} \rangle}{\partial e} \frac{\partial e}{\partial \hat{\Omega}_\phi} \nonumber\\*
    & =  \frac{p^{3/2}}{3\sqrt{1-e^2}} - \frac{(33+5 e^2)\sqrt{p}}{122\sqrt{1-e^2}}  + \mathcal{O}\!\left(\frac{1}{\sqrt{p}}\right),
\end{align}\end{subequations}
such that \mbox{${\partial \langle z_{(1)} \rangle}/{\partial \hat{\Omega}_\phi} = \mathcal{O}(1/\sqrt{p})$}, which is two  orders smaller than the $\mathcal{O}(p^{3/2})$ scaling one would naively expect.
Consequently, our relative 10PN expression for the redshift can only give us access to the constants of motion at 9PN. These are given in terms of $(p,e)$ as
\begin{subequations}\label{eq:PN_expansion_E1_L1_Jr1}\begin{align}\label{seq:PN_expansion_E1}
    \hat{E}_{(1)} &= \frac{1-e^2}{6p}\Bigg\{1 - \frac{3(1-e^2)}{p} + \mathcal{O}\!\left(\frac{1}{p^2}\right)\Bigg\}\,,\\\label{seq:PN_expansion_L1}
    \hat{L}_{(1)} &= \frac{1}{\sqrt{p}}\Bigg\{- \frac{17}{12} + \frac{e^2}{6}  +\mathcal{O}\!\left(\frac{1}{p}\right)\Bigg\}\,,\\
    \label{seq:PN_expansion_Jr1}
    \hat{J}_{r(1)} &= - \frac{1}{3} \sqrt{\frac{p}{1-e^2}}\Bigg\{1 + \mathcal{O}\!\left(\frac{1}{p}\right)
    \Bigg\}\,,
\end{align}\end{subequations}
and in terms of $(y,\lambda)$ as 
\begin{subequations}\begin{align}
    \hat{E}_{(1)} &= \frac{y}{6} \Bigg\{1+ y \left(-1 - \frac{2}{ \lambda}\right) + \mathcal{O}(y^2) \Bigg\},\\
    \hat{L}_{(1)} &= \sqrt{\frac{\lambda}{y}}\Bigg\{- \frac{1}{6}- \frac{5}{4 \lambda}  + \mathcal{O}(y) \Bigg\},\\
    \hat{J}_{r(1)} &= - \frac{1}{3\sqrt{y}} \Bigg\{1 + \mathcal{O}(y) \Bigg\}.
\end{align}\end{subequations}
The complete 9PN expressions,  in terms  of both $(p,e)$ and  $(y,\lambda)$, are provided in the ancillary file \texttt{PN\_expressions.csv}~\cite{supp}.

\begin{figure}[h!] 
   \centering
   \hspace{10pt}
    \includegraphics[width=0.92\linewidth,trim = {20pt 22.5pt 30pt 0pt},clip]{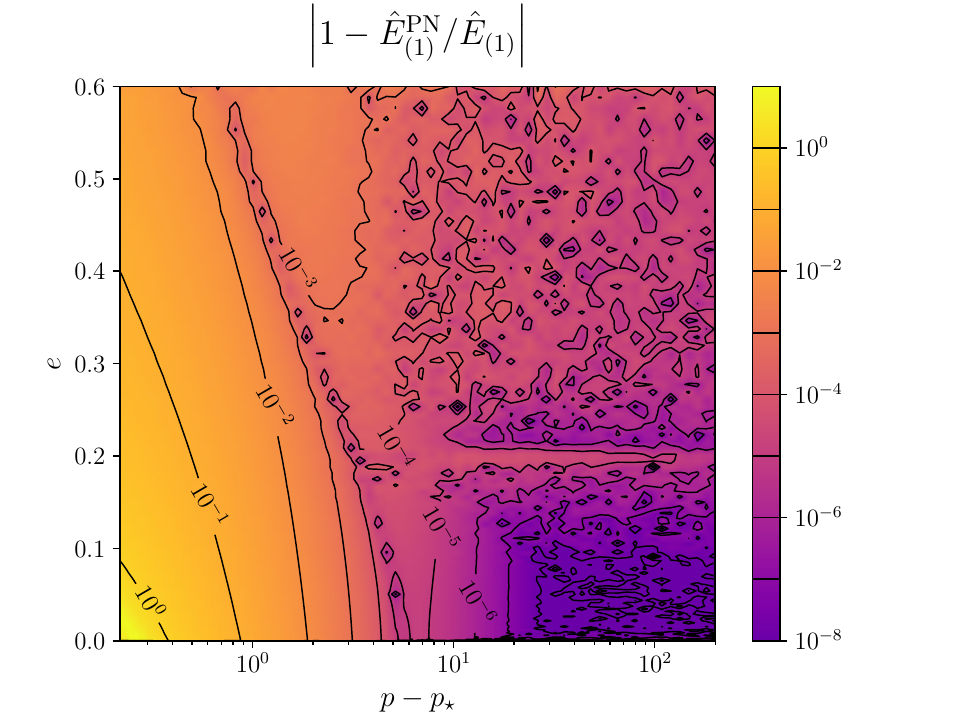}
    \includegraphics[width=0.92\linewidth,trim = {20pt 22.5pt 30pt -10pt},clip]{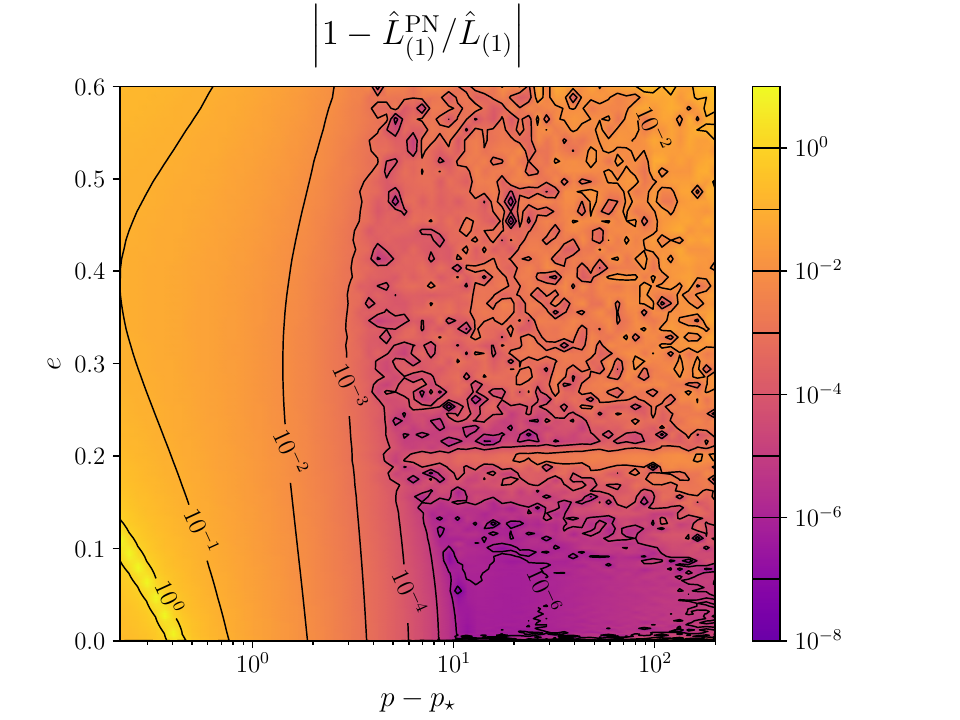}
    \includegraphics[width=0.92\linewidth,trim = {20pt 5pt 30pt -10pt},clip]{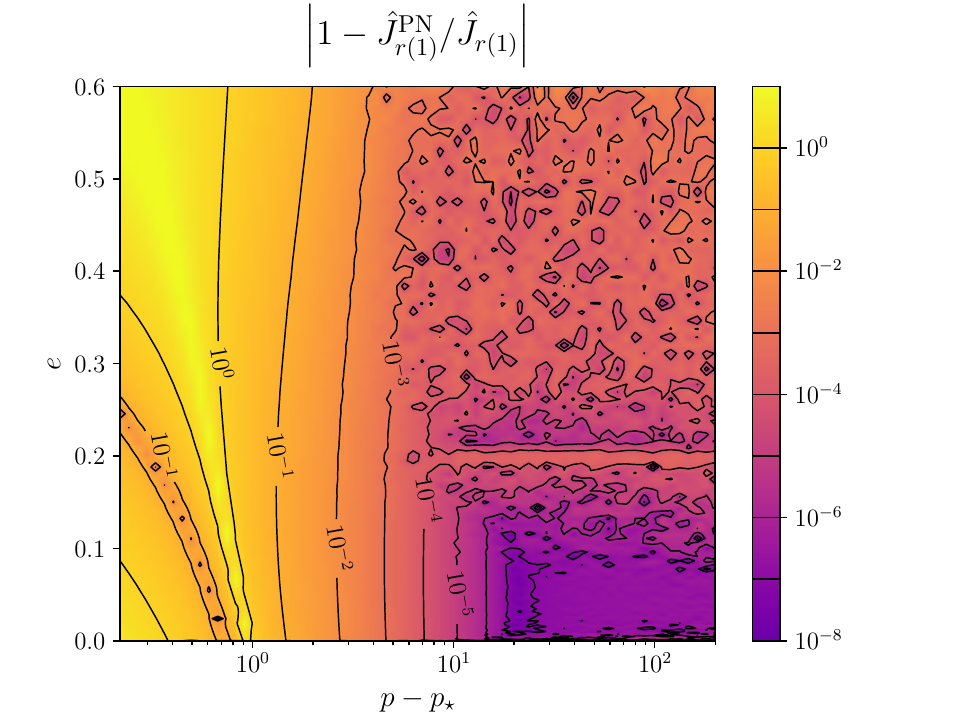}
    \caption{
    \label{fig:comparison_ELJ1} Comparison between our analytical (9PN) and numerical results for the 1SF corrections to the energy (top), angular momentum (middle), and radial action (bottom). The 1SF corrections diverge along the isofrequency curve $p_\mathrm{iso}(e)$, leading to a relative error of exactly 1 along the curve. This corresponds precisely to the last $10^0$ contour at the bottom-left corner of each panel. Furthermore, the relative errors diverge due to zero-crossings in the 1SF corrections. For $L_{(1)}$ and $J_{r(1)}$, the zero-crossings, and thus the maximum relative errors, occur at $p > p_\mathrm{iso}$. For $E_{(1)}$, they occur at $p < p_\mathrm{iso}$.
    \vspace{-30pt}
    }
\end{figure}

In Fig.~\ref{fig:comparison_ELJ1}, we plot the relative differences between our derived 9PN expressions and our numerical results for $\hat{E}_{(1)}$, $\hat{L}_{(1)}$, and $\hat{J}_{r(1)}$. We find relative errors $\lesssim 10^{-2}$ for $p \gtrsim p_{\star} + 1$ in the case of $\hat E_{(1)}$, for $p\gtrsim p_{\star} +2$ in the case of $\hat L_{(1)}$, and for $p \gtrsim p_{\star} + 3$ in the case of $\hat J_{r(1)}$.  While the agreement generally improves at larger $p$, the relative errors start to grow again for $p \gtrsim 100$ and $e \gtrsim 0.2$. Like in the case of the redshift, this is due to numerical error in our data rather than a loss of accuracy in the PN expressions. The numerical error here is significantly larger than for the redshift due to the large cancellations described in Sec.~\ref{sec:1SFcorrections}. 
When taking these errors into account, the PN results for $\hat{E}_{(1)}$, $\hat{L}_{(1)}$, and $\hat{J}_{(r1)}$ lie within the estimated error bars of our numerical results for $p - p_{\star} \geq 2.7904$, $p - p_{\star} \geq 3.7861$, and $p - p_{\star} \geq 4.1915$, respectively. 
We have checked that analogous comparisons in terms of $(y,\lambda)$ yield significantly worse agreement.

\begin{figure}
    \centering
    \includegraphics[width=0.99\linewidth]{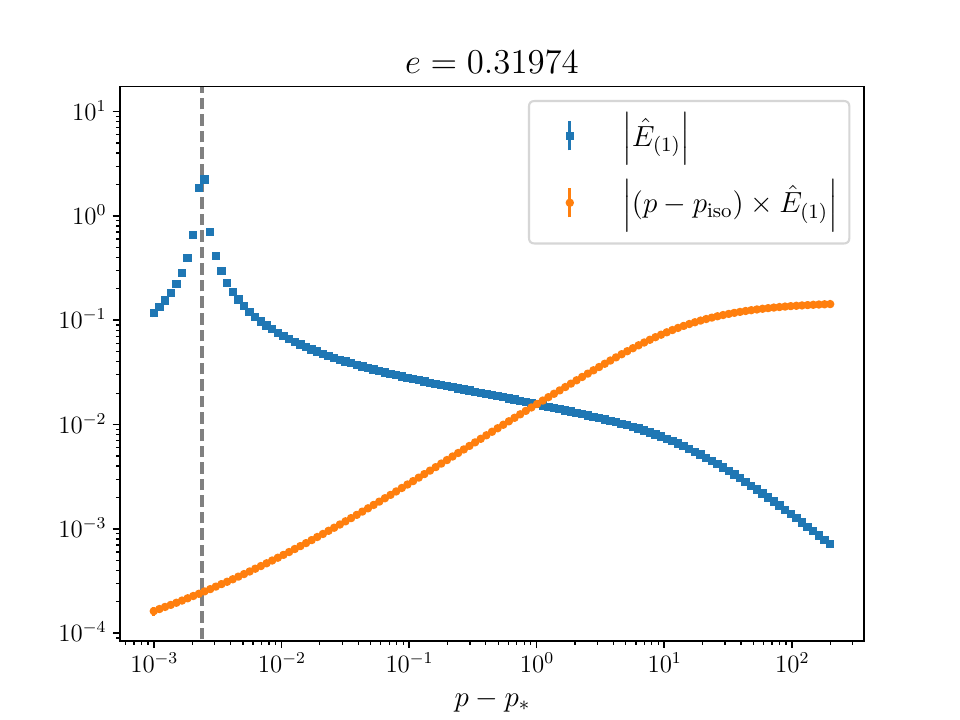}
    \caption{The 1SF correction to the energy for $e=0.31974$ as a function of $p-p_{\star}$. We see that~$\hat{E}_{(1)}$ diverges when it crosses the isofrequency curve at $p-p_\star = p_{\rm iso}-p_\star\approx 0.0023852$ (vertical dashed line). If we instead divide by the pole \mbox{$(p - p_{\rm iso})^{-1}$}, we get a smooth result.}
    \label{fig:E1_e0.15}
\end{figure}

Our use of a fixed-frequencies expansion also has another downside, besides numerical cancellations: $\hat\Omega_A$ are not valid phase-space coordinates along the isofrequency degeneracy curve discussed in Sec.~\ref{sec:geoExact} (though they are valid coordinates in each of the two regions to the left and right of that curve). Because the inverse Jacobian $(\partial \hat\Omega_A/\partial\pi_B)^{-1}$ diverges along that curve, the 1SF corrections to the constants of motion become ill defined there in a fixed-frequencies expansion. This is illustrated in  Fig.~\ref{fig:E1_e0.15}, where we see that the correction $\hat E_{(1)}$ diverges in the approach to the isofrequency degeneracy.

\section{Applications}
\label{sec:applications}

In this section, we discuss the utility of the constants of motion in identifying special locations in the parameter space and in waveform generation. Specifically, we compute the location of circular orbits in the $(p,e)$ plane, the correction to the constants of motion along the separatrix, and corrections at the innermost stable circular orbit (ISCO). We then highlight the simplifications to waveform generation that arise from using balance laws for $E$ and $L$.

\subsection{Circular orbits}
\label{subsec:circ}
Given the freedom in defining $(p,e)$ in the perturbed system, it is not guaranteed that $e=0$ will continue to specify a circular orbit. It is equally unclear what value $e$ takes in genuine quasicircular inspirals, which are characterized by the metric's dependence on only one phase variable and frequency, $(\phi_p,\Omega)$, rather than two~\cite{Miller:2020bft}.

We can define a (stable) circular orbit by a vanishing radial action---namely $\hat{J}_r = \hat{J}_{r(0)} + \varepsilon \, \hat{J}_{r(1)} =0$. This condition is preserved throughout a 1PA quasicircular evolution: from the evolution equation~(320) in Ref.~\cite{Lewis:2025ydo}, one can easily show $\dd J_r/\dd t=0$ in the quasicircular case, due to the absence of any dependence on the radial phase~$\varphi_r$. Expressed in terms of $(p,e)$, $\hat J_r=0$ determines a curve $e_{\odot}(p; \varepsilon)$, along which any circular (or quasicircular) orbit must lie. However, it will be more useful to work with the square of $e$--namely $e^2_{\odot}(p;\varepsilon)$---as we explain below. 

Of course, since $e^2_{\odot}=0$ at geodesic order, $e^2_{\odot}(p;\varepsilon)$ must vanish when $\varepsilon\to0$, but thanks to our expression for the radial action, we can now control the nonvanishing 1SF term in $e^2_{\odot}$. We obtain the correction numerically by injecting the ansatz 
\begin{equation}\label{eq:ecirc_ansatz}
e^2_{\odot}(p;\varepsilon) = \varepsilon \, \delta e_{\odot}^2(p) +  \mathcal{O}(\varepsilon^2) 
\end{equation}
into the radial action.  After replacement, we are left to solve 
\begin{equation}\label{eq:circ condition}
    \hat{J}_{r(0)}(p,0) + \varepsilon \, \delta e_{\odot}^2(p) \frac{\p \hat{J}_{r(0)}}{\p e^2}(p,0) + \varepsilon \, \hat{J}_{r(1)}(p,0) = \mathcal{O}(\varepsilon^2) \,.
\end{equation}
The zeroth-order term vanishes by construction of the quasi-Keplerian variables.
The derivative appearing in the subleading term is given analytically by
\begin{align}
    \left.\frac{\p \hat{J}_{r(0)}}{\p e^2}\right|_{p}(p,0) &= \frac{p^{3/2}\sqrt{p-6}}{2(p-2)\sqrt{p-3}}\,,
\end{align}
allowing us to immediately solve Eq.~\eqref{eq:circ condition} to find
\begin{align}\label{eq:delta_e2_inTermsOf_Jr1}
    \delta e_{\odot}^2(p) &= - \frac{2(p-2)\sqrt{p-3}}{p^{3/2}\sqrt{p-6}} \hat{J}_{r(1)}(p,0)\,.
\end{align}
Here we see the motivation for working with $e^2_{\odot}$ rather than $e_{\odot}$: $e_{\odot}$ scales with $\sqrt{\varepsilon}$ rather than $\varepsilon$. This is a consequence of the small-$e$ behavior of $\hat J_{r(0)}$, which has no linear term in a small-$e$ expansion.

\begin{figure}
    \centering
  
    \includegraphics[width=0.95\linewidth]{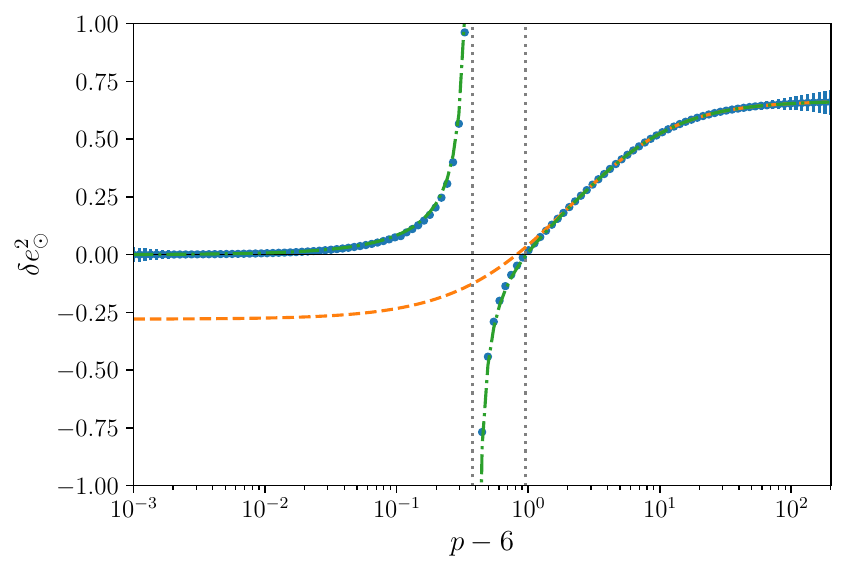}
    \caption{\justifying Plot of $\delta e_{\odot}^2(p)$ using various schemes. The blue dots and error bars are obtained by evaluating Eq.~\eqref{eq:delta_e2_inTermsOf_Jr1} using our numerical results for $\hat{J}_{r(1)}^\odot(p)$. The green dot-dashed line is obtained by replacing $\rho(x)$ with its 9PN expression in Eq.~\eqref{eq:e2circ_rho_p}, whereas the orange dashed line is obtained by using a nonresummed 9PN expansion \eqref{eq:delta_e2_PN} for $\delta e_{\odot}^2(p)$. The vertical gray dotted lines correspond to, from left to right, 
    $p=p_\text{iso}\approx 6.38$ and $p = p_\text{sign} \approx 6.95$. Thus, for $p_\text{iso}<p<p_\text{sign}$, we have $\delta e_{\odot}^2 (p)<0$, and $e_{\odot}$ is an imaginary number.
    }
    \label{fig:delta_e2}
\end{figure}

Although Eq.~\eqref{eq:delta_e2_inTermsOf_Jr1} is exact, it suffers from hidden divergences in $\hat{J}_{r(1)}(p,0)$. To make these explicit, we build on previous works \cite{Barack:2010ny,Bini:2016qtx} which
have obtained the circular link between \mbox{$x = [(m_1+m_2)\Omega_\phi]^{2/3}$} and $W = (\Omega_r/\Omega_\phi)^2$. The link reads \mbox{$W^{\odot}(x) = 1-6x + \varepsilon \rho(x) + \mathcal{O}(\varepsilon^2)$}, where $\rho(x)$ is provided numerically, is smooth in the domain $0<x \le 1/6$, and converges to a finite value $\rho(1/6)\approx 0.83413$~\cite{Barack:2010ny}.
In Appendix~\ref{app:circular}, we show that $\hat{J}_{r(1)}(p,0)$ can be related to the $\rho(x)$ function via
\begin{align}\label{eq:Jr1_circ_inTermsOf_rho}
    \hat{J}_{r(1)}(p,0) =  \frac{p^{3/2} (p-6)^{3/2} \bigl[p \,\rho(1/p) - 4\bigr]}{3 \sqrt{p-3} (86 -39 p +4 p^2)} \,.
\end{align}
Thus, we find that the eccentricity for circular orbits can be reexpressed as
\begin{align}\label{eq:e2circ_rho_p}
    \delta e_{\odot}^2(p) = \  \frac{2(p-6)(p-2)}{3(86-39p+4p^2)}\bigl[4-p \,\rho(1/p)\bigr] \,,
\end{align}
where the singular behavior is now fully explicit. We see that $\delta e_{\odot}^2(p)$ has a pole for \mbox{$p_\text{iso}= (39+ \sqrt{145})/8 \approx 6.38$,} which is exactly the value of $p$ corresponding to $e=0$ on the (geodesic) isofrequency curve \cite{Warburton:2013yj}. Moreover, $\delta e_{\odot}^2(p)$ passes through zero at $p=p_\text{sign}\approx 6.95$, where $p_\text{sign}$ is the solution to  $4-p\, \rho(1/p) =0$. This means that $\delta e_{\odot}^2$ changes sign at $p_\text{sign}$, becoming negative for $p_\text{iso} <p < p_\text{sign}$ and $0< \varepsilon \ll 1$, such that $e_{\odot}(p)$ becomes purely imaginary in that range. This points to the occasionally counterintuitive nature of $e$ in the fixed-frequencies gauge. For clarity, we have plotted the behavior of $\delta e_{\odot}^2(p)$ in Fig.~\ref{fig:delta_e2}.

Given the expression \eqref{eqn:OmegasofPE} for the orbital frequencies, we are now able to immediately find the frequencies for circular orbits. Injecting Eq.~\eqref{eq:e2circ_rho_p} into Eq.~\eqref{eqn:OmegasofPE}  and expanding to 1SF order, we find
\begin{subequations}\label{eq:Omega_phi_Omega_r_circ_p}\begin{align}\label{seq:Omega_phi_circ_p}
    \hat\Omega_\phi^{\odot}(p) &= \frac{1}{p^{3/2}}+ \varepsilon\, \frac{(22-10p+p^2)}{p^{3/2}(86-39p+4p^2)}\bigl[p \,\rho\left(1/p\right) - 4\bigr], \\
\label{seq:Omega_r_circ_p}
     \hat\Omega_r^{\odot}(p) &= \frac{\sqrt{p-6}}{p^{2}} \nonumber\\*
    & \ + \varepsilon\, \frac{(-266+165p-32p^2+2p^3)}{2p^{2}(86-39p+4p^2)\sqrt{p-6}}\bigl[p \,\rho\left(1/p\right) - 4\bigr].  
\end{align}
\end{subequations}
Notice that the leading-order term in $\hat{\Omega}_r^{\odot}(p)$ vanishes for $p=6$, whereas the subleading term is divergent, which indicates a breakdown of the perturbative expansion around that point.  We therefore will find it much more fruitful to work with the square of  the radial frequency, 
\begin{align}
\label{eq:Omega_r_circ_squared_p}
     &(\hat\Omega_r^{\odot})^2(p) = \frac{{p-6}}{p^{4}} \nonumber\\*
    & \qquad \qquad  + \varepsilon\, \frac{(-266+165p-32p^2+2p^3)}{p^{4}(86-39p+4p^2)}\bigl[p \,\rho\left(1/p\right) - 4\bigr],
\end{align}
where the subleading term is finite for $p=6$.
Moreover, from Eq.~\eqref{seq:Omega_phi_circ_p}, 
one immediately deduces
\begin{align}\label{eq:ycirc_p}
    y_{\odot}(p)=\frac{1}{p}+\varepsilon\,\frac{2(22-10p+p^2)}{3p(86-39p+4p^2)}\bigl[p \,\rho\left(1/p\right) - 4\bigr] \,. 
\end{align}
The expression for $\lambda_{\odot}(p) = \frac{3 y_{\odot}}{  \hat\Omega_\phi^{\odot}/  \hat\Omega_r^{\odot} -1}$ is straightforward to obtain but too long to be worth displaying here.  

In the spirit of the fixed-frequency decomposition, we perform the following expansion
\begin{subequations}\label{eq:z_E_L_Omegar_circ_y_expansion}
    \begin{align}
    z^{\odot}(y)&= z_{(0)}^{\odot} (y)+\varepsilon\,z_{(1)}^{\odot}(y) + \mathcal{O}(\varepsilon^2)\,,\\
    \hat{E}^{\odot}(y)&= \hat{E}_{(0)}^{\odot}(y)+\varepsilon\,\hat{E}_{(1)}^{\odot}(y) + \mathcal{O}(\varepsilon^2)\,, \\
    \hat{L}^{\odot}(y)&= \hat{L}_{(0)}^{\odot}(y)+\varepsilon\,\hat{L}_{(1)}^{\odot}(y) + \mathcal{O}(\varepsilon^2)\,, \\
    \hat{\Omega}_r^{\odot}(y)&= \hat{\Omega}_{r(0)}^{\odot}(y)+\varepsilon\,\hat{\Omega}_{r(1)}^{\odot}(y) + \mathcal{O}(\varepsilon^2) \,.
\end{align}
\end{subequations}
Note that, by definition, the radial action is vanishing to all orders for circular orbits. At geodesic order, we have the well-known expressions 
\begin{subequations}\label{eq:z0_E0_L0_Omegar0_circ_y}
    \begin{align}
        z_{(0)}^{\odot}(y)&=\sqrt{1-3y} \,,\\
        \hat{E}_{(0)}^{\odot}(y) &= \frac{1-2y}{\sqrt{1-3y}} \,, \\
        \hat{L}_{(0)}^{\odot}(y) &= \frac{1}{\sqrt{y(1-3y)}}\,,\\
        \hat{\Omega}_{r(0)}^{\odot}(y) &=  y^{3/2}\sqrt{1-6y}\,.
    \end{align}
\end{subequations}
At 1SF order, we need to invert the relation \eqref{seq:Omega_phi_circ_p} to read off
\begin{subequations}\begin{align}
    \delta p_{\odot}(y) &= \frac{2(1-10y+22y^2)}{3 y^2(4-39y+86y^2)}\big[\rho(y)-4y\big] \,,\\*
    \delta e^2_{\odot}(y) &= \frac{2(1-6y)(1-2y)}{3y(4-39y+86y^2)}\big[4y-\rho(y)\big]\,.
\end{align}\end{subequations}
The results are then obtained numerically and provided in the ancillary file \texttt{data\_circular.csv}, which contains, for each data point, the values of the following quantities: $y$, $\hat\Omega_\phi$, $\hat{\Omega}_{r(0)}$, $\langle z_{(0)}^{\odot} \rangle$, $\hat{E}_{(0)}^{\odot}$, $\hat{L}_{(0)}^{\odot}$, $\hat{\Omega}_{r(1)}^{\odot}$, $\langle z_{(1)}^{\odot} \rangle$, $\hat{E}_{(1)}^{\odot}$, $\hat{L}_{(1)}^{\odot}$, $\delta p_{\odot}$, and $\delta e_{\odot}^2$. Absolute error estimates are also provided.

We now proceed to determine these circular links as PN expansions. First, we compute the self-force corrections to the circular squared eccentricity $e_{\odot}^2$ in terms of $p$. Using our 9PN expansion of $\hat{J}_{r(1)}(p,0)$, or equivalently, using the PN expansion for $\rho(x)$ provided in Eq.~(9) of Ref.~\cite{Bini:2016qtx}, we can approximate the location of circular orbits as
\begin{align}\label{eq:delta_e2_PN}
    e^2_{\odot} &= \varepsilon \,\Bigg\{\frac{2}{3} - \frac{7}{6 p} + \mathcal{O}\left(\frac{1}{p^2}\right) \Bigg\} + \mathcal{O}(\varepsilon^2) \,,
\end{align}
where the complete 9PN expression is given in the ancillary file \texttt{PN\_expressions.csv}~\cite{supp}.

Similarly, it is straightforward to PN-expand the frequencies $\hat\Omega_\phi^{\odot}(p)$ and $\hat\Omega_r^{\odot}(p)$ provided in Eqs.~\eqref{eq:Omega_phi_Omega_r_circ_p}; then, eliminating $p$, we can find the gauge-invariant relation between the two frequencies. The latter is most conveniently represented through the fixed-frequency expansion of \mbox{$\lambda^{\odot}(y)=\lambda_{(0)}^{\odot}+ \varepsilon \,\lambda_{(1)}^{\odot}+ \mathcal{O}(\varepsilon^2)$;} we find that
\begin{subequations}
\begin{align}
    \lambda_{(0)}^{\odot}(y)&= 1 - \frac{9}{2}y - \frac{9}{4}y^2 +  \mathcal{O}(y^3) \,,\\*
    \lambda_{(1)}^{\odot}(y) &= - \frac{2}{3} + \frac{7}{3}y+ y^2\left(\frac{379}{12}- \frac{41}{32}\pi^2\!\right) + \mathcal{O}(y^3)\,,
\end{align}
\end{subequations}
where the complete 9PN expressions are given in the ancillary file \texttt{PN\_expressions.csv}~\cite{supp}. After appropriate rescaling, we find agreement at geodesic and 1SF orders with the 3PN-accurate expression~(7.4a) of Ref.~\cite{Trestini:2025yyc}, which includes contributions from the tail that enters the 4PN dynamics.

It is possible to deduce other gauge-invariant links for circular orbits. For instance, the energy is related to the angular momentum through \mbox{$ \hat{E}^{\odot}(\hat{L})= \hat{E}_{(0)}^{\odot}(\hat{L})+\varepsilon \,\hat{E}_{(1)}^{\odot}(\hat{L}) + \mathcal{O}(\varepsilon^2)$.} We find that
\begin{subequations}
\begin{align}
    \hat{E}_{(0)}^{\odot}(\hat{L})&= - \frac{1}{2}\hat{L}^{-2}- \frac{9}{8}\hat{L}^{-4} - \frac{81}{16}\hat{L}^{-6}+ \mathcal{O}(\hat{L}^{-8})\,,\\
    \hat{E}_{(1)}^{\odot}(\hat{L}) &=  \frac{1}{2}\hat{L}^{-2} + \hat{L}^{-4}+ \frac{11}{2}\hat{L}^{-6}+ \mathcal{O}(\hat{L}^{-8})\,,
\end{align}
\end{subequations}
where the complete 9PN expressions are given in the ancillary file \texttt{PN\_expressions.csv}~\cite{supp}. After rescaling by the correct prefactor, we found perfect agreement with the 4PN circular link in Eq.~(5.3) of Ref.~\cite{Damour:2014jta} at geodesic and 1SF orders.

\subsection{Homoclinic orbits on the separatrix}
\label{subsec:sep}

We now turn our attention to the separatrix between bound and plunging orbits. Orbits on this separatrix correspond to homoclinic trajectories that are maximally ``zoom-whirly'', with a single radial excursion (the ``zoom'') to apastron and an infinite ``whirl'' phase in approaching and departing periastron~\cite{Bombelli:1991eg,Dean:1999jr,Levin:2008mq,Perez-Giz:2008ajn}; a point with orbital radius precisely \emph{at} this periastron is a fixed point, corresponding to an unstable circular orbit. The homoclinic orbits necessarily have an infinite radial period, taking infinite time to leave the fixed point in the past and infinite time to reach it in the future. Hence, they have vanishing radial frequency, and we can use $\Omega_r=0$ to define the location of the separatrix, to all orders in the mass ratio. 
At geodesic order, the condition $\Omega_r=0$ corresponds to $p=6+2e$. Since we have chosen the quasi-Keplerian parameters $(p,e)$ to be geodesically related to the frequencies, the separatrix remains at $p=6+2e$ at all SF orders. Denoting points on the separatrix with a star, as in $(p_\star,e_\star)$, we parametrize the separatrix location as $p_{\star}(e) = 6 + 2e$. The azimuthal frequency can then be obtained as a pure function of the eccentricity, $\Omega_\phi^{\star}(e) = \Omega_\phi(p_{\star}(e),e)$. Computing this explicitly from Eq.~\eqref{eqn:OmegasofPE} is in fact nontrivial because the limit of $\Omega_\phi(p,e)$ as  $\delta  \equiv p - p_{\star}(e) \rightarrow 0$ is singular.  To examine this limit, we first apply the relations between elliptic functions provided in Eqs.~\eqref{eq:Pi_rule} and \eqref{eq:Pi_rule_applications}, then use the near-separatrix behavior of the  elliptic integral function of the third kind, $\Pi(\cdot|\cdot)$, provided in Eq.~(3.5)~of~Ref.~\cite{Lhost:2024jmw}, as well as the limits 
\begin{subequations}
 \begin{align}
    K\left(\frac{1}{1+z} \right) &\underset{z \rightarrow 0^+}{\sim} - \frac{1}{2} \ln \left(\frac{z}{16}\right) \,,\\
    E\left(\frac{1}{1+z} \right) &\underset{z \rightarrow 0^+}{\longrightarrow} 1 \,.
\end{align}\end{subequations}
Using the expansions of the elliptic integrals, we can determine that all three periods diverge logarithmically:
\begin{subequations}
    \begin{align}
    T_r &\underset{\delta \rightarrow 0^+}{\sim}  \frac{4(3+e)^2}{\sqrt{e}(1+e)^{3/2}}\ln\left(\frac{1}{\delta}\right) \,,\\
    \Phi & \underset{\delta \rightarrow 0^+}{\sim} \sqrt{\frac{2(3+e)}{e}}\ln\left(\frac{1}{\delta}\right) \,,\\
    \mathcal{T}_r &\underset{\delta \rightarrow 0^+}{\sim} \sqrt{\frac{8(3-e)(3+e)^3}{e (1+e)^3}}\ln\left(\frac{1}{\delta}\right)\,,
\end{align}
\end{subequations}
implying that $\Omega_r$ vanishes in this limit (as expected). On the other hand, the geodesic redshift $\langle z_0 \rangle = \hat{T}_r/\hat{\mathcal{T}}_r$ has a finite near-separatrix limit:
\begin{align}\label{eq:z0sep_geo}
\langle z_{(0)}^{\star} \rangle = \sqrt{\frac{3-e}{2(3+e)}} = \sqrt{\frac{6}{p}- \frac{1}{2}}= \sqrt{1-3y} \,.
\end{align}
Finally, the azimuthal frequency $\hat{\Omega}_\phi = \Phi/\hat{\mathcal{T}}_r$ also has a finite limit on the separatrix~\cite{Lhost:2024jmw}:
\begin{align}\label{eq:ysep}
    y^{\star}  = \left(\hat{\Omega}_\phi^{\star}\right)^{2/3} = \frac{1+e^\star}{6+2e_\star} = \frac{1}{2} - \frac{2}{p^\star}\,.
\end{align}
This expression is identical to that of an unstable circular (geodesic) orbit: $y=1/\hat r+\mathcal{O}(\varepsilon)$, where the radius of the periastron $\hat r_{\rm min} = p/(1+e)$ is evaluated for $p=p_{\star}=6+2e$ on the separatrix.
Finally, note that the vanishing of the radial frequency immediately leads to $\lambda^{\star}(e) = 0$. 

The geodesic energy, angular momentum, and radial action all admit a finite limit on the separatrix. The geodesic energy and angular momentum reduce to the usual geodesic expressions for circular orbits~\cite{Chandrasekhar:1985kt} with (effective) radius $\hat{r}_\text{min}$. 
In the case of the  radial action, using the previously established limits for the elliptic integrals, we find a complicated expression  that depends on Legendre's elliptic integrals of the third kind $\Pi(\cdot|\cdot)$ as well as the functions $\Lambda(e)$ and $\Theta(e)$  introduced in~Ref.~\cite{Lhost:2024jmw}. Alternatively, by directly integrating Eq.~\eqref{eq:J0r} with \mbox{$p=p_{\star}=6+2e$,} we are able to find a more tractable expression,\footnote{This integral was performed in App. A of Ref.~\cite{Perez-Giz:2008ajn} for Kerr spacetime; one recovers the Schwarzschild result of Eq.~\eqref{seq:Jr0_sep} by setting $a=0$.}
which agrees numerically with the former. The energy, angular momentum, and radial action on the separatrix finally read 
\begin{subequations}\label{eq:Esep_Lsep_Jrsep_geo}
\begin{align}
\label{seq:E0_sep}
\hat{E}_{(0)}^{\star} &= \sqrt{\frac{8}{(3+e)(3-e)}} \,,\\
\label{seq:L0_sep}
\hat{L}_{(0)}^{\star} &{}={} \frac{2(3+e)}{\sqrt{(1+e)(3-e)}} \,,\\
\label{seq:Jr0_sep}
    \hat{J}_{r{(0)}}^{\star} &= \frac{2}{\pi\sqrt{9-e^2}}\Bigg\{\frac{7+e^2}{\sqrt{1-e^2}}\arctan\left(\sqrt{\frac{2e}{1-e}}\right)\\*
    &\quad\quad - 4\sqrt{2}\,\text{arctanh}\left(\sqrt{\frac{e}{1+e}}\right) - (3+e)\sqrt\frac{2e}{1+e}\Bigg\} \nn \,.
\end{align}
\end{subequations}
It is immediate to reexpress these in terms of either $p$ or $y$ using Eq.~\eqref{eq:ysep}. 

We now use the results of this paper to compute the self-force corrections to the redshift, energy, angular momentum, and radial action on the separatrix. These invariants can be decomposed as usual into a (fixed-frequencies) expansion in the mass ratio:
\begin{subequations}\label{eq:sepExpansions}\begin{align}
      \langle z^{\star}\rangle(e) &=\langle z_{(0)}^{\star}\rangle(e) + \varepsilon \, \langle z_{(1)}^{\star}\rangle(e) + \mathcal{O}(\varepsilon^2)\,, \\*
      \label{eq:EsepExpansion}
     \hat{E}^{\star}(e) &= \hat{E}_{(0)}^{\star}(e) + \varepsilon\, \hat{E}_{(1)}^{\star}(e) + \mathcal{O}(\varepsilon^2) \,,\\* \label{eq:LsepExpansion}
    \hat{L}^{\star}(e) &= \hat{L}_{(0)}^{\star}(e) + \varepsilon\, \hat{L}_{(1)}^{\star}(e) + \mathcal{O}(\varepsilon^2) \,, \\*
   \hat{J}_{r}^{\star}(e) &=\hat{J}_{r(0)}^{\star}(e) + \varepsilon \, \hat{J}_{r(1)}^{\star}(e) + \mathcal{O}(\varepsilon^2)\,.
\end{align}\end{subequations}
The 1SF corrections are estimated by fitting for the first-order terms in Eq.~\eqref{eq:sepExpansions} using our values for $\langle z_{(1)}\rangle$, $\hat{E}_{(1)}$, $\hat{L}_{(1)}$, and $\hat{J}_{r(1)}$ near the separatrix. These values are reported in
the ancillary file \texttt{data\_separatrix.csv}~\cite{supp}, which includes the following quantities: $p$, $e$, $\hat{\Omega}_\phi$, $y$, $\langle z_{(0)}^{\star} \rangle$, $\hat{E}_{(0)}^{\star}$, $\hat{L}_{(0)}^{\star}$, $\hat{J}_{r(0)}^{\star}$, $\langle z_{(1)}^{\star} \rangle$, $\hat{E}_{(1)}^{\star}$, $\hat{L}_{(1)}^{\star}$, and $\hat{J}_{r(1)}^{\star}$. Absolute error estimates are also provided.

Finally, we note that in an effective-potential description, points on the separatrix correspond to orbits whose energies are at the peak of the effective potential. That textbook description of the geodesic dynamics extends straightforwardly to 1SF order: The condition $g^{\alpha\beta}p_\alpha p_\beta=(m_2)^2$, together with $p_t = E + \varepsilon \delta p_t$ and $p_\phi = L + \varepsilon \delta p_\phi$, implies the (adimensionalized) radial equation
\begin{equation}\label{rdot=E-V}
\left(\frac{d\hat r_p}{d\hat\tau}\right)^{\!2} = \hat E^2-V(\hat r_p,\hat E,\hat L,\varepsilon),    
\end{equation}
with the effective potential
\begin{equation}
V = f_p\left(1+\frac{\hat L^2}{\hat r_p^2}\right)
+ 2\varepsilon\left(f_p\frac{\hat L\delta \hat p_\phi}{\hat r_p^2}-\hat E\delta \hat p_t\right). 
\end{equation}
Here, the dimensionless quantities are $\hat \tau:=\tau/m_1$, $\hat r_p:=r_p/m_1$, $f_p:=1-2/\hat r_p$, $\delta\hat p_t:=\delta p_t/m_2$ and $\delta\hat p_\phi:=\delta p_\phi/(m_1m_2)$. In writing $V=V(\hat r_p,\hat E,\hat L,\varepsilon)$, we have also used the fact that if we adopt $(r_p,\phi_p,E,L)$ as phase-space coordinates, then $\delta \hat p_t=\delta \hat p_t(\hat r_p,\hat E,\hat L)$ and $\delta \hat p_\phi=\delta \hat p_\phi(\hat r_p,\hat E,\hat L)$, because the dynamics is independent of $\phi_p$. The quantities $\delta \hat p_t$ and $\delta\hat p_\phi$ can be extracted from Ref.~\cite{Lewis:2025ydo}, but the explicit expressions are not needed for the general discussion here.

Just as in the usual geodesic description, Eq.~\eqref{rdot=E-V} implies 
\begin{equation}\label{E^2=V}
\hat E^2=V(\hat r_{\rm min/max},\hat E,\hat L,\varepsilon) 
\end{equation}
at turning points and on circular orbits. Differentiating Eq.~\eqref{rdot=E-V} with respect to $\hat\tau$ and restricting to the conservative sector (where $E$ and $L$ are constants), we also get
\begin{equation}\label{rddot=dVdr}
\frac{d^2\hat r_p}{d\hat\tau^2} = -\frac{1}{2}\frac{\partial V}{\partial \hat r_p},    
\end{equation}
again just as in the geodesic case. Unstable circular orbits correspond to maxima of the effective potential, where $\frac{\partial V}{\partial \hat r_p}=0$ picks out circular orbits and $\frac{\partial^2 V}{\partial r_p^2}<0$ ensures they are unstable. The two relations~\eqref{E^2=V} and $\frac{\partial V}{\partial \hat r_p}=0$, with $\frac{\partial^2 V}{\partial r_p^2}<0$, determine $\hat E_{\rm UCO}(\hat L,\varepsilon)$ and $\hat r_{\rm UCO}(\hat L,\varepsilon)$. The homoclinic orbits are eccentric trajectories with $\hat E^\star(\hat L,\varepsilon)=\hat E_{\rm UCO}(\hat L,\varepsilon)$ and $\hat r_{\rm min}(\hat L,\varepsilon)=\hat r_{\rm UCO}(\hat L,\varepsilon)$, such that they precisely reach the peak of the effective potential at their periapsis, in the infinite past and future.

\begin{figure}
    \centering
    \includegraphics[width=0.98\linewidth]{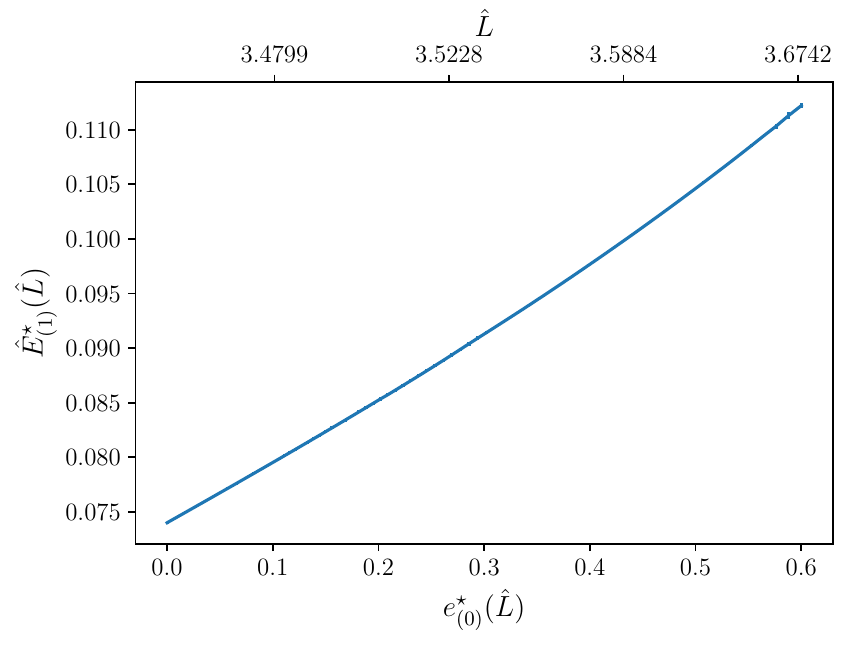}
    \caption{A plot of the separatrix curve $\hat{E}^\star_{(1)}(\hat{L})$ in terms of both the angular momentum, $\hat{L}$, and the eccentricity which is geodesically linked to the angular momentum, $e_{(0)}^\star(\hat{L})$; see Eq.~\eqref{seq:e0starL} for its explicit definition. Error bars are included based on the estimated numerical uncertainty in the data.}
    \label{fig:E1ofL}
\end{figure}

The relationship $\hat r_{\rm min}(\hat L)$ is gauge dependent, since a gauge transformation trivially shifts the orbital radius. But $\hat E^\star(\hat L,\varepsilon)$ is a gauge-invariant relationship between invariant quantities, meaning it provides an alternative description of the separatrix location $\Omega_r=0$. We can write this curve as 
\begin{align} \label{eq:ESepofL}
    \hat{E}^{\star}(\hat{L},\varepsilon) = \hat{E}^{\star}_{(0)}(\hat{L}) + \varepsilon \, \hat{E}^{\star}_{(1)}(\hat{L}).
\end{align}
The geodesic curve, $\hat{E}^{\star}_{(0)}(\hat{L})$, is given by the 0SF relations in Eq.~\eqref{eq:Esep_Lsep_Jrsep_geo}. In principle $E^{\star}_{(1)}(\hat{L})$ can be computed directly from the perturbed effective potential, but there are several far simpler methods of computing it from the first-order quantities in Eq.~\eqref{eq:sepExpansions}; these methods are summarized in Appendix~\ref{app:correction-to-sep}. We find that they all give the same numerical result (up to numerical errors and uncertainty). The correction $E^{\star}_{(1)}(\hat{L})$, which we can call the separatrix shift (the shift in its location in the $\hat P_A$ plane), is plotted in Fig.~\ref{fig:E1ofL}. 

\subsection{Innermost stable circular orbit}
\label{sec:1SF-ISCO}

We now specialize to the circular case and verify that we are able to recover known results for the shift in the frequency of the innermost stable circular orbit (ISCO), as well as the values of constants of motion at the ISCO. Most of these quantities are highly singular near the ISCO: limits do not necessarily commute, Taylor expansions are often singular when expanding around $(p,e)=(6,0)$ and different orders of operations can yield different results. Thus, we here follow the usual definition in the literature: we first restrict to the curve of circular orbits, then take the limit toward the ISCO. Precisely at the ISCO, the curve of circular orbits (defined by $J_r=0$) intersects the separatrix (defined by $\Omega_r=0$), and we can equivalently define the ISCO by $\Omega_r=0=J_r$.

At geodesic order, $(p,e)=(6,0)$ and Eq.~\eqref{eq:ysep} immediately yields $y_\text{ISCO} = 1/6 + \mathcal{O}(\varepsilon)$. The redshift, energy, angular momentum, and radial action read, respectively, $z_{(0)}^\text{ISCO}=1/\sqrt{2}$, $\hat{E}_{(0)}^\text{ISCO} = 2\sqrt{2}/3$, $\hat{L}_{(0)}^\text{ISCO} = 2\sqrt{2}$ and $\hat{J}_{r(0)}^\text{ISCO}=0$.

To extend these results to 1SF order, we first want to find the values of $(p,e)$ at the ISCO, which we denote $(p^\text{ISCO}, e^\text{ISCO})$. We will first restrict to the curve of circular orbits ($\hat{J}_r=0$), such that $p$ and $e$ are related by Eqs.~\eqref{eq:ecirc_ansatz}--\eqref{eq:e2circ_rho_p}. On that curve, the radial frequency is a function only of $p$, denoted $\Omega_r^{\odot}(p)$; see Eq.~\eqref{seq:Omega_r_circ_p}. To find the ISCO, which is the marginally stable circular orbit, we must find $p$ such that $\Omega_r^{\odot}=0$. However, we must account for the fact that our expansion for $\Omega_r^{\odot}(p)$ is ill behaved around $p=6$:  the geodesic term in Eq.~\eqref{seq:Omega_r_circ_p} scales as $\sqrt{p-6}$, such that only the right-handed limit is physically well defined and vanishing, whereas the subleading term scales as $1/\sqrt{p-6}$ and thus blows up. Since we are working perturbatively and performing a Taylor expansion, we want to work with regular quantities. Fortunately, we have found in Eq.~\eqref{eq:Omega_r_circ_squared_p} that $(\Omega_r^{\odot})^2$ is regular around $p=6$, so we will instead solve for $(\Omega_r^{\odot})^2=0$. This choice is in line with Refs.~\cite{Barack:2010ny, Barack:2010tm, Barack:2011ed}, which solve for $W = \Omega_r^2/\Omega_\phi^2=0$.
To do this in practice, we write the following ansatz for the value of the semilatus rectum on the ISCO: 
\begin{equation}
p_\text{ISCO} = 6 + \,\varepsilon \,\delta p_\text{ISCO} + \mathcal{O}(\varepsilon^2).
\end{equation}
Plugging the ansatz for $p_\text{ISCO}$ into Eq.~\eqref{eq:Omega_r_circ_squared_p}, reexpanding in $\varepsilon$ and solving for $p_\text{ISCO}$, we find that
\begin{align}
    p_\mathrm{ISCO} &= 6 + 2\varepsilon\bigl[3 \,\rho(1/6)-2\bigr] + \mathcal{O}(\varepsilon^2)\,.
\end{align}
Injecting this value into Eq.~\eqref{eq:ycirc_p}, we finally find that the frequency at the ISCO reads
\begin{align}\label{eq:yISCO}
    y_\text{ISCO} &= \frac{1}{6} + \varepsilon\,\frac{3 \,\rho(1/6)-2}{18} + \mathcal{O}(\varepsilon^2) \,.
\end{align}
This is of course in agreement with the expression we would have found by directly solving \mbox{$W=(\Omega_r^{\odot}/\Omega_\phi^{\odot})^2=0$} in Eq.~\eqref{eq:W_circ}.

Finally, we compute the 1SF corrections to the constants of motion at the ISCO. We inject the expression~\eqref{eq:yISCO} for $y_\text{ISCO}$ into Eqs.~\eqref{eq:z_E_L_Omegar_circ_y_expansion} and \eqref{eq:z0_E0_L0_Omegar0_circ_y}, and expand in small $\varepsilon$. We find that
\begin{subequations}\begin{align}
    z^\text{ISCO} &= \frac{1}{\sqrt{2}} + \varepsilon  \, z_{(1)}^\text{ISCO} + \mathcal{O}(\varepsilon^2)  \,, \\*
    \hat{E}^\text{ISCO} &= \frac{2\sqrt{2}}{3} + \varepsilon  \, \hat{E}_{(1)}^\text{ISCO} + \mathcal{O}(\varepsilon^2) \,, \\*
    \hat{L}^\text{ISCO} &= 2 \sqrt{3}+ \varepsilon \,\hat{L}_{(1)}^\text{ISCO} + \mathcal{O}(\varepsilon^2) \,,
\end{align}\end{subequations}
where \mbox{$\hat{E}_{(1)}^\text{ISCO} = \hat{E}_{(1)}^{\odot}(6)$,} \mbox{$\hat{L}_{(1)}^\text{ISCO} = \hat{L}^{\odot}_{(1)}(6)$,}  and 
\begin{align}
    z_{(1)}^\text{ISCO} = \frac{1}{6\sqrt{2}}\left[2-3 \rho\left(\frac{1}{6}\right)\right]+  z_{(1)}^{\odot} (6)\,.
\end{align}
By definition, the radial action is vanishing to all orders on the ISCO.

To numerically evaluate the ISCO frequency shift, we obtain $\rho(1/6)$ by computing $\partial_p \langle z_{(1)}\rangle(p,0)$, $\partial_e^2 \langle z_{(1)}\rangle(p,0)$, and $\rho(1/p)$ near the separatrix [using Eq.~\eqref{eq:Jr1_circ_inTermsOf_rho}] and fitting for its value on the ISCO; see Appendix \ref{app:circular-numerics} for further details. From this procedure we obtain
\begin{align}
    \rho(1/6) = 0.8340(2) \,.
\end{align}
Decomposing the reduced ISCO frequency in powers of the mass ratio--- namely, as 
\begin{equation}
\hat\Omega_\text{ISCO} = \hat\Omega_{(0)}^\text{ISCO}+ \varepsilon\, \hat\Omega_{(1)}^\text{ISCO} + \mathcal{O}(\varepsilon^2), 
\end{equation}
we then find that the relative ISCO frequency shift reads 
\begin{align}
    \frac{\hat\Omega_{(1)}^\text{ISCO}}{\hat\Omega_{(0)}^\text{ISCO}} = \frac{3}{2}\,\rho(1/6)-1 = 0.2510(2) \,,
\end{align}
which is in agreement\footnote{The value reported in Ref.~\cite{Isoyama:2014mja} is $C_\Omega -1 =  \hat{\Omega}^\mathrm{ISCO}_{(1)}/\hat{\Omega}^\mathrm{ISCO}_{(0)} = 0.25101539(4)$.}  with the high-accuracy calculations reported in Ref.~\cite{Isoyama:2014mja}.
Similarly to $\rho$, we follow the methods described in Appendix \ref{app:circular-numerics} to fit for the near-ISCO behavior, leading to
\begin{subequations}
    \begin{align}
     z_{(1)}^\mathrm{ISCO} &= + 0.08885(6) \,,  \\*
    \hat{E}_{(1)}^\mathrm{ISCO} &= +0.019424700(6) \,,
    \\*
    \hat{L}_{(1)}^\mathrm{ISCO} &= -0.80219089(3) \,.
\end{align}
\end{subequations}
After properly accounting for the different conventions, we find very good agreement with the highly accurate redshift for circular orbits used in the 1PA waveform model~\cite{Wardell:2021fyy}; good agreement is also found with Refs.~\cite{LeTiec:2011dp,Leather:2025nhu}

\subsection{Waveform generation}
\label{subsec:waveform_gen}
The most significant application of the conserved quantities is in waveform models that employ balance laws, which express the binary evolution in terms of fluxes of energy and angular momentum out to infinity and into the primary black hole. Here, we outline the construction of such models at 1PA order. We also explain the critical assumption they would currently need to rely on.

As described in Sec.~\ref{sec:formalism}, the general framework for a 1PA waveform model is well understood. If the waveform is decomposed in spin-weighted harmonics, as in 
\begin{equation}
h_+-ih_\times = \frac{m_1}{r}\sum_{\ell m}h_{\ell m}(u){}_{-2}Y_{\ell m}(\theta,\phi), 
\end{equation}
then the expansions~\eqref{eq:metric multiscale expansion} and \eqref{eq:Fourier series} imply that each $(\ell, m)$ mode of the waveform is given by
\begin{multline}
    h_{\ell m} = \sum_{n=-\infty}^\infty\Bigl[ \varepsilon h^{(1)}_{\ell m n}(\pi_A)
    +\varepsilon^2 h^{(2)}_{\ell m n}(\pi_A)\\+{\cal O}(\varepsilon^3)\Bigr]e^{-i(m\varphi_\phi + n\varphi_r)},
\end{multline}
where we have adopted the mode labels $k_A=(n,m)$ common in the literature~\cite{Chapman-Bird:2025xtd}. The time evolution of the phase-space coordinates $(\varphi_A,\pi_A)$, and hence the waveform's time dependence, is governed by the ordinary differential equations~\eqref{eq:phidot} and \eqref{eq:pidot}. In those evolution equations, the 0PA dynamics are well understood and efficiently evaluated in terms of gravitational-wave fluxes~\cite{Sago:2005fn,Hughes:2005qb}. But historically, the 1PA terms $\Omega^{(1)}_A$ and $F_A^{(1)}$ have only been written in terms of the local self-forces $f^\alpha_{(1)}$ and $f^\alpha_{(2)}$ acting on the particle~\cite{Pound:2021qin,Mathews:2025nyb} (or equivalent local expressions in terms of a pseudo-Hamiltonian~\cite{Lewis:2025ydo}).

In Refs.~\cite{Lewis:2025ydo,Nasipak:2025tby} (building on Ref.~\cite{Fujita:2016igj}), we emphasized that the  conservative sector of the complete 1PA dynamics, as represented by the frequency correction $\Omega^{(1)}_A$, can be most straightforwardly computed in terms of the redshift~$\langle z_{(1)}\rangle$. However, this leaves the dissipative 1PA term, $F^{(1)}_A$, as a complicated mix of local 1SF and 2SF quantities; as an example, see Eq. (A10) of Ref.~\cite{Miller:2020bft} for the simple case of quasicircular orbits. Even in that simple case, the local 1PA expressions have never been evaluated. Instead, the complicated local quantities are replaced by simple fluxes. Here, we explain that replacement, generalizing the argument from the quasicircular case in Refs.~\cite{Wardell:2021fyy,Albertini:2022rfe}.

The argument begins from exact, fully nonlinear balance laws at infinity and at the primary's horizon. At future null infinity, the Bondi mass and angular momentum, $P^{\infty}_A=(E^\infty,L^\infty)$, satisfy well-known, exact balance laws~\cite{Compere:2019gft}:
\begin{equation}
\frac{dP^{\infty}_A}{du} = -{\cal F}^{\infty}_A(u,\varepsilon),
\end{equation}
where ${\cal F}^{\infty}_A$ are the fluxes of energy and angular momentum to infinity. Similarly, at the primary black hole's horizon, the black hole mass and spin, $P^{\rm BH}_A=(m_1,s_1)$, satisfy exact balance laws~\cite{Ashtekar:2004cn,Chandrasekaran:2018aop}:
\begin{equation}
\frac{dP^{\rm BH}_A}{dv} = {\cal F}^{\cal H}_A(v,\varepsilon),
\end{equation}
where ${\cal F}^{\cal H}_A$ are the fluxes of energy and angular momentum down the horizon.

These exact balance laws for $P^\infty_A$ and $P^{\rm BH}_A$ translate into approximate ones for the orbital quantities $P_A$, subject to the following unproved assumption: on average, the total (Bondi) mass and angular momentum are equal, up to numerically small differences, to the binary's total \emph{mechanical} energy and angular momentum. Here, by mechanical energy, we mean the sum of the black hole's mass and the orbital energy, and likewise for the mechanical angular momentum. In short, the assumption is
\begin{equation}\label{eq:sum of charges}
P_A + \bigl\langle P^{\rm BH}_A\bigr\rangle = \bigl\langle P_A^{\infty}\bigr\rangle. 
\end{equation}
This relationship is required to hold on each time slice $s=\text{constant}$, where we recall that $s$ reduces to $v$ at the horizon, $t$ at the particle, and $u$ at future null infinity. Taking a time derivative of Eq.~\eqref{eq:sum of charges} and appealing to the exact balance laws above, we obtain 
\begin{equation}\label{eq:Pdot=-F}
\frac{dP_A}{ds} = -{\cal F}_A,
\end{equation}
where we have defined the total \emph{torus-averaged} fluxes
\begin{equation}
    {\cal F}_A := \bigl\langle{\cal F}^{\infty}_A\bigr\rangle + \bigl\langle{\cal F}^{\cal H}_A\bigr\rangle.
\end{equation}

The relationship~\eqref{eq:sum of charges} holds exactly at 0PA order~\cite{Dolan:2012jg, Merlin:2016boc, vanDeMeent:2017oet, Toomani:2021jlo}. However, at least for quasicircular orbits, it is known to fail at 1PA order~\cite{Pound:2019lzj}, specifically at 4PN. In Ref.~\cite{Trestini:2025nzr}, one of us showed concretely that, after orbit-averaging, the Bondi mass differs from the mechanical energy by a 4PN Schott term related to dissipative tail effects. This will be extended to a fully relativistic result in Ref.~\cite{Grant:InPrep}, implying a corrected relationship 
\begin{equation}\label{eq:sum of charges with Schott}
P_A + \bigl\langle P^{\rm BH}_A\bigr\rangle + \delta P^{\rm Schott}_A = \bigl\langle P_A^{\infty}\bigr\rangle. 
\end{equation}
In principle, an exact 1PA formula for $dP_A/ds$ can be obtained by differentiating this equality, but $\delta P^{\rm Schott}_A$, which depends on our choice of spacetime foliation, is not yet known for eccentric orbits.  
On the other hand, the results in Ref.~\cite{Pound:2019lzj} suggest that the Schott term is numerically small (in addition to carrying a relative factor of $\varepsilon$) and it has been neglected in all 1PA waveform models; see Ref.~\cite{Albertini:2022rfe} for a tentative assessment of the impact of this small error on gravitational-wave phasing. 

If we proceed cavalierly under the assumption that $\delta P^{\rm Schott}_A$ is numerically small, then the evolution equations~\eqref{eq:phidot} and \eqref{eq:pidot} can be replaced with
\begin{align}
    \frac{d\varphi_A}{dt} &= \Omega_A^{(0)}( \hat P_B) + \varepsilon\Omega_A^{(1)}(\hat P_B),\label{eq:phidot FC}\\
    \frac{d \hat P_A}{dt} &= -\varepsilon\left[{\cal F}^{(0)}_A(\hat P_B) + \varepsilon{\cal F}^{(1)}_A (\hat P_B)\right].\label{eq:Pdot FC}
\end{align}
Here, given the definition of $s$, we freely use $t$ at the particle with the understanding that it is replaced by $u$ in the waveform. The 0PA forcing function ${\cal F}^{(0)}_A(\hat P_B)$ is the standard 0PA flux (normalized by $m_2$) computed from the first-order waveform mode amplitudes $h^{(1)}_{\ell mn}$ (at the horizon and infinity)~\cite{Pound:2021qin}, and ${\cal F}^{(1)}_A$ is computed from products of the amplitudes $h^{(1)}_{\ell mn}$ and $h^{(2)}_{\ell mn}$.\footnote{In principle this includes 1PA fluxes at the primary's horizon, but they have not yet been included in 1PA waveform models. We again refer to Ref.~\cite{Albertini:2022rfe} for a discussion of how this omission impacts the waveform phase.} The frequency correction $\Omega^{(1)}_A(\hat P_B)$ is given by Eq.~\eqref{eq:Omega1}.

Even if the field equations are solved on a grid of $(p,e)$ values, one can still work directly with the evolution equations~\eqref{eq:phidot FC} and \eqref{eq:Pdot FC} by adopting the ``fixed constants'' gauge discussed in Sec.~\ref{sec:1SF}. In this gauge the quasi-Keplerian parameters, denoted $\pi^{\rm FC}_A$, are related to $(E,L)$ by the Schwarzschild-geodesic relationship: $P_A(\pi^{\rm FC}_{B},\varepsilon)=P^{(0)}_A(\pi^{\rm FC}_{B})$. Hence, while evolving $P_A$, one can immediately find stored data at corresponding values of $\pi^{\rm FC}_A$.

Alternatively, from Eq.~\eqref{eq:Pdot FC}, we can immediately derive evolution equations for our fixed-frequency variables; recall these satisfy $\Omega_A(\pi_B,\varepsilon)=\Omega^{(0)}_A(\pi_B)$. Given the chain rule 
$\dot{\pi}_A = (\partial\pi_A/\partial P_B) \dot{P}_B$,
we have
\begin{subequations}\begin{align}
    \frac{d\varphi_A}{dt} &= \Omega_A^{(0)}(\pi_B),\\
    \frac{d\pi_A}{dt} &= -\varepsilon\frac{\partial\pi_A}{\partial P_B}\Bigl\{{\cal F}^{(0)}_B(\pi_C) \nonumber\\*
    &\qquad\ \quad + \varepsilon\left[{\cal F}^{(1)}_B (\pi_C) + P^{(1)}_{C}\partial_{P^{(0)}_C}{\cal F}^{(0)}_B\right]\Bigr\}\,,
\end{align}\end{subequations}
where the Jacobian is given by
\begin{align}\label{eq:dpidP}
    \frac{\partial\pi_A}{\partial P_B} &= \begin{pmatrix}
        \frac{\partial E}{\partial p} & \frac{\partial E}{\partial e}\\
        \frac{\partial L}{\partial p} & \frac{\partial L}{\partial e}
    \end{pmatrix}^{\!-1}= \frac{1}{\mathrm{det}\left|\frac{\partial P_A}{\partial\pi_B}\right|}\!\begin{pmatrix}
        \frac{\partial L}{\partial p} & -\frac{\partial E}{\partial e}\\
        -\frac{\partial L}{\partial p}& \frac{\partial E}{\partial e}
    \end{pmatrix}\!.
\end{align}
This matrix can be expanded for small $\varepsilon$ as
\begin{equation}
    \frac{\partial\pi_A}{\partial P_B} =  \frac{\partial \pi_A}{\partial P^{(0)}_B} - \varepsilon \frac{\partial \pi_A}{\partial P^{(0)}_C} \frac{\partial P^{(1)}_C}{\partial \pi_D} \frac{\partial \pi_D}{\partial P^{(0)}_B} +{\cal O}(\varepsilon^2), 
\end{equation}
where it is understood that $\partial \pi_A/\partial P^{(0)}_B$ represents the inverse of $\partial P^{(0)}_B/\partial \pi_A$. 
However, leaving the Jacobian in \emph{unexpanded} form is likely to lead to a more accurate waveform model, in analogy with the ``resummed'' models for quasicircular binaries in Refs.~\cite{Mathews:2025txc,Honet:2025gge,Honet:2025lmk}. More specifically, it is likely important to leave the determinant denominator in Eq.~\eqref{eq:dpidP} intact, as it preserves the Jacobian's singularity at the \emph{corrected} separatrix rather than at the background one.

Regardless of whether the Jacobian $\partial\pi_A/\partial P_B$ is expanded, we note that since the corrections $P^{(1)}_A$ are already computed as first derivatives of the redshift, the Jacobian requires \emph{second} derivatives of the redshift. In App.~\ref{app:hessian}, we derive a formula which can be used to express $\partial P_A^{(1)}/\partial \pi_B$ explicitly in terms of $\partial^2 \langle z_1\rangle/(\partial \pi_A \partial \pi_B)$.

\section{Discussion and future applications}
\label{sec:conclusion}

In this paper we have carried out the first comprehensive calculations of the energy, angular momentum, and radial action of nonspinning binaries at 1SF order beyond the geodesic approximation, providing analytical results up to 9PN and numerical results across the parameter space up to eccentricity $e=0.5$. These data and formulas are provided in the Supplemental Material~\cite{supp}. Although, for simplicity, we have restricted our calculations to a nonspinning primary, the formalism~\cite{Lewis:2025ydo} as well as the numerical~\cite{vandeMeent:2015lxa,Nasipak:2025tby} and analytical~\cite{Bini:2016dvs,Munna:2023wce} methods are already well developed for eccentric orbits around a Kerr primary.

Our primary motivation for studying these constants of motion is described in Sec.~\ref{subsec:waveform_gen}: to utilize them in 1PA waveform generation, capitalizing on the streamlined form that the waveform takes when written in terms of balance laws. Without access to the constants of motion, 1PA waveform models require calculation of local 2SF forces, which have never been carried out. Balance laws evade that requirement, as demonstrated in the quasicircular case~\cite{Wardell:2021fyy}. 

Actually implementing the 1PA waveform-generation scheme outlined in Sec.~\ref{subsec:waveform_gen} will require the energy and angular momentum fluxes contributed by the second-order metric perturbation $h^{(2)}_{\alpha\beta}$. Such results are currently only available for quasicircular orbits around a Schwarzschild black hole~\cite{Warburton:2021kwk,Warburton:2024xnr}. Extending that calculation to more generic binary configurations will require extending the current 2SF methods (detailed in Refs.~\cite{Miller:2020bft,Spiers:2023mor,Miller:2023ers,Upton:2025bja,Cunningham:2024dog}). References~\cite{Upton:2021oxf,Toomani:2021jlo,Leather:2023dzj,Wei:2025lva,Spiers:2023cip,Osburn:2022bby,PanossoMacedo:2024pox,Bourg:2024cgh,Dolan:2021ijg,Dolan:2023enf,Wardell:2024yoi} describe work toward that goal. 

As explained in Sec.~\ref{subsec:waveform_gen}, a truly complete 1PA waveform would also require calculation of the 1PA Schott corrections $\delta P^{\rm Schott}_A$ to the constants of motion. Although these terms are known for quasicircular orbits~\cite{Trestini:2025nzr,Grant:InPrep}, they are not yet known for eccentric orbits. Moreover, 1PA fluxes through the primary's horizon, as well as recently discovered ``memory distortion'' terms in the fluxes to infinity~\cite{Cunningham:2024dog}, must also be included for a genuinely complete 1PA waveform. These missing pieces merit further study, but we expect all of them to be numerically small relative to other 1PA terms.

In lieu of 2SF flux results for eccentric orbits, there is a closer-at-hand opportunity to develop 1PA waveform models: combining SF and PN information. As demonstrated in Refs.~\cite{Honet:2025gge,Honet:2025lmk} for quasicircular orbits, writing the waveform in terms of balance laws allows for straightforward hybridization of SF and PN information~\cite{Damour:2014jta,Blanchet:2023bwj,Cho:2022syn,Marsat:2014xea,Henry:2022ccf}, and the resulting waveform model might be sufficiently accurate for high-precision EMRI science with LISA (though more work is needed to assess this, particularly for more generic binary configurations), as well as being highly accurate across the full range of possible mass ratios. To see how this hybridization works, note that in both SF and PN theory, we can write the waveform in an invariant form:
\begin{align}
    h_+-ih_\times &= \frac{1}{r}\sum_{\ell m} h_{\ell m}(m_A,{\cal P}_A,\varphi_A) {}_{-2}Y_{\ell m}(\theta,\phi),\\
    \frac{d\varphi_A}{ds} &= \Omega_A(m_B,{\cal P}_B),\\
    \frac{d{\cal P}_A}{ds} &= -{\cal F}_A(m_B,{\cal P}_B),
\end{align}
as follows from the general arguments of Sec.~\ref{subsec:waveform_gen}, where we have defined $m_{A}=(m_1,m_2)$ and the corrected constants of motion ${\cal P}_A = P_A+\delta P^{\rm Schott}_A$. For completeness, we also note that the masses evolve according to the fluxes through each black hole's horizon ${\cal H}_A$,
\begin{equation}
    \frac{dm_A}{ds} = {\cal F}^{{\cal H}_A}_A(m_B,{\cal P}_B),
\end{equation}
as do the spins (which cannot remain zero at 1PA order but which we elide here for simplicity). Hybridization can then be carried out immediately at the level of the invariant functions $\Omega_A(m_B,{\cal P}_B)$, ${\cal F}_A(m_B,{\cal P}_B)$, and $h_{\ell m}(m_A,\varphi_A,{\cal P}_A)$. This is easily done using composite expansions: for a function $f(m_A,\varphi_A,{\cal P}_A)$, given an expansion $f^{\rm SF}$ in powers of the mass ratio and a PN expansion $f^{\rm PN}$, we write
\begin{equation}
    f = f^{\rm SF} + f^{\rm PN} - f^{\text{SF-PN}}
\end{equation}
where $f^{\text{SF-PN}}$ is $f^{\rm PN}$ reexpanded in powers of the mass ratio to the same order as $f^{\rm SF}$, serving to remove otherwise doubly counted information. In the present case, one could, for instance, use the 3PN fluxes for non-spinning eccentric orbits~\cite{Arun:2007rg, Arun:2007sg, Arun:2009mc, Trestini:2025yyc}.

One could also calculate inputs for such a hybrid model using PM results for scattering orbits. This could be achieved using the ``boundary to bound'' link between scattering data and bound-orbit data~\cite{Damour:1985,Kalin:2019rwq,Kalin:2019inp,Saketh:2021sri,Cho:2021arx,Gonzo:2023goe,Adamo:2024oxy,Gonzo:2024xjk,Khalaf:2025jpt}: the frequencies $\Omega_A(m_B,{\cal P}_B)$ and fluxes ${\cal F}_A(m_B,{\cal P}_B)$ for bound orbits can be obtained directly from the scattering angle, elapsed coordinate time, and total emitted fluxes for scattering orbits; the first two of these can also be obtained from the single relationship $J_r(m_A, {\cal P}_B)$ via the first law of of black hole binary mechanics~\cite{Gonzo:2024xjk}. This prospect of utilizing PM results is particularly promising, because 2SF terms in the PM fluxes should soon be available~\cite{Driesse:2024xad,Driesse:2024feo,Bern:2024adl,Bern:2025zno,Dlapa:2025biy,Bern:2025wyd}, complementing PN expansions of the 2SF fluxes. However, this approach will require extending boundary-to-bound relationships, which (in their current form) break down beyond 0SF order due to nonlocal-in-time tail effects~\cite{Cho:2021arx,Dlapa:2024cje}; alternatively, the program embarked on in Ref.~\cite{Bini:2019nra} can sidestep this breakdown through a separate treatment of nonlocal terms.

We expect such hybridizations, whether using only SF and PN information or SF, PN, and PM information all together, to be especially valuable in the case of a spinning, Kerr primary. 2SF flux data for eccentric orbits will be particularly challenging to generate in that case, while, as pointed out at the beginning of this section, all the other 1PA information can be readily computed within self-force theory by employing the same tools we have used in this paper~\cite{vandeMeent:2015lxa,Nasipak:2025tby,Bini:2016dvs,Munna:2023wce}.

An open question in these constructions, and in waveform generation more broadly, is which choice of phase-space gauge is most useful: the fixed-constants gauge, the fixed-frequencies gauge, a fixed-action-variables gauge, or some other gauge? 
Reference~\cite{Mathews:2025nyb} showed that the 1PA waveform is ultimately invariant under this choice up to 2PA residuals, and the argument there is easily extended to $n$PA. But in that same reference, one of us emphasized that the presence of the $(n+1)$PA residuals between the different gauges will lead to numerically different waveforms.  There is some indication~\cite{Wardell:2021fyy} that a fixed-frequencies gauge provides optimal accuracy, because it expresses the waveform in terms of the waveform's own frequencies. The fixed-frequencies gauge also has the advantage of offering a clean, all-orders identification of the separatrix. Due to the simplification discussed below Eq.~\eqref{eq:<z1>}, it additionally provides the most straightforward path to calculating the SF-corrected relationships between frequencies and constants of motion (our main reason for utilizing it in this paper). However, for eccentric orbits, it has the disadvantage of becoming singular at the isofrequency degeneracy curve, and it can cause a loss of precision due to large cancellations in Jacobians, as described at the end of Sec.~\ref{sec:1SFcorrections}. The relative merits of the fixed-constants gauge, in contrast, have been little explored. Ultimately, to assess the practical utility and accuracy of different phase-space gauges, waveform models in different gauges should be constructed and compared against one another (and compared to numerical relativity). We refer to Refs.~\cite{VanDeMeent:2018cgn,Drummond:2023wqc,Piovano:2024yks,Mathews:2025nyb,Lewis:2025ydo} for further discussion.

Another possible use for the 1SF constants of motion is in the construction of EOB models. The Hamiltonian in action-angle coordinates is a gauge-invariant representation of the conservative sector of the dynamics, simply equal to the energy as a function of the action variables: $H=E(J_r,L,m_1,m_2)$. Such a representation was employed in the original motivating discussion for EOB~\cite{Buonanno:1998gg}, and one could build EOB models directly in action-angle coordinates. The invariant 1SF contribution to the Hamiltonian could then be immediately incorporated over the full range of orbital parameters. Even if such an approach is not pursued, 1SF data for generic eccentricities should offer a path to improving SF-informed EOB models for eccentric binaries, such as the TEOB model in Ref.~\cite{Nagar:2022fep}, which have historically been limited to low-order small-eccentricity expansions of 1SF results~\cite{LeTiec:2015kgg,Akcay:2015pjz,Bini:2015bfb}.

Outside of these direct uses in waveform models, conserved quantities provide useful, invariant ways of analyzing the dynamics. For example, as originally outlined by Le Tiec in Ref.~\cite{LeTiec:2015kgg}, 1SF definitions of $(E,L,J_r)$ provide invariant characterizations of surfaces in phase space, such as circular orbits (defined by $J_r=0$) and the separatrix (defined by $\Omega_r=0$). Using these definitions, in this paper we have computed the location of circular orbits in the $(p,e)$ plane, as an invariant curve $e_{\odot}(p)$, in the process highlighting subtleties in the definition of eccentricity for accelerated orbits. Likewise, we have computed the location of the separatrix in the $(E,L)$ plane, as an invariant curve $E_{\star}(L)$ [noting that its location $p_{\star}(e)$ is unchanged from the geodesic case as a consequence of our definition of $p$ and $e$]. Another curve one might analyze, as already noted by Le Tiec, is the isofrequency degeneracy curve defined by ${\rm det}|\partial\Omega_A/\partial P_B|=0$. We leave this to future work. 

\begin{acknowledgments}

D.T. thanks Benjamin Leather for providing data associated to Ref.~\cite{Leather:2025nhu} and Leor Barack for an insightful conversation about the ISCO shift.
This work used the \texttt{Black Hole Perturbation Toolkit}~\cite{BHPToolkit}, \texttt{PNpedia}~\cite{Trestini_PNpedia_2025} and the \texttt{pybhpt} package~\cite{pybhpt-0.9.8}.

We acknowledge the support of the ERC Consolidator/UKRI Frontier Research Grant GWModels (selected by the ERC and funded by UKRI [grant number EP/Y008251/1]). AP additionally acknowledges the support of a Royal Society University Research Fellowship.
\end{acknowledgments}

\appendix

\section{Elliptic integrals}
\label{app:elliptic}

We recall the definitions and properties of elliptic integrals, which are used to explicitly express the geodesic frequencies in terms of $(p,e)$ in Sec.~\ref{sec:geo}.

We adopt the conventions of \texttt{Wolfram Language} when defining elliptic integral functions~\cite{EllipticIntegrals} (these differ from the definition of the NIST Digital Library
of Mathematical Functions by the power of the last argument~\cite{NIST_elliptic}). Consequently, Legendre's complete elliptic integrals of the first, second, and third kind, respectively, take the forms 
\begin{subequations}\label{eq:K_E_Pi_def}\begin{align}
    \label{seq:K_def}
   \dK(m)  &= \int_0^{\pi/2} \frac{\dd\theta}{\sqrt{1-m \sin^2\theta}} \,,
    \\
    \label{seq:E_def}
   \dE(m) &= \int_0^{\pi/2} \dd\theta\,{\sqrt{1-m \sin^2\theta}} \,,
    \\
    \label{seq:Pi_def}
    \Pi(n| m) &= \int_0^{\pi/2} \frac{\dd\theta}{(1-n\sin^2\theta)\sqrt{1-m \sin^2\theta}} \,.
\end{align}\end{subequations}

In order to perform the PN expansions of the frequencies, we will require the following small-$m$ expansions~\cite{EllipticIntegrals,Witzany:2024ttz}: 
\begin{subequations}\label{eq:PNseries_E_K_Pi}\begin{align}
   \dK(m) &= \frac{\pi}{2} \sum_{i=0}^\infty \frac{[(1/2)_i]^2 m^i}{(i!)^2}\,,\\
   \dE(m) &= \frac{\pi}{2} \sum_{i=0}^\infty \frac{(-1/2)_i\,(1/2)_i m^i}{(i!)^2} \,,\\
    \Pi(n|m) &=  \frac{\pi}{2} \sum_{i=0}^\infty \frac{[(1/2)_i]^2 }{(i!)^2}\Bigg[ \frac{i!}{ n^i\sqrt{1-n} (1/2)_i } \nonumber\\*
    &  \qquad\qquad\qquad- \frac{2i}{n} \sum_{j=0}^{i-1} \frac{(1-i)_j}{(3/2)_j}\left(1- \frac{1}{n}\right)^j\Bigg] m^i \,, 
\end{align}\end{subequations}
where $(a)_i = \Gamma(a+n)/\Gamma(a) = a(a+1)\cdots(a+i-1)$ is the Pochhammer symbol. In practice, \texttt{Mathematica} is able to directly perform the expansions of the associated functions $\texttt{EllipticK}$ and \texttt{EllipticE} using the \texttt{Series} function, but this does not work with \texttt{EllipticPi}; it was necessary to replace the latter by  its representation in terms of the Appell hypergeometric function of two variables $F_1(\cdots)$, denoted as \texttt{AppellF1} in \texttt{Mathematica}. This relation reads $\Pi(n|m) = \frac{\pi}{2} F_1\left(\frac{1}{2};1,\frac{1}{2};1;n,m\right)$, and one can then proceed with the small-$m$ expansion of the latter using the $\texttt{Series}$ function.

We also report the following relation between elliptic integrals, which is given in Eq. (17.7.17) of Ref.~\cite{AbramowitzStegun64}:
\begin{align}\label{eq:Pi_rule}
    \Pi(n|m) = \frac{n(m-1)}{(n-1)(n-m)}\Pi \!\left(\!\frac{n-m}{n-1} \Big| m\!\right) - \frac{m}{n-m}K(m) \,.
\end{align}
This formula allows us to reconcile different formulations for the fundamental frequencies that are present in the literature by noticing that, for $m = 4e/(p-6+2e)$, we have
\begin{subequations}\label{eq:Pi_rule_applications}\begin{align}
    n = - \frac{2e}{1-e} &\Rightarrow \frac{n-m}{m-1} = \frac{2e (p-4)}{(1+e)(p-6+2e)} \,,\\
    n = \frac{4e}{p+2e-2} &\Rightarrow  \frac{n-m}{m-1} = \frac{16e}{(p-2-2e)(p-6+2e)} \,.
\end{align}\end{subequations}

\section{Circular orbits from periastron advance}
\label{app:circular}

In this Appendix, we derive the circular link between $p$ and $e$ at postgeodesic order in terms of $\rho(x)$, and we deduce the relation between $\rho(1/p)$ and $J_{r(1)}(p,0)$, which is used in the study of circular orbits in Sec.~\ref{sec:applications}.

First, we recall that $y=(m_1 \Omega_\phi)^{2/3}$ was defined in Eq.~\eqref{eq:def_y_lambda}, and we recall the definition $x = [(m_1+m_2)\Omega_\phi]^{2/3}$. We then find the relation
\begin{align} 
x = y(1+\varepsilon)^{2/3} = y \left[1+ \frac{2}{3} \varepsilon + \mathcal{O}(\varepsilon^2)\right]\,.
\end{align}
Following Refs.~\cite{Barack:2010tm,Barack:2010ny,Barack:2011ed}, we also define
\begin{align}
    W  = K^{-2} = \left(\frac{\hat\Omega_r}{\hat\Omega_\phi}\right)^{2}\,.
\end{align}
On a circular orbit, the two frequencies are not independent; they satisfy a circular link $\hat{\Omega}_r = \hat{\Omega}_r^{\odot}(\hat\Omega_\phi)$. This link can be expressed in terms of $W$, which reads \cite{Barack:2010ny} 
\begin{align}\label{eq:W_circ}
   W^{\odot} = \left(\!\frac{\hat{\Omega}_r^{\odot}(\hat{\Omega}_\phi)}{\hat{\Omega}_\phi} \!\right)^2 & =  1 - 6 x + \varepsilon \rho(x) + \mathcal{O}(\varepsilon^2) \nonumber\\*
   & = 1- 6 y + \varepsilon \bigl[\rho(y) - 4 y \bigr] + \mathcal{O}(\varepsilon^2),
\end{align}
where $\rho(x)$ is given (i) numerically in Table II of Ref.~\cite{Barack:2010ny} for $x \in [1/80,1/6]$ and in the Supplemental Material of~\cite{vandeMeent:2016hel} for $x \in [1/1000,1/10]$; and  (ii) as a PN series in Eq.~(9) of Ref.~\cite{Bini:2016qtx}.

At geodesic order, $p_{\odot} = 1/y + \mathcal{O}(\varepsilon)$ and $e_{\odot}^2 = \mathcal{O}(\varepsilon)$, so we introduce the ansatz
\begin{align}
p_{\odot} &= 1/y + \varepsilon \, \delta p_{\odot} +\mathcal{O}(\varepsilon^2),\\
e_{\odot}^2 &= \varepsilon \, \delta e_{\odot}^2 +\mathcal{O}(\varepsilon^2).
\end{align}
We plug this ansatz into the exact expression of $(\hat\Omega_\phi, \hat\Omega_r)$ in terms of $(p,e)$ and Taylor-expand in $\varepsilon$; see App.~\ref{app:elliptic} for how to do this in practice. Thanks to these expressions, we find that
\begin{subequations}\begin{align}
   y = \hat\Omega_\phi^{2/3}(p_{\odot},e_{\odot}^2) &= y  + \varepsilon \mathcal{A}(y, \delta p, \delta e),  \\
    W^{\odot} =K^{-2}(p_{\odot},e_{\odot}^2)  &= 1-6y + \varepsilon \,\mathcal{B}(y, \delta p, \delta e), 
\end{align}\end{subequations}
where we have ignored $\mathcal{O}(\varepsilon^2)$ terms and where
\begin{subequations}\begin{align}
    \mathcal{A}(y, \delta p, \delta e) &= - y^2 \delta p + \frac{y(-1+10 y -22 y^2)}{1-8y +12 y^2} \delta e_{\odot}^2  \,,\\
    \mathcal{B}(y, \delta p, \delta e) &= 6 y^2 \delta p + \frac{3 y^2}{-2+12 y} \delta e_{\odot}^2 \,.
\end{align}\end{subequations}
Equating this to Eq.~\eqref{eq:W_circ}, we obtain the constraints
\begin{subequations}\begin{align}
    \mathcal{A}(y, \delta p, \delta e) &= 0 \,, \,\\
    \mathcal{B}(y, \delta p, \delta e) &= \rho(y) - 4 y  \,,
\end{align}\end{subequations}
which translate to 
\begin{subequations}\begin{align}
    \delta p_{\odot} &= - \frac{2(1-10y +22 y^2)\big[4y - \rho(y)\big]}{3 y^2 (4- 39 y + 86 y^2)} \,,\\*
    \delta e_{\odot}^2 &= \frac{2(1-2y)(1-6y)\big[4y - \rho(y)\big]}{3 y (4- 39 y + 86 y^2)} \,.
\end{align}\end{subequations}
Thus, using $y = 1/p + \mathcal{O}(\varepsilon)$, we have
\begin{align}
    \delta e_{\odot}^2(p) = \frac{2(p-6)(p-2)}{3(86-39p+4p^2)}\bigl[4-p \,\rho(1/p)\bigr] \,.
\end{align}
Injecting this expression into Eq.~\eqref{eq:delta_e2_inTermsOf_Jr1} finally leads to 
\begin{align}
    J_{r(1)}^{\odot}(p) =  \frac{p^{3/2} (p-6)^{3/2} \bigl[p \,\rho(1/p) - 4\bigr] }{3 \sqrt{p-3} (86 -39 p +4 p^2)} \,.
\end{align}
Using our PN expansion for $J_{r}^{(1)}(p,0)$ and the PN expansion of $\rho(x)$ provided in Ref.~\cite{Bini:2016qtx}, we have checked that the previous result is consistent through 9PN order. 

\section{Numerical evaluation of circular limits}
\label{app:circular-numerics}

We describe our numerical approach to computing 1SF corrections along circular orbits and on the ISCO, as reported in Secs.~\ref{sec:1SFcorrections} and \ref{sec:1SF-ISCO}. While frequency derivatives are well defined in the circular limit, the Jacobian elements $\partial e/\partial \hat\Omega_r$ and $\partial e/\partial \hat\Omega_\phi$ are singular at $e=0$. Thus, care must be taken when numerically evaluating derivatives in this limit. First, we expand all quantities about $e=0$ and find
\begin{subequations}\begin{align}
    \frac{\partial p}{\partial \hat\Omega_r} &= \frac{4p^3\sqrt{p-6}(p^2-10p+22)}{3(4p^2-39p+86)} + \mathcal{O}(e^2),
    \\
    \frac{\partial p}{\partial \hat\Omega_\phi} &= -\frac{2p^{5/2}(2p^3-32p^2+165p-266)}{3(4p^2-39p+86)} + \mathcal{O}(e^2),
    \\
    \frac{\partial e}{\partial \hat\Omega_r} &= -\frac{2p^2(p-6)^{3/2}(p-2)}{3(4p^2-39p+86)e} + \mathcal{O}(e),
    \\
    \frac{\partial e}{\partial \hat\Omega_\phi} &= \frac{2p^{3/2}(p-6)(p-2)(p-8)}{3(4p^2-39p+86)e} + \mathcal{O}(e).
\end{align}\end{subequations}
We then expand the redshift in small eccentricity:
\begin{align}\label{eq:z1_small_e_expansion}
    \langle z_{(1)}\rangle(p,e) = z_{(1)}^{[0]}(p) + e^2\,z^{[2]}_{(1)}(p)  + \mathcal{O}(e^4)\,,
\end{align}
where $z_{(1)}^{[0]}(p) = \langle z_{(1)}\rangle(p,0)$ and
\begin{align}
    z^{[2]}_{(1)}(p)  = \frac{1}{2} \left.\frac{\partial^2 \langle z_{(1)} \rangle}{\partial e^2}\right\vert_{e=0} \,.
\end{align}
In Eq.~\eqref{eq:z1_small_e_expansion}, we have  leveraged ${\partial \langle z_{(1)} \rangle}/{\partial e} = 0$ for $e=0$. Note that, for this work, we directly compute $z_{(1)}^{[0]}(p)$, but we obtain $z_{(1)}^{[2]}(p)$ via finite differencing.
 
The 1SF corrections to the constants of motion can then be expressed in terms of radial derivatives of the redshift at zero eccentricity and the eccentricity corrections $z^{[2]}_{(1)}$ (or, alternatively, first derivatives of the redshift with respect to $e^2$), leading to 
\begin{subequations}\begin{align}
    \hat{E}_{(1)}^{\odot}(p) &= \frac{1}{2} z_{(1)}^{[0]}(p) +\frac{p(p-2)}{3(4p^2-39p+86)}\ \frac{d z_{(1)}^{[0]}}{d p} \nonumber\\
    &\quad + \frac{4(p-6)(p-2)}{3(4p^2-39p+86)}\ z^{[2]}_{(1)} (p) \,,
    \\
    \hat{L}_{(1)}^{\odot}(p) &=
    \frac{p^{5/2}(2p^3-32p^2+165p-266)}{3(4p^2-39p+86)}\ \frac{d z_{(1)}^{[0]}}{d p} \nonumber\\
    &\quad - \frac{2p^{3/2}(p-6)(p-2)(p-8)}{3(4p^2-39p+86)}\ z^{[2]}_{(1)} (p) \,,
    \\ \label{eqn:Jr1-circ}
    \hat{J}_{r(1)}^{\odot}(p) &= -\frac{2p^3\sqrt{p-6} (p^2-10p+22)}{3(4p^2-39p+86)}\ \frac{d z_{(1)}^{[0]}}{d p} \nonumber\\
    &\quad +\frac{2p^2{(p-6)^{3/2}} (p-2)}{3(4p^2-39p+86)}\ z^{[2]}_{(1)}(p)  \,.
\end{align}\end{subequations}

Further simplifications need to be made when approaching the ISCO. Expanding in $\delta_0 \equiv \delta\vert_{e=0} = p-6$ leads to the near-ISCO ansatz  
\begin{subequations}\label{eqn:near-circ-near-sep-ansatz}\begin{align}
    z^{[0]}_{(1)}(p) &= a_0 + a_1 \delta_0 + a_2 \delta_0^2 + \mathcal{O}(\delta_0^3),
    \\*
    z^{[2]}_{(1)}(p) &= b_{-1} \delta_0^{-1}  + b_0 + b_1 \delta_0 + \mathcal{O}(\delta_0^2),
\end{align}\end{subequations}
\begin{subequations}\label{eqn:near-circ-near-sep-ansatz}\begin{align}
    z^{[0]}_{(1)}(p) &= a_0 + a_1 (p-6) + a_2 (p-6)^2 + \mathcal{O}\left[(p-6)^3\right],
    \\*
    z^{[2]}_{(1)}(p) &= b_{-1} \delta_0^{-1}  + b_0 + b_1 (p-6) + \mathcal{O}\left[(p-6)^2\right],
\end{align}\end{subequations}
where the coefficients $a_n$ and $b_n$ are not known exactly. In this work, we compute these coefficients by fitting the expansions in Eq.~\eqref{eqn:near-circ-near-sep-ansatz} to numerical values of $z^{[0]}_{(1)}(p)$ and $z^{[2]}_{(1)}(p)$ near the ISCO. Inserting Eq.~\eqref{eqn:near-circ-near-sep-ansatz} into our expression for the radial action \eqref{eqn:Jr1-circ}, we find
\begin{align}
    \hat{J}_{r(1)}^\mathrm{ISCO} &= -24 (b_{-1} + 3 a_1)\sqrt{\delta_0} 
    \\* \notag
    & \quad  - 2(34 b_1 + 12 b_0 + 63 a_1 + 72 a_2)\delta_0^{3/2} 
    + \mathcal{O}(\delta_0^{5/2})\,.
\end{align}
Recalling from Eq.~\eqref{eq:Jr1_circ_inTermsOf_rho} that $\hat{J}_{r(1)}^{\odot}(p)/(p-6)^{3/2}$ must be regular and finite as $p \rightarrow6$, we conclude that \mbox{$b_{-1} = -3 a_1$}; thus, the first term vanishes. This conclusion is consistent with our numerics. Consequently, at leading order we have
\begin{subequations}\begin{align}
    \hat{E}_{(1)}^\mathrm{ISCO} &= \frac{1}{2}a_0 + 2a_1,
    \\*
    \hat{L}_{(1)}^\mathrm{ISCO} &= 12 \sqrt{6} a_1,
    \\*
    \hat{J}_{r(1)}^\mathrm{ISCO} &= 2(39 a_1 -72 a_2 - 12 b_0)(p-6)^{3/2},
\end{align}\end{subequations}
with the postgeodesic correction to the radial action vanishing exactly on the ISCO. Note that these expressions are consistent with the relations in Ref.~\cite{LeTiec:2011dp} between the redshift, the binding energy, and the angular momentum in the circular limit, once we have reexpanded in $x = [(m_1+m_2)\Omega_\phi]^{2/3}$ in terms of $p$ to 1SF order. 

From this, we can also express $\rho(1/6)$ and $\Omega^\mathrm{ISCO}_{(1)}$ in terms of these expansion terms,
\begin{subequations}\begin{align}
    \rho(1/6) &= \frac{2}{3} + \sqrt{2}(4b_0 -13 a_1 + 24 a_2),
    \\
    \frac{\Omega^\mathrm{ISCO}_{(1)}}{\Omega^\mathrm{ISCO}_{(0)}} &= \frac{3}{\sqrt{2}}(4b_0 -13 a_1 + 24 a_2).
\end{align}\end{subequations}
By fitting near-ISCO redshift values, we obtain the following numerical estimates for the expansion coefficients:
\begin{align}
    a_0 \;&= \phantom{-}0.14801375(1) \,, \nonumber
    \\
    a_1 \;&= -0.027291088(3) \,, \nonumber
    \\
    a_2 \;&= \phantom{-}0.0073291(1) \,,
    \\
    b_{-1} \;&= \phantom{-} 0.081873(2) \,, \nonumber
    \\
    b_{0} \;&= -0.10309(3) \,.\nonumber
\end{align}
We obtain these results using a least-squares fit with \texttt{scipy.optimize}.\texttt{curve\_fit} within the Python  \textsc{SciPy} library \cite{2020SciPy-NMeth}. We fit to expansions with the functional forms
\begin{align}
    f(\delta_0) = \sum_{n=n_0}^N A_n \delta_0^n,
\end{align}
where $n_0 = -1$ for $z^{[2]}_{(1)}(p)$ and $\rho(1/p)$, and $n_0=0$ for all other quantities. The fit is weighted by the estimated numerical errors of the input data. We produce a family of fits by varying the number of coefficients $A_n$ and the range of input grid data that we fit over. In particular, we vary $4 \leq N - n_0 \leq 16$ and grid values in the range $\delta_\mathrm{min} \leq \delta_0 \leq 5$, though we do not include data from grid points with estimated relative errors $>100$. We take $5 \times \sigma_{nn}$ to be the estimated error in the fit parameter $A_n$, where $\sigma_{nm}^2$ is the covariance matrix produced from the fit. From the set of fits, we select the three fits that produce the smallest estimated errors $5 \times \sigma_{n_0 n_0}$ for $A_{n_0}$. If the variance of these three estimated values of $A_{n_0}$ is smaller than their estimated errors, then we use the fit with the smallest error. If the variance is greater, then we take the mean and variance of these three fits to get the values and errors of the fit coefficients $A_n$.

\section{Corrections to the separatrix curve}
\label{app:correction-to-sep}

Here we present multiple methods for deriving the first-order correction to the separatrix curve $\hat{E}^{\star}(\hat{L})$ discussed in Sec.~\ref{subsec:sep}. First, we can invert Eq.~\eqref{eq:LsepExpansion} with the ansatz
\begin{align}
    e^{\star}(L) = e^{\star}_{(0)}(\hat{L}) + \varepsilon \, e^{\star}_{(1)}(\hat{L}),
\end{align}
which leads to the relations 
\begin{subequations}\label{eq:e1e0starL}\begin{align}\label{seq:e0starL}
    e^{\star}_{(0)}(\hat{L}) &= \frac{\hat{L}^2-12+2\sqrt{\hat{L}^2\left(\hat{L}^2-12\right)}}{\hat{L}^2+4}\,,\\\label{seq:e1starL}
    e^{\star}_{(1)}(\hat{L}) &= - \left.\frac{\hat{L}^{\star}_{(1)}}{\dd\hat{L}^{\star}_{(0)}/\dd e}\right|_{e=e_{(0)}^{\star}(\hat{L})}.
\end{align}\end{subequations}
Inserting these into Eq.~\eqref{eq:EsepExpansion}, we find
\begin{align}
    \hat{E}^\star_{(1)}(\hat{L}) = E^{\star}_{(1)}(e_{(0)}^{\star}(\hat{L})) - \left[\frac{\dd\hat{E}^{\star}_{(0)}/\dd e}{\dd\hat{L}^{\star}_{(0)}/ \dd e} \hat{L}^{\star}_{(1)}\right]_{e=e_{(0)}^{\star}(\hat{L})},
\end{align}
where $\dd\hat{E}^{\star}_{(0)}/\dd e$ and $\dd\hat{L}^{\star}_{(0)}/\dd e$ can be computed from Eqs.~\eqref{seq:E0_sep} and \eqref{seq:L0_sep}, respectively.

Alternatively, we can consider the perturbations to $\pi_A = (p,e)$ at fixed $\hat{P}_B =(\hat{E},\hat{L})$, such that
\begin{align}
    \pi_A(\hat{P}_B, \epsilon) = \pi_A^{(0)}(\hat{P}_B) + \varepsilon \pi_A^{(1)}(\hat{P}_B),
\end{align}
with
\begin{align}
    \pi_A^{(1)}(\hat{P}_B) = -\frac{\partial \pi_A^{(0)}}{\partial \hat{P}_B} \hat{P}_B^{(1)}.
\end{align}
Demanding that $p = 6 + 2e$ on the separatrix leads to the relation
\begin{align}
    p^{\star}_{(0)}(\hat{P}_B) + \varepsilon p^{\star}_{(1)}(\hat{P}_B) = 6 + 2e^{\star}_{(0)}(\hat{P}_B) +  \varepsilon  \,2e^{\star}_{(1)}(\hat{P}_B).
\end{align}
Expanding in terms of $\hat{E}^{\star}(\hat{L}) = \hat{E}^{\star}_{(0)}(\hat{L})+\varepsilon \, \hat{E}^{\star}_{(1)}(\hat{L})$, we then find
\begin{align}
    \hat{E}^{\star}_{(1)}(\hat{L}) = \frac{p^{\star}_{(1)}(\hat{E}^{\star}_{(0)}(\hat{L}),\hat{L}) - 2 e^{\star}_{(1)}(\hat{E}^{\star}_{(0)}(\hat{L}),\hat{L})}{2 \partial_E e^{\star}_{(0)}(\hat{E}^{\star}_{(0)}(\hat{L}),\hat{L}) - \partial_E p^{\star}_{(0)}(\hat{E}^{\star}_{(0)}(\hat{L}),\hat{L})}.
\end{align}

The final method comes from working in the fixed-constants gauge, rather than the fixed-frequencies gauge used throughout this work. Recall that in this gauge, our numerical $(p,e)$ data grid is instead geodesically linked to $(\hat{E},\hat{L})$---i.e., $\hat{P}_A(\pi_B, \epsilon) = \hat{P}_{A(0)}(\pi_B)$. Then, we can compute
\begin{align}
    \hat{\Omega}_r(\hat{E},\hat{L}) = \hat{\Omega}_{r(0)}(\hat{E}, \hat{L}) + \varepsilon \, \hat{\Omega}_{r(1)}(\hat{E},\hat{L}),
\end{align}
with the first-order correction given by Eq.~\eqref{eq:Omega1}. Inserting Eq.~\eqref{eq:ESepofL} into the expression above, and demanding that $\hat{\Omega}_r$ vanishes on the separatrix, we then find
\begin{align}
    \hat{E}^{\star}_{(1)}(\hat{L}) = -\frac{\hat{\Omega}_{r(1)}(\hat{E}^{\star}_{(0)}(\hat{L}), \hat{L})}{ \partial_{\hat{E}} \hat{\Omega}_{r(0)}(\hat{E}^{\star}_{(0)}(\hat{L}), \hat{L})}.
\end{align}

\section{Relations between partial derivatives}
\label{app:hessian}

Here, we derive relations between (i) the partial derivatives of the conserved energy, angular momentum, and radial action with respect to $p$ and $e$, and (ii)~the second-order partial derivatives of the redshift $\langle z_1\rangle$ with respect to $p$ and $e$. These formulas are needed in one of the formulations   for waveform generation described in Sec.~\ref{subsec:waveform_gen}. 
 
We know the explicit expressions for $\hat\Omega_A(\pi_B)$, but not for the inverses $\pi_A(\hat\Omega_B)$. However, it is straightforward to compute, by matrix inversion, the first partial derivatives of $(p,e)$ with respect to $(\Omega_r, \Omega_\phi)$ as functions of $(p,e)$; see Eq.~\eqref{eq:dpidP}. Here, we extend this method to the case of \textit{second} derivatives with respect to $p$ and $e$.

Using the chain rule, we first obtain the identity  
\begin{align}
    0 = \frac{\p^2 \pi_A}{\p \pi_B \p \pi_C} = \frac{\p \hat{\Omega}_D}{\p \pi_C} \frac{\p \hat{\Omega}_E}{\p \pi_B} \frac{\p^2 \pi_A}{\p \hat{\Omega}_D \p \hat{\Omega}_E} + \frac{\p \pi_A}{\p \hat{\Omega}_E} \frac{\p^2 \hat{\Omega}_E}{\p \pi_C \p \pi_B}.
\end{align}  
Contracting both sides with $(\p \pi_C/\p \hat{\Omega}_D)\times(\p \pi_B/\p \hat{\Omega}_E)$, we find that
\begin{align}
    \frac{\p^2 \pi_A}{\p \hat{\Omega}_B \partial \hat{\Omega}_C} = - \frac{\p \pi_A}{\p \hat{\Omega}_F} \frac{\p \pi_D}{\p \hat{\Omega}_B} \frac{\p \pi_E}{\p \hat{\Omega}_C} \frac{\p^2 \hat{\Omega}_F}{\p \pi_D \p \pi_E}.
\end{align}
For any function $f(p,e)$, we can use this identity to obtain
\begin{align}\label{eq:Hessian id}
\frac{\p^2 f}{\p \hat{\Omega}_B \partial \Omega_C} = \frac{\p \pi_D}{\p \hat{\Omega}_B} \frac{\p \pi_E}{\p \hat{\Omega}_C}\Bigg( \frac{\p^2 f}{\p \pi_D \p \pi_E} - \frac{\p f}{\p \pi_A} \frac{\p \pi_A}{\p \hat{\Omega}_F} \frac{\p^2 \hat{\Omega}_F}{\p \pi_D \p \pi_E}\Bigg).
\end{align}
 
The 1SF correction to the energy, angular momentum, and radial action are given in terms of derivatives $\partial\langle z_{(1)}\rangle/\partial \hat\Omega_A$ in Eq.~\eqref{eqn:E1J1Jr1SF}. 
Using the identity~\eqref{eq:Hessian id} with $f=\langle z_{(1)}\rangle$, we can write their derivatives with respect to $(p,e)$ as
\begin{widetext}\begin{subequations}\begin{align}
    \frac{\p \hat{E}_{(1)}}{\p \pi_A} &= -\frac{\hat{\Omega}_C}{2} \frac{\p \pi_B}{\p \hat{\Omega}_C}\Bigg(\frac{\p^2 \langle z_{(1)} \rangle}{\p \pi_A \p \pi_B} - \frac{\p \langle z_{(1)} \rangle}{\p \pi_D} \frac{\p \pi_D}{\p \hat{\Omega}_E} \frac{\p^2 \hat{\Omega}_E}{\p \pi_A \p \pi_B} \Bigg)  \,,\\*
    \frac{\p \hat{L}_{(1)}}{\p \pi_A}  &= - \frac{1}{2}  \frac{\p \pi_B}{\p \hat{\Omega}_\phi}\Bigg(\frac{\p^2 \langle z_{(1)} \rangle}{\p \pi_A \p \pi_B} - \frac{\p \langle z_{(1)} \rangle}{\p \pi_D} \frac{\p \pi_D}{\p \hat{\Omega}_E} \frac{\p^2 \hat{\Omega}_E}{\p \pi_A \p \pi_B} \Bigg)  \,,\\*
    \frac{\p \hat{J}_{r(1)}}{\p \pi_A}  &= - \frac{1}{2}  \frac{\p \pi_B}{\p \hat{\Omega}_r}\Bigg(\frac{\p^2 \langle z_{(1)} \rangle}{\p \pi_A \p \pi_B} - \frac{\p \langle z_{(1)} \rangle}{\p \pi_D} \frac{\p \pi_D}{\p \hat{\Omega}_E} \frac{\p^2 \hat{\Omega}_E}{\p \pi_A \p \pi_B} \Bigg) \,.
\end{align}\end{subequations}\end{widetext}

\bibliography{references}

@article{Fujita:2009bp,
    author = "Fujita, Ryuichi and Hikida, Wataru",
    title = "{Analytical solutions of bound timelike geodesic orbits in Kerr spacetime}",
    eprint = "0906.1420",
    archivePrefix = "arXiv",
    primaryClass = "gr-qc",
    doi = "10.1088/0264-9381/26/13/135002",
    journal = "Class. Quant. Grav.",
    volume = "26",
    pages = "135002",
    year = "2009"
}

@article{Fujita:2016igj,
    author = "Fujita, Ryuichi and Isoyama, Soichiro and Le Tiec, Alexandre and Nakano, Hiroyuki and Sago, Norichika and Tanaka, Takahiro",
    title = "{Hamiltonian Formulation of the Conservative Self-Force Dynamics in the Kerr Geometry}",
    eprint = "1612.02504",
    archivePrefix = "arXiv",
    primaryClass = "gr-qc",
    reportNumber = "KUNS-2648, YITP-16-122",
    doi = "10.1088/1361-6382/aa7342",
    journal = "Class. Quant. Grav.",
    volume = "34",
    number = "13",
    pages = "134001",
    year = "2017"
}

@article{Witzany:2024ttz,
    author = "Witzany, Vojt{\v{e}}ch and Skoup{\'y}, Viktor and Stein, Leo C. and Tanay, Sashwat",
    title = "{Actions of spinning compact binaries: Spinning particle in Kerr matched to dynamics at 1.5 post-Newtonian order}",
    eprint = "2411.09742",
    archivePrefix = "arXiv",
    primaryClass = "gr-qc",
    doi = "10.1103/PhysRevD.111.044032",
    journal = "Phys. Rev. D",
    volume = "111",
    number = "4",
    pages = "044032",
    year = "2025"
}

@article{Akcay:2015pza,
    author = "Akcay, Sarp and Le Tiec, Alexandre and Barack, Leor and Sago, Norichika and Warburton, Niels",
    title = "{Comparison Between Self-Force and Post-Newtonian Dynamics: Beyond Circular Orbits}",
    eprint = "1503.01374",
    archivePrefix = "arXiv",
    primaryClass = "gr-qc",
    doi = "10.1103/PhysRevD.91.124014",
    journal = "Phys. Rev. D",
    volume = "91",
    number = "12",
    pages = "124014",
    year = "2015"
}

@article{Blanchet:2017rcn,
    author = "Blanchet, Luc and Le Tiec, Alexandre",
    title = "{First Law of Compact Binary Mechanics with Gravitational-Wave Tails}",
    eprint = "1702.06839",
    archivePrefix = "arXiv",
    primaryClass = "gr-qc",
    doi = "10.1088/1361-6382/aa79d7",
    journal = "Class. Quant. Grav.",
    volume = "34",
    number = "16",
    pages = "164001",
    year = "2017"
}

@article{LeTiec:2011ab,
    author = "Le Tiec, Alexandre and Blanchet, Luc and Whiting, Bernard F.",
    title = "{The First Law of Binary Black Hole Mechanics in General Relativity and Post-Newtonian Theory}",
    eprint = "1111.5378",
    archivePrefix = "arXiv",
    primaryClass = "gr-qc",
    doi = "10.1103/PhysRevD.85.064039",
    journal = "Phys. Rev. D",
    volume = "85",
    pages = "064039",
    year = "2012"
}

@article{Blanchet:2012at,
    author = "Blanchet, Luc and Buonanno, Alessandra and Le Tiec, Alexandre",
    title = "{First law of mechanics for black hole binaries with spins}",
    eprint = "1211.1060",
    archivePrefix = "arXiv",
    primaryClass = "gr-qc",
    doi = "10.1103/PhysRevD.87.024030",
    journal = "Phys. Rev. D",
    volume = "87",
    number = "2",
    pages = "024030",
    year = "2013"
}

@article{Trestini:2025yyc,
    author = "Trestini, David",
    title = "{Constants of motion and fundamental frequencies for elliptic orbits at fourth post-Newtonian order}",
    eprint = "2511.10735",
    archivePrefix = "arXiv",
    primaryClass = "gr-qc",
    doi = "10.1088/1361-6382/ae60dc",
    journal = "Class. Quant. Grav.",
    volume = "43",
    number = "9",
    pages = "095009",
    year = "2026"
}

@article{Warburton:2013yj,
    author = "Warburton, Niels and Barack, Leor and Sago, Norichika",
    title = "{Isofrequency pairing of geodesic orbits in Kerr geometry}",
    eprint = "1301.3918",
    archivePrefix = "arXiv",
    primaryClass = "gr-qc",
    doi = "10.1103/PhysRevD.87.084012",
    journal = "Phys. Rev. D",
    volume = "87",
    number = "8",
    pages = "084012",
    year = "2013"
}

@article{Lhost:2024jmw,
    author = "Lhost, Guillaume and Comp{\`e}re, Geoffrey",
    title = "{Approach to the separatrix with eccentric orbits}",
    eprint = "2412.04249",
    archivePrefix = "arXiv",
    primaryClass = "gr-qc",
    doi = "10.21468/SciPostPhysCore.8.3.059",
    journal = "SciPost Phys. Core",
    volume = "8",
    pages = "059",
    year = "2025"
}

@software{Trestini_PNpedia_2025,
author = {Trestini, David},
doi = {10.5281/zenodo.15002834},
month = mar,
title = {{PNpedia}},
url = {https://github.com/davidtrestini/PNpedia},
version = {1.0.0},
year = {2025}
}

@article{Damour:2014jta,
    author = {Damour, Thibault and Jaranowski, Piotr and Sch{\"a}fer, Gerhard},
    title = "{Nonlocal-in-time action for the fourth post-Newtonian conservative dynamics of two-body systems}",
    eprint = "1401.4548",
    archivePrefix = "arXiv",
    primaryClass = "gr-qc",
    doi = "10.1103/PhysRevD.89.064058",
    journal = "Phys. Rev. D",
    volume = "89",
    number = "6",
    pages = "064058",
    year = "2014"
}

@article{Bernard:2017ktp,
    author = "Bernard, Laura and Blanchet, Luc and Faye, Guillaume and Marchand, Tanguy",
    title = "{Center-of-Mass Equations of Motion and Conserved Integrals of Compact Binary Systems at the Fourth Post-Newtonian Order}",
    eprint = "1711.00283",
    archivePrefix = "arXiv",
    primaryClass = "gr-qc",
    doi = "10.1103/PhysRevD.97.044037",
    journal = "Phys. Rev. D",
    volume = "97",
    number = "4",
    pages = "044037",
    year = "2018"
}

@article{Arun:2007rg,
    author = "Arun, K. G. and Blanchet, Luc and Iyer, Bala R. and Qusailah, Moh'd S. S.",
    title = "{Tail effects in the 3PN gravitational wave energy flux of compact binaries in quasi-elliptical orbits}",
    eprint = "0711.0250",
    archivePrefix = "arXiv",
    primaryClass = "gr-qc",
    doi = "10.1103/PhysRevD.77.064034",
    journal = "Phys. Rev. D",
    volume = "77",
    pages = "064034",
    year = "2008"
}

@article{Arun:2009mc,
    author = "Arun, K. G. and Blanchet, Luc and Iyer, Bala R. and Sinha, Siddhartha",
    title = "{Third post-Newtonian angular momentum flux and the secular evolution of orbital elements for inspiralling compact binaries in quasi-elliptical orbits}",
    eprint = "0908.3854",
    archivePrefix = "arXiv",
    primaryClass = "gr-qc",
    doi = "10.1103/PhysRevD.80.124018",
    journal = "Phys. Rev. D",
    volume = "80",
    pages = "124018",
    year = "2009"
}

@article{Arun:2007sg,
    author = "Arun, K. G. and Blanchet, Luc and Iyer, Bala R. and Qusailah, Moh'd S. S.",
    title = "{Inspiralling compact binaries in quasi-elliptical orbits: The Complete 3PN energy flux}",
    eprint = "0711.0302",
    archivePrefix = "arXiv",
    primaryClass = "gr-qc",
    doi = "10.1103/PhysRevD.77.064035",
    journal = "Phys. Rev. D",
    volume = "77",
    pages = "064035",
    year = "2008"
}

@article{Blanchet:2013haa,
    author = "Blanchet, Luc",
    title = "{Post-Newtonian Theory for Gravitational Waves}",
    eprint = "1310.1528",
    archivePrefix = "arXiv",
    primaryClass = "gr-qc",
    doi = "https://link.springer.com/article/10.1007/s41114-024-00050-z",
    journal = "Living Rev. Rel.",
    volume = "27",
    pages = "4",
    year = "2024"
}

@article{VanDeMeent:2018cgn,
    author = "Van De Meent, Maarten and Warburton, Niels",
    title = "{Fast Self-forced Inspirals}",
    eprint = "1802.05281",
    archivePrefix = "arXiv",
    primaryClass = "gr-qc",
    doi = "10.1088/1361-6382/aac8ce",
    journal = "Class. Quant. Grav.",
    volume = "35",
    number = "14",
    pages = "144003",
    year = "2018"
}

@article{vandeMeent:2016hel,
    author = "van de Meent, Maarten",
    title = "{Self-force corrections to the periapsis advance around a spinning black hole}",
    eprint = "1610.03497",
    archivePrefix = "arXiv",
    primaryClass = "gr-qc",
    doi = "10.1103/PhysRevLett.118.011101",
    journal = "Phys. Rev. Lett.",
    volume = "118",
    number = "1",
    pages = "011101",
    year = "2017"
}

@article{Munna:2022gio,
    author = "Munna, Christopher and Evans, Charles R.",
    title = "{High-order post-Newtonian expansion of the redshift invariant for eccentric-orbit nonspinning extreme-mass-ratio inspirals}",
    eprint = "2203.13832",
    archivePrefix = "arXiv",
    primaryClass = "gr-qc",
    doi = "10.1103/PhysRevD.106.044004",
    journal = "Phys. Rev. D",
    volume = "106",
    number = "4",
    pages = "044004",
    year = "2022"
}

@article{LeTiec:2015kgg,
    author = "Le Tiec, Alexandre",
    title = "{First Law of Mechanics for Compact Binaries on Eccentric Orbits}",
    eprint = "1506.05648",
    archivePrefix = "arXiv",
    primaryClass = "gr-qc",
    doi = "10.1103/PhysRevD.92.084021",
    journal = "Phys. Rev. D",
    volume = "92",
    number = "8",
    pages = "084021",
    year = "2015"
}

@article{Barack:2010ny,
    author = "Barack, Leor and Damour, Thibault and Sago, Norichika",
    title = "{Precession effect of the gravitational self-force in a Schwarzschild spacetime and the effective one-body formalism}",
    eprint = "1008.0935",
    archivePrefix = "arXiv",
    primaryClass = "gr-qc",
    doi = "10.1103/PhysRevD.82.084036",
    journal = "Phys. Rev. D",
    volume = "82",
    pages = "084036",
    year = "2010"
}

@article{Barack:2010tm,
    author = "Barack, Leor and Sago, Norichika",
    title = "{Gravitational self-force on a particle in eccentric orbit around a Schwarzschild black hole}",
    eprint = "1002.2386",
    archivePrefix = "arXiv",
    primaryClass = "gr-qc",
    doi = "10.1103/PhysRevD.81.084021",
    journal = "Phys. Rev. D",
    volume = "81",
    pages = "084021",
    year = "2010"
}

@misc{BHPToolkit,
  title = {{Black Hole Perturbation Toolkit}},
  howpublished = {(\href{http://bhptoolkit.org/}{bhptoolkit.org})},
  note = {accessed on 25/11/2025}
}

@article{Bini:2016qtx,
    author = "Bini, Donato and Damour, Thibault and Geralico, andrea",
    title = "{New gravitational self-force analytical results for eccentric orbits around a Schwarzschild black hole}",
    eprint = "1601.02988",
    archivePrefix = "arXiv",
    primaryClass = "gr-qc",
    doi = "10.1103/PhysRevD.93.104017",
    journal = "Phys. Rev. D",
    volume = "93",
    number = "10",
    pages = "104017",
    year = "2016"
}

@inbook{NIST_elliptic,
    author = "Carlson, Bille C.",
    title = "{\it NIST Digital Library of Mathematical Functions}",
    publisher = "Cambridge University Press",
    year = "2010",
    chapter = "19 (``Elliptic Integrals'')",
    note = "\url{https://dlmf.nist.gov/19.2}"
}

@article{LISAConsortiumWaveformWorkingGroup:2023arg,
    author = "Afshordi, Niayesh and others",
    collaboration = "LISA Consortium Waveform Working Group",
    title = "{Waveform modelling for the Laser Interferometer Space Antenna}",
    eprint = "2311.01300",
    archivePrefix = "arXiv",
    primaryClass = "gr-qc",
    doi = "10.1007/s41114-025-00056-1",
    journal = "Living Rev. Rel.",
    volume = "28",
    number = "1",
    pages = "9",
    year = "2025"
}

@article{Barack:2018yvs,
    author = "Barack, Leor and Pound, Adam",
    title = "{Self-force and radiation reaction in general relativity}",
    eprint = "1805.10385",
    archivePrefix = "arXiv",
    primaryClass = "gr-qc",
    doi = "10.1088/1361-6633/aae552",
    journal = "Rept. Prog. Phys.",
    volume = "82",
    number = "1",
    pages = "016904",
    year = "2019"
}

@book{Baumgarte:2010ndz,
    author = "Baumgarte, Thomas W. and Shapiro, Stuart L.",
    title = "{Numerical Relativity: Solving Einstein's Equations on the Computer}",
    doi = "10.1017/CBO9781139193344",
    publisher = "Cambridge University Press",
    year = "2010"
}

@article{Damour:2016gwp,
    author = "Damour, Thibault",
    title = "{Gravitational scattering, post-Minkowskian approximation and Effective One-Body theory}",
    eprint = "1609.00354",
    archivePrefix = "arXiv",
    primaryClass = "gr-qc",
    doi = "10.1103/PhysRevD.94.104015",
    journal = "Phys. Rev. D",
    volume = "94",
    number = "10",
    pages = "104015",
    year = "2016"
}

@article{Buonanno:1998gg,
    author = "Buonanno, A. and Damour, T.",
    title = "{Effective one-body approach to general relativistic two-body dynamics}",
    eprint = "gr-qc/9811091",
    archivePrefix = "arXiv",
    reportNumber = "IHES-P-98-74",
    doi = "10.1103/PhysRevD.59.084006",
    journal = "Phys. Rev. D",
    volume = "59",
    pages = "084006",
    year = "1999"
}

@article{Buonanno:2000ef,
    author = "Buonanno, Alessandra and Damour, Thibault",
    title = "{Transition from inspiral to plunge in binary black hole coalescences}",
    eprint = "gr-qc/0001013",
    archivePrefix = "arXiv",
    reportNumber = "IHES-P-99-90, GRP-99-521",
    doi = "10.1103/PhysRevD.62.064015",
    journal = "Phys. Rev. D",
    volume = "62",
    pages = "064015",
    year = "2000"
}

@unpublished{Berti:2025hly,
    author = "Berti, Emanuele and others",
    editor = "Berti, Emanuele and Cardoso, Vitor and Carullo, Gregorio",
    title = "{Black hole spectroscopy: from theory to experiment}",
    eprint = "2505.23895",
    archivePrefix = "arXiv",
    primaryClass = "gr-qc",
    month = "5",
    year = "2025"
}

@article{Honet:2025gge,
    author = {Honet, Lo{\"\i}c and Pound, Adam and Comp{\`e}re, Geoffrey},
    title = "{Hybrid waveform model for asymmetric spinning binaries: Self-force meets post-Newtonian theory}",
    eprint = "2510.16114",
    archivePrefix = "arXiv",
    primaryClass = "gr-qc",
    doi = "10.1103/rhwy-59y2",
    journal = "Phys. Rev. D",
    volume = "113",
    number = "6",
    pages = "064035",
    year = "2026"
}

@unpublished{Honet:2025lmk,
    author = {Honet, Lo{\"\i}c and Mathews, Josh and Comp{\`e}re, Geoffrey and Pound, Adam and Wardell, Barry and Piovano, Gabriel Andres and van de Meent, Maarten and Warburton, Niels},
    title = "{Spin-aligned inspiral waveforms from self-force and post-Newtonian theory}",
    eprint = "2510.16112",
    archivePrefix = "arXiv",
    primaryClass = "gr-qc",
    month = "10",
    year = "2025"
}

@article{Lewis:2025ydo,
    author = "Lewis, Jack and Kakehi, Takafumi and Pound, Adam and Tanaka, Takahiro",
    title = "{Postadiabatic dynamics and waveform generation in self-force theory: An invariant pseudo-Hamiltonian framework}",
    eprint = "2507.08081",
    archivePrefix = "arXiv",
    primaryClass = "gr-qc",
    doi = "10.1103/jllr-kl86",
    journal = "Phys. Rev. D",
    volume = "113",
    number = "6",
    pages = "064046",
    year = "2026"
}

@article{Gonzo:2024xjk,
    author = "Gonzo, Riccardo and Lewis, Jack and Pound, Adam",
    title = "{First Law of Binary Black Hole Scattering}",
    eprint = "2409.03437",
    archivePrefix = "arXiv",
    primaryClass = "gr-qc",
    doi = "10.1103/s85p-gh7b",
    journal = "Phys. Rev. Lett.",
    volume = "135",
    number = "13",
    pages = "131401",
    year = "2025"
}

@article{Blanco:2022mgd,
    author = "Blanco, Francisco M. and Flanagan, \'Eanna \'E.",
    title = "{Particle Motion under the Conservative Piece of the Self-Force is Hamiltonian}",
    eprint = "2205.01667",
    archivePrefix = "arXiv",
    primaryClass = "gr-qc",
    doi = "10.1103/PhysRevLett.130.051201",
    journal = "Phys. Rev. Lett.",
    volume = "130",
    number = "5",
    pages = "051201",
    year = "2023"
}

@article{Mathews:2025txc,
    author = "Mathews, Josh and Wardell, Barry and Pound, Adam and Warburton, Niels",
    title = "{Postadiabatic self-force waveforms: Slowly spinning primary and precessing secondary}",
    eprint = "2510.16113",
    archivePrefix = "arXiv",
    primaryClass = "gr-qc",
    doi = "10.1103/ph3p-mscl",
    journal = "Phys. Rev. D",
    volume = "113",
    number = "6",
    pages = "064034",
    year = "2026"
}

@article{Wardell:2021fyy,
    author = "Wardell, Barry and Pound, Adam and Warburton, Niels and Miller, Jeremy and Durkan, Leanne and Le Tiec, Alexandre",
    title = "{Gravitational Waveforms for Compact Binaries from Second-Order Self-Force Theory}",
    eprint = "2112.12265",
    archivePrefix = "arXiv",
    primaryClass = "gr-qc",
    doi = "10.1103/PhysRevLett.130.241402",
    journal = "Phys. Rev. Lett.",
    volume = "130",
    number = "24",
    pages = "241402",
    year = "2023"
}

@unpublished{Pound:2021qin,
    author = "Pound, Adam and Wardell, Barry",
    title = "{Black hole perturbation theory and gravitational self-force}",
    eprint = "2101.04592",
    archivePrefix = "arXiv",
    primaryClass = "gr-qc",
    doi = "10.1007/978-981-15-4702-7\_38-1",
    month = "1",
    year = "2021"
}

@article{Mathews:2025nyb,
    author = "Mathews, Josh and Pound, Adam",
    title = "{Postadiabatic waveform-generation framework for asymmetric precessing binaries}",
    eprint = "2501.01413",
    archivePrefix = "arXiv",
    primaryClass = "gr-qc",
    doi = "10.1103/rbkb-qnxv",
    journal = "Phys. Rev. D",
    volume = "112",
    number = "10",
    pages = "104078",
    year = "2025"
}

@article{Burke:2023lno,
    author = "Burke, Ollie and Piovano, Gabriel Andres and Warburton, Niels and Lynch, Philip and Speri, Lorenzo and Kavanagh, Chris and Wardell, Barry and Pound, Adam and Durkan, Leanne and Miller, Jeremy",
    title = "{Assessing the importance of first postadiabatic terms for small-mass-ratio binaries}",
    eprint = "2310.08927",
    archivePrefix = "arXiv",
    primaryClass = "gr-qc",
    doi = "10.1103/PhysRevD.109.124048",
    journal = "Phys. Rev. D",
    volume = "109",
    number = "12",
    pages = "124048",
    year = "2024"
}

@article{Hinderer:2008dm,
    author = "Hinderer, Tanja and Flanagan, Eanna E.",
    title = "{Two timescale analysis of extreme mass ratio inspirals in Kerr. I. Orbital Motion}",
    eprint = "0805.3337",
    archivePrefix = "arXiv",
    primaryClass = "gr-qc",
    doi = "10.1103/PhysRevD.78.064028",
    journal = "Phys. Rev. D",
    volume = "78",
    pages = "064028",
    year = "2008"
}

@article{Schmidt:2002qk,
    author = "Schmidt, Wolfram",
    title = "{Celestial mechanics in Kerr space-time}",
    eprint = "gr-qc/0202090",
    archivePrefix = "arXiv",
    doi = "10.1088/0264-9381/19/10/314",
    journal = "Class. Quant. Grav.",
    volume = "19",
    pages = "2743",
    year = "2002"
}

@article{Grant:2024ivt,
    author = "Grant, Alexander M.",
    title = "{Flux-balance laws for spinning bodies under the gravitational self-force}",
    eprint = "2406.10343",
    archivePrefix = "arXiv",
    primaryClass = "gr-qc",
    doi = "10.1103/PhysRevD.111.084015",
    journal = "Phys. Rev. D",
    volume = "111",
    number = "8",
    pages = "084015",
    year = "2025"
}

@article{Witzany:2019nml,
    author = "Witzany, Vojt{\v{e}}ch",
    title = "{Hamilton-Jacobi equation for spinning particles near black holes}",
    eprint = "1903.03651",
    archivePrefix = "arXiv",
    primaryClass = "gr-qc",
    doi = "10.1103/PhysRevD.100.104030",
    journal = "Phys. Rev. D",
    volume = "100",
    number = "10",
    pages = "104030",
    year = "2019"
}

@article{Witzany:2018ahb,
    author = "Witzany, Vojt{\v{e}}ch and Steinhoff, Jan and Lukes-Gerakopoulos, Georgios",
    title = "{Hamiltonians and canonical coordinates for spinning particles in curved space-time}",
    eprint = "1808.06582",
    archivePrefix = "arXiv",
    primaryClass = "gr-qc",
    doi = "10.1088/1361-6382/ab002f",
    journal = "Class. Quant. Grav.",
    volume = "36",
    number = "7",
    pages = "075003",
    year = "2019"
}

@article{Ramond:2024ozy,
    author = "Ramond, Paul",
    title = "{On the integrability of extended test body dynamics around black holes}",
    eprint = "2402.02670",
    archivePrefix = "arXiv",
    primaryClass = "gr-qc",
    doi = "10.1088/1361-6382/adb197",
    journal = "Class. Quant. Grav.",
    volume = "42",
    number = "6",
    pages = "065019",
    year = "2025"
}

@article{Pompili:2023tna,
    author = "Pompili, Lorenzo and others",
    title = "{Laying the foundation of the effective-one-body waveform models SEOBNRv5: Improved accuracy and efficiency for spinning nonprecessing binary black holes}",
    eprint = "2303.18039",
    archivePrefix = "arXiv",
    primaryClass = "gr-qc",
    doi = "10.1103/PhysRevD.108.124035",
    journal = "Phys. Rev. D",
    volume = "108",
    number = "12",
    pages = "124035",
    year = "2023"
}

@article{Ramos-Buades:2023ehm,
    author = "Ramos-Buades, Antoni and Buonanno, Alessandra and Estell{\'e}s, H{\'e}ctor and Khalil, Mohammed and Mihaylov, Deyan P. and Ossokine, Serguei and Pompili, Lorenzo and Shiferaw, Mahlet",
    title = "{Next generation of accurate and efficient multipolar precessing-spin effective-one-body waveforms for binary black holes}",
    eprint = "2303.18046",
    archivePrefix = "arXiv",
    primaryClass = "gr-qc",
    doi = "10.1103/PhysRevD.108.124037",
    journal = "Phys. Rev. D",
    volume = "108",
    number = "12",
    pages = "124037",
    year = "2023"
}

@article{Leather:2025nhu,
    author = "Leather, Benjamin and Buonanno, Alessandra and van de Meent, Maarten",
    title = "{Inspiral-merger-ringdown waveforms with gravitational self-force results within the effective-one-body formalism}",
    eprint = "2505.11242",
    archivePrefix = "arXiv",
    primaryClass = "gr-qc",
    doi = "10.1103/6qc3-xn17",
    journal = "Phys. Rev. D",
    volume = "112",
    number = "4",
    pages = "044012",
    year = "2025"
}

@article{Rifat:2019ltp,
    author = "Rifat, Nur E. M. and Field, Scott E. and Khanna, Gaurav and Varma, Vijay",
    title = "{Surrogate model for gravitational wave signals from comparable and large-mass-ratio black hole binaries}",
    eprint = "1910.10473",
    archivePrefix = "arXiv",
    primaryClass = "gr-qc",
    doi = "10.1103/PhysRevD.101.081502",
    journal = "Phys. Rev. D",
    volume = "101",
    number = "8",
    pages = "081502",
    year = "2020"
}

@article{Ajith:2009bn,
    author = "Ajith, P. and others",
    title = "{Inspiral-merger-ringdown waveforms for black-hole binaries with non-precessing spins}",
    eprint = "0909.2867",
    archivePrefix = "arXiv",
    primaryClass = "gr-qc",
    doi = "10.1103/PhysRevLett.106.241101",
    journal = "Phys. Rev. Lett.",
    volume = "106",
    pages = "241101",
    year = "2011"
}

@article{Varma:2018mmi,
    author = "Varma, Vijay and Field, Scott E. and Scheel, Mark A. and Blackman, Jonathan and Kidder, Lawrence E. and Pfeiffer, Harald P.",
    title = "{Surrogate model of hybridized numerical relativity binary black hole waveforms}",
    eprint = "1812.07865",
    archivePrefix = "arXiv",
    primaryClass = "gr-qc",
    doi = "10.1103/PhysRevD.99.064045",
    journal = "Phys. Rev. D",
    volume = "99",
    number = "6",
    pages = "064045",
    year = "2019"
}

@article{Santamaria:2010yb,
    author = "Santamaria, L. and others",
    title = "{Matching post-Newtonian and numerical relativity waveforms: systematic errors and a new phenomenological model for non-precessing black hole binaries}",
    eprint = "1005.3306",
    archivePrefix = "arXiv",
    primaryClass = "gr-qc",
    reportNumber = "LIGO-P1000048, AEI-2010-122",
    doi = "10.1103/PhysRevD.82.064016",
    journal = "Phys. Rev. D",
    volume = "82",
    pages = "064016",
    year = "2010"
}

@article{Hannam:2013oca,
    author = {Hannam, Mark and Schmidt, Patricia and Boh{\'e}, Alejandro and Haegel, Le{\"\i}la and Husa, Sascha and Ohme, Frank and Pratten, Geraint and P{\"u}rrer, Michael},
    title = "{Simple Model of Complete Precessing Black-Hole-Binary Gravitational Waveforms}",
    eprint = "1308.3271",
    archivePrefix = "arXiv",
    primaryClass = "gr-qc",
    doi = "10.1103/PhysRevLett.113.151101",
    journal = "Phys. Rev. Lett.",
    volume = "113",
    number = "15",
    pages = "151101",
    year = "2014"
}

@article{London:2017bcn,
    author = "London, Lionel and Khan, Sebastian and Fauchon-Jones, Edward and Garc{\'\i}a, Cecilio and Hannam, Mark and Husa, Sascha and Jim{\'e}nez-Forteza, Xisco and Kalaghatgi, Chinmay and Ohme, Frank and Pannarale, Francesco",
    title = "{First higher-multipole model of gravitational waves from spinning and coalescing black-hole binaries}",
    eprint = "1708.00404",
    archivePrefix = "arXiv",
    primaryClass = "gr-qc",
    doi = "10.1103/PhysRevLett.120.161102",
    journal = "Phys. Rev. Lett.",
    volume = "120",
    number = "16",
    pages = "161102",
    year = "2018"
}

@article{Pratten:2020ceb,
    author = "Pratten, Geraint and others",
    title = "{Computationally efficient models for the dominant and subdominant harmonic modes of precessing binary black holes}",
    eprint = "2004.06503",
    archivePrefix = "arXiv",
    primaryClass = "gr-qc",
    doi = "10.1103/PhysRevD.103.104056",
    journal = "Phys. Rev. D",
    volume = "103",
    number = "10",
    pages = "104056",
    year = "2021"
}

@article{Dean:1999jr,
    author = "Dean, Bruce",
    title = "{Phase-plane analysis of perihelion precession and Schwarschild orbital dynamics}",
    doi = "10.1119/1.19194",
    journal = "Am. J. Phys.",
    volume = "67",
    pages = "78--86",
    year = "1999"
}

@unpublished{Bern:2025wyd,
    author = "Bern, Zvi and Herrmann, Enrico and Roiban, Radu and Ruf, Michael S. and Smirnov, Alexander V. and Smith, Sid and Zeng, Mao",
    title = "{Scattering Amplitudes and Conservative Binary Dynamics at $O(G^5)$ without Self-Force Truncation}",
    eprint = "2512.23654",
    archivePrefix = "arXiv",
    primaryClass = "hep-th",
    month = "12",
    year = "2025"
}

@article{Levin:2008mq,
    author = "Levin, Janna and Perez-Giz, Gabe",
    title = "{A Periodic Table for Black Hole Orbits}",
    eprint = "0802.0459",
    archivePrefix = "arXiv",
    primaryClass = "gr-qc",
    doi = "10.1103/PhysRevD.77.103005",
    journal = "Phys. Rev. D",
    volume = "77",
    pages = "103005",
    year = "2008"
}

@article{Perez-Giz:2008ajn,
    author = "Perez-Giz, Gabe and Levin, Janna",
    title = "{Homoclinic Orbits around Spinning Black Holes II: The Phase Space Portrait}",
    eprint = "0811.3815",
    archivePrefix = "arXiv",
    primaryClass = "gr-qc",
    doi = "10.1103/PhysRevD.79.124014",
    journal = "Phys. Rev. D",
    volume = "79",
    pages = "124014",
    year = "2009"
}

@article{Blanchet:2023bwj,
    author = "Blanchet, Luc and Faye, Guillaume and Henry, Quentin and Larrouturou, Fran{\c{c}}ois and Trestini, David",
    title = "{Gravitational-Wave Phasing of Quasicircular Compact Binary Systems to the Fourth-and-a-Half Post-Newtonian Order}",
    eprint = "2304.11185",
    archivePrefix = "arXiv",
    primaryClass = "gr-qc",
    reportNumber = "DESY-23-043",
    doi = "10.1103/PhysRevLett.131.121402",
    journal = "Phys. Rev. Lett.",
    volume = "131",
    number = "12",
    pages = "121402",
    year = "2023"
}

@article{Cho:2022syn,
    author = "Cho, Gihyuk and Porto, Rafael A. and Yang, Zixin",
    title = "{Gravitational radiation from inspiralling compact objects: Spin effects to the fourth post-Newtonian order}",
    eprint = "2201.05138",
    archivePrefix = "arXiv",
    primaryClass = "gr-qc",
    reportNumber = "DESY-22-004, ET-0001A-22, DESY-22-004; ET-0001A-22",
    doi = "10.1103/PhysRevD.106.L101501",
    journal = "Phys. Rev. D",
    volume = "106",
    number = "10",
    pages = "L101501",
    year = "2022"
}

@article{Marsat:2014xea,
    author = "Marsat, Sylvain",
    title = "{Cubic order spin effects in the dynamics and gravitational wave energy flux of compact object binaries}",
    eprint = "1411.4118",
    archivePrefix = "arXiv",
    primaryClass = "gr-qc",
    doi = "10.1088/0264-9381/32/8/085008",
    journal = "Class. Quant. Grav.",
    volume = "32",
    number = "8",
    pages = "085008",
    year = "2015"
}

@article{Henry:2022ccf,
    author = "Henry, Quentin",
    title = "{Complete gravitational-waveform amplitude modes for quasicircular compact binaries to the 3.5PN order}",
    eprint = "2210.15602",
    archivePrefix = "arXiv",
    primaryClass = "gr-qc",
    doi = "10.1103/PhysRevD.107.044057",
    journal = "Phys. Rev. D",
    volume = "107",
    number = "4",
    pages = "044057",
    year = "2023"
}

@article{Bombelli:1991eg,
    author = "Bombelli, Luca and Calzetta, Esteban",
    title = "{Chaos around a black hole}",
    reportNumber = "GTCRG-91-12",
    doi = "10.1088/0264-9381/9/12/004",
    journal = "Class. Quant. Grav.",
    volume = "9",
    pages = "2573--2599",
    year = "1992"
}

@article{Bern:2025zno,
    author = "Bern, Zvi and Herrmann, Enrico and Roiban, Radu and Ruf, Michael S. and Smirnov, Alexander V. and Smirnov, Vladimir A. and Zeng, Mao",
    title = "{Second-Order Self-Force Potential-Region Binary Dynamics at O(G5) in Supergravity}",
    eprint = "2509.17412",
    archivePrefix = "arXiv",
    primaryClass = "hep-th",
    doi = "10.1103/jmby-htz9",
    journal = "Phys. Rev. Lett.",
    volume = "136",
    number = "8",
    pages = "081401",
    year = "2026"
}

@article{Bern:2024adl,
    author = "Bern, Zvi and Herrmann, Enrico and Roiban, Radu and Ruf, Michael S. and Smirnov, Alexander V. and Smirnov, Vladimir A. and Zeng, Mao",
    title = "{Amplitudes, supersymmetric black hole scattering at $ \mathcal{O}\left({G}^5\right) $, and loop integration}",
    eprint = "2406.01554",
    archivePrefix = "arXiv",
    primaryClass = "hep-th",
    doi = "10.1007/JHEP10(2024)023",
    journal = "JHEP",
    volume = "10",
    pages = "023",
    year = "2024"
}

@article{Bini:2019nra,
    author = "Bini, Donato and Damour, Thibault and Geralico, Andrea",
    title = "{Novel approach to binary dynamics: application to the fifth post-Newtonian level}",
    eprint = "1909.02375",
    archivePrefix = "arXiv",
    primaryClass = "gr-qc",
    doi = "10.1103/PhysRevLett.123.231104",
    journal = "Phys. Rev. Lett.",
    volume = "123",
    number = "23",
    pages = "231104",
    year = "2019"
}

@article{Khalaf:2025jpt,
    author = "Khalaf, Majed and Shen, Chia-Hsien and Telem, Ofri",
    title = "{Bound-unbound universality and the all-order semi-classical wave function in Schwarzschild}",
    eprint = "2503.23317",
    archivePrefix = "arXiv",
    primaryClass = "gr-qc",
    doi = "10.1007/JHEP10(2025)063",
    journal = "JHEP",
    volume = "10",
    pages = "063",
    year = "2025"
}

@article{Damour:1985,
    author = {T. Damour and N. Deruelle},
    title = {General relativistic celestial mechanics of binary systems I. {T}he post-{N}ewtonian motion}, 
    journal = {Ann. Inst. Henri Poincar´e A},
    volume = {43}, 
    pages = {107},
    year = {1985}
}

@article{Kalin:2019inp,
    author = {K\"alin, Gregor and Porto, Rafael A.},
    title = "{From boundary data to bound states. Part II. Scattering angle to dynamical invariants (with twist)}",
    eprint = "1911.09130",
    archivePrefix = "arXiv",
    primaryClass = "hep-th",
    reportNumber = "DESY 19-201, SLAC-PUB-17487, UUITP-46/19, DESY-19-201",
    doi = "10.1007/JHEP02(2020)120",
    journal = "JHEP",
    volume = "02",
    pages = "120",
    year = "2020"
}

@article{Kalin:2019rwq,
    author = {K\"alin, Gregor and Porto, Rafael A.},
    title = "{From Boundary Data to Bound States}",
    eprint = "1910.03008",
    archivePrefix = "arXiv",
    primaryClass = "hep-th",
    reportNumber = "DESY 19-167, UUITP-40/19, DESY-19-167",
    doi = "10.1007/JHEP01(2020)072",
    journal = "JHEP",
    volume = "01",
    pages = "072",
    year = "2020"
}

@article{Saketh:2021sri,
    author = "Saketh, M. V. S. and Vines, Justin and Steinhoff, Jan and Buonanno, Alessandra",
    title = "{Conservative and radiative dynamics in classical relativistic scattering and bound systems}",
    eprint = "2109.05994",
    archivePrefix = "arXiv",
    primaryClass = "gr-qc",
    doi = "10.1103/PhysRevResearch.4.013127",
    journal = "Phys. Rev. Res.",
    volume = "4",
    number = "1",
    pages = "013127",
    year = "2022"
}

@article{Cho:2021arx,
    author = {Cho, Gihyuk and K\"alin, Gregor and Porto, Rafael A.},
    title = "{From boundary data to bound states. Part III. Radiative effects}",
    eprint = "2112.03976",
    archivePrefix = "arXiv",
    primaryClass = "hep-th",
    reportNumber = "DESY 21-212",
    doi = "10.1007/JHEP04(2022)154",
    journal = "JHEP",
    volume = "04",
    pages = "154",
    year = "2022",
    note = "[Erratum: JHEP 07, 002 (2022)]"
}

@article{Gonzo:2023goe,
    author = "Gonzo, Riccardo and Shi, Canxin",
    title = "{Boundary to bound dictionary for generic Kerr orbits}",
    eprint = "2304.06066",
    archivePrefix = "arXiv",
    primaryClass = "hep-th",
    doi = "10.1103/PhysRevD.108.084065",
    journal = "Phys. Rev. D",
    volume = "108",
    number = "8",
    pages = "084065",
    year = "2023"
}

@article{Adamo:2024oxy,
    author = "Adamo, Tim and Gonzo, Riccardo and Ilderton, Anton",
    title = "{Gravitational bound waveforms from amplitudes}",
    eprint = "2402.00124",
    archivePrefix = "arXiv",
    primaryClass = "hep-th",
    doi = "10.1007/JHEP05(2024)034",
    journal = "JHEP",
    volume = "05",
    pages = "034",
    year = "2024"
}

@article{Dlapa:2024cje,
    author = {Dlapa, Christoph and K\"alin, Gregor and Liu, Zhengwen and Porto, Rafael A.},
    title = "{Local in Time Conservative Binary Dynamics at Fourth Post-Minkowskian Order}",
    eprint = "2403.04853",
    archivePrefix = "arXiv",
    primaryClass = "hep-th",
    reportNumber = "DESY 24-029",
    doi = "10.1103/PhysRevLett.132.221401",
    journal = "Phys. Rev. Lett.",
    volume = "132",
    number = "22",
    pages = "221401",
    year = "2024"
}

@article{Bern:2022jvn,
    author = "Bern, Zvi and Parra-Martinez, Julio and Roiban, Radu and Ruf, Michael S. and Shen, Chia-Hsien and Solon, Mikhail P. and Zeng, Mao",
    title = "{Scattering amplitudes and conservative dynamics at the fourth post-Minkowskian order}",
    doi = "10.22323/1.416.0051",
    journal = "PoS",
    volume = "LL2022",
    pages = "051",
    year = "2022"
}

@article{Sasaki:2003xr,
    author = "Sasaki, Misao and Tagoshi, Hideyuki",
    title = "{Analytic black hole perturbation approach to gravitational radiation}",
    eprint = "gr-qc/0306120",
    archivePrefix = "arXiv",
    doi = "10.12942/lrr-2003-6",
    journal = "Living Rev. Rel.",
    volume = "6",
    pages = "6",
    year = "2003"
}

@article{Driesse:2024xad,
    author = "Driesse, Mathias and Jakobsen, Gustav Uhre and Mogull, Gustav and Plefka, Jan and Sauer, Benjamin and Usovitsch, Johann",
    title = "{Conservative Black Hole Scattering at Fifth Post-Minkowskian and First Self-Force Order}",
    eprint = "2403.07781",
    archivePrefix = "arXiv",
    primaryClass = "hep-th",
    reportNumber = "HU-EP-24/08-RTG, CERN-TH-2024-032",
    doi = "10.1103/PhysRevLett.132.241402",
    journal = "Phys. Rev. Lett.",
    volume = "132",
    number = "24",
    pages = "241402",
    year = "2024"
}

@article{Driesse:2024feo,
    author = "Driesse, Mathias and Jakobsen, Gustav Uhre and Klemm, Albrecht and Mogull, Gustav and Nega, Christoph and Plefka, Jan and Sauer, Benjamin and Usovitsch, Johann",
    title = "{Emergence of Calabi{\textendash}Yau manifolds in high-precision black-hole scattering}",
    eprint = "2411.11846",
    archivePrefix = "arXiv",
    primaryClass = "hep-th",
    reportNumber = "HU-EP-24/32-RTG, QMUL-PH-24-26, BONN-TH-2024-15, TUM-HEP-1532/24",
    doi = "10.1038/s41586-025-08984-2",
    journal = "Nature",
    volume = "641",
    number = "8063",
    pages = "603--607",
    year = "2025"
}

@article{Dlapa:2025biy,
    author = {Dlapa, Christoph and K{\"a}lin, Gregor and Liu, Zhengwen and Porto, Rafael A.},
    title = "{Local-in-Time Conservative Binary Dynamics at Fifth Post-Minkowskian and First Self-Force Orders}",
    eprint = "2506.20665",
    archivePrefix = "arXiv",
    primaryClass = "hep-th",
    reportNumber = "DESY 25-089",
    doi = "10.1103/215k-27sj",
    journal = "Phys. Rev. Lett.",
    volume = "135",
    number = "25",
    pages = "251401",
    year = "2025"
}

@article{Albanesi:2025txj,
    author = "Albanesi, Simone and Gamba, Rossella and Bernuzzi, Sebastiano and Fontbut{\'e}, Joan and Gonzalez, Alejandra and Nagar, Alessandro",
    title = "{Effective-one-body modeling for generic compact binaries with arbitrary orbits}",
    eprint = "2503.14580",
    archivePrefix = "arXiv",
    primaryClass = "gr-qc",
    doi = "10.1103/3snf-w1x7",
    journal = "Phys. Rev. D",
    volume = "112",
    number = "12",
    pages = "L121503",
    year = "2025"
}

@unpublished{Iglesias:2025tnt,
    author = "Iglesias, Hector and Durkan, Leanne and Shoemaker, Deirdre",
    title = "{Hybridization of second-order gravitational self-force and numerical relativity waveforms for quasi-circular and non-spinning black hole binaries}",
    eprint = "2510.11685",
    archivePrefix = "arXiv",
    primaryClass = "gr-qc",
    reportNumber = "UT-WI-11-2025",
    month = "10",
    year = "2025"
}

@article{Antonelli:2019fmq,
    author = "Antonelli, Andrea and van de Meent, Maarten and Buonanno, Alessandra and Steinhoff, Jan and Vines, Justin",
    title = "{Quasicircular inspirals and plunges from nonspinning effective-one-body Hamiltonians with gravitational self-force information}",
    eprint = "1907.11597",
    archivePrefix = "arXiv",
    primaryClass = "gr-qc",
    doi = "10.1103/PhysRevD.101.024024",
    journal = "Phys. Rev. D",
    volume = "101",
    number = "2",
    pages = "024024",
    year = "2020"
}

@article{Damour:2009sm,
    author = "Damour, Thibault",
    title = "{Gravitational Self Force in a Schwarzschild Background and the Effective One Body Formalism}",
    eprint = "0910.5533",
    archivePrefix = "arXiv",
    primaryClass = "gr-qc",
    doi = "10.1103/PhysRevD.81.024017",
    journal = "Phys. Rev. D",
    volume = "81",
    pages = "024017",
    year = "2010"
}

@book{Chandrasekhar:1985kt,
    author = "Chandrasekhar, Subrahmanyan",
    title = "{{The Mathematical Theory of Black Holes}}",
    publisher = "{{Oxford University Press, Inc.}}",
    year = 1983
}

@article{Bini:2013rfa,
    author = "Bini, Donato and Damour, Thibault",
    title = "{High-order post-Newtonian contributions to the two-body gravitational interaction potential from analytical gravitational self-force calculations}",
    eprint = "1312.2503",
    archivePrefix = "arXiv",
    primaryClass = "gr-qc",
    doi = "10.1103/PhysRevD.89.064063",
    journal = "Phys. Rev. D",
    volume = "89",
    number = "6",
    pages = "064063",
    year = "2014"
}

@article{Bini:2013zaa,
    author = "Bini, Donato and Damour, Thibault",
    title = "{Analytical determination of the two-body gravitational interaction potential at the fourth post-Newtonian approximation}",
    eprint = "1305.4884",
    archivePrefix = "arXiv",
    primaryClass = "gr-qc",
    doi = "10.1103/PhysRevD.87.121501",
    journal = "Phys. Rev. D",
    volume = "87",
    number = "12",
    pages = "121501",
    year = "2013"
}

@article{Huerta:2017kez,
    author = "Huerta, E. A. and others",
    title = "{Eccentric, nonspinning, inspiral, Gaussian-process merger approximant for the detection and characterization of eccentric binary black hole mergers}",
    eprint = "1711.06276",
    archivePrefix = "arXiv",
    primaryClass = "gr-qc",
    doi = "10.1103/PhysRevD.97.024031",
    journal = "Phys. Rev. D",
    volume = "97",
    number = "2",
    pages = "024031",
    year = "2018"
}

@article{Kavanagh:2015lva,
    author = "Kavanagh, Chris and Ottewill, Adrian C. and Wardell, Barry",
    title = "{Analytical high-order post-Newtonian expansions for extreme mass ratio binaries}",
    eprint = "1503.02334",
    archivePrefix = "arXiv",
    primaryClass = "gr-qc",
    doi = "10.1103/PhysRevD.92.084025",
    journal = "Phys. Rev. D",
    volume = "92",
    number = "8",
    pages = "084025",
    year = "2015"
}

@article{Barausse:2011dq,
    author = "Barausse, Enrico and Buonanno, Alessandra and Le Tiec, Alexandre",
    title = "{The complete non-spinning effective-one-body metric at linear order in the mass ratio}",
    eprint = "1111.5610",
    archivePrefix = "arXiv",
    primaryClass = "gr-qc",
    doi = "10.1103/PhysRevD.85.064010",
    journal = "Phys. Rev. D",
    volume = "85",
    pages = "064010",
    year = "2012"
}

@article{Damour:2007xr,
    author = "Damour, Thibault and Nagar, Alessandro",
    title = "{Faithful effective-one-body waveforms of small-mass-ratio coalescing black-hole binaries}",
    eprint = "0705.2519",
    archivePrefix = "arXiv",
    primaryClass = "gr-qc",
    doi = "10.1103/PhysRevD.76.064028",
    journal = "Phys. Rev. D",
    volume = "76",
    pages = "064028",
    year = "2007"
}

@article{Taracchini:2014zpa,
    author = "Taracchini, Andrea and Buonanno, Alessandra and Khanna, Gaurav and Hughes, Scott A.",
    title = "{Small mass plunging into a Kerr black hole: Anatomy of the inspiral-merger-ringdown waveforms}",
    eprint = "1404.1819",
    archivePrefix = "arXiv",
    primaryClass = "gr-qc",
    doi = "10.1103/PhysRevD.90.084025",
    journal = "Phys. Rev. D",
    volume = "90",
    number = "8",
    pages = "084025",
    year = "2014"
}

@article{Nagar:2022fep,
    author = "Nagar, Alessandro and Albanesi, Simone",
    title = "{Toward a gravitational self-force-informed effective-one-body waveform model for nonprecessing, eccentric, large-mass-ratio inspirals}",
    eprint = "2207.14002",
    archivePrefix = "arXiv",
    primaryClass = "gr-qc",
    doi = "10.1103/PhysRevD.106.064049",
    journal = "Phys. Rev. D",
    volume = "106",
    number = "6",
    pages = "064049",
    year = "2022"
}

@article{Long:2024ltn,
    author = "Long, Oliver and Whittall, Christopher and Barack, Leor",
    title = "{Black hole scattering near the transition to plunge: Self-force and resummation of post-Minkowskian theory}",
    eprint = "2406.08363",
    archivePrefix = "arXiv",
    primaryClass = "gr-qc",
    doi = "10.1103/PhysRevD.110.044039",
    journal = "Phys. Rev. D",
    volume = "110",
    number = "4",
    pages = "044039",
    year = "2024"
}

@article{Buonanno:2024byg,
    author = "Buonanno, Alessandra and Mogull, Gustav and Patil, Raj and Pompili, Lorenzo",
    title = "{Post-Minkowskian Theory Meets the Spinning Effective-One-Body Approach for Bound-Orbit Waveforms}",
    eprint = "2405.19181",
    archivePrefix = "arXiv",
    primaryClass = "gr-qc",
    reportNumber = "HU-EP-24/16-RTG",
    doi = "10.1103/PhysRevLett.133.211402",
    journal = "Phys. Rev. Lett.",
    volume = "133",
    number = "21",
    pages = "211402",
    year = "2024"
}

@article{Damour:2025uka,
    author = "Damour, Thibault and Nagar, Alessandro and Placidi, Andrea and Rettegno, Piero",
    title = "{Novel Lagrange-multiplier approach to the effective-one-body dynamics of binary systems in post-Minkowskian gravity}",
    eprint = "2503.05487",
    archivePrefix = "arXiv",
    primaryClass = "gr-qc",
    doi = "10.1103/41sd-g2gb",
    journal = "Phys. Rev. D",
    volume = "113",
    number = "2",
    pages = "024042",
    year = "2026"
}

@online{EllipticIntegrals,
  author = {{Wolfram Research, Inc.}},
  title = {{`Elliptic Integrals'}},
  year = 1988,
  note = {\url{functions.wolfram.com/EllipticIntegrals/}},
  urldate = {2022}
}

@online{EllipticK,
  author = {{Wolfram Research}},
  title = {\texttt{EllipticK}},
  year = 1988,
  url = {https://reference.wolfram.com/language/ref/EllipticK.html},
  urldate = {2022}
}

@online{EllipticPi,
  author = {{Wolfram Research}},
  title = {\texttt{EllipticPi}},
  year = 1988,
  url = {https://reference.wolfram.com/language/ref/EllipticPi.html},
  urldate = {2022}
}

@article{Hughes:2021exa,
    author = "Hughes, Scott A. and Warburton, Niels and Khanna, Gaurav and Chua, Alvin J. K. and Katz, Michael L.",
    title = "{Adiabatic waveforms for extreme mass-ratio inspirals via multivoice decomposition in time and frequency}",
    eprint = "2102.02713",
    archivePrefix = "arXiv",
    primaryClass = "gr-qc",
    doi = "10.1103/PhysRevD.103.104014",
    journal = "Phys. Rev. D",
    volume = "103",
    number = "10",
    pages = "104014",
    year = "2021",
    note = "[Erratum: Phys.Rev.D 107, 089901 (2023)]"
}

@article{Katz:2021yft,
    author = "Katz, Michael L. and Chua, Alvin J. K. and Speri, Lorenzo and Warburton, Niels and Hughes, Scott A.",
    title = "{Fast extreme-mass-ratio-inspiral waveforms: New tools for millihertz gravitational-wave data analysis}",
    eprint = "2104.04582",
    archivePrefix = "arXiv",
    primaryClass = "gr-qc",
    doi = "10.1103/PhysRevD.104.064047",
    journal = "Phys. Rev. D",
    volume = "104",
    number = "6",
    pages = "064047",
    year = "2021"
}

@article{Chapman-Bird:2025xtd,
    author = "Chapman-Bird, Christian E. A. and others",
    title = "{Efficient waveforms for asymmetric-mass eccentric equatorial inspirals into rapidly spinning black holes}",
    eprint = "2506.09470",
    archivePrefix = "arXiv",
    primaryClass = "gr-qc",
    doi = "10.1103/scbp-75pf",
    journal = "Phys. Rev. D",
    volume = "112",
    number = "10",
    pages = "104023",
    year = "2025"
}

@article{LeTiec:2014oez,
    author = "Le Tiec, Alexandre",
    title = "{The Overlap of Numerical Relativity, Perturbation Theory and Post-Newtonian Theory in the Binary Black Hole Problem}",
    eprint = "1408.5505",
    archivePrefix = "arXiv",
    primaryClass = "gr-qc",
    doi = "10.1142/S0218271814300225",
    journal = "Int. J. Mod. Phys. D",
    volume = "23",
    number = "10",
    pages = "1430022",
    year = "2014"
}

@article{Akcay:2015pjz,
    author = "Akcay, Sarp and van de Meent, Maarten",
    title = "{Numerical computation of the effective-one-body potential $q$ using self-force results}",
    eprint = "1512.03392",
    archivePrefix = "arXiv",
    primaryClass = "gr-qc",
    doi = "10.1103/PhysRevD.93.064063",
    journal = "Phys. Rev. D",
    volume = "93",
    number = "6",
    pages = "064063",
    year = "2016"
}

@article{Bini:2015bfb,
    author = "Bini, Donato and Damour, Thibault and Geralico, Andrea",
    title = "{Confirming and improving post-Newtonian and effective-one-body results from self-force computations along eccentric orbits around a Schwarzschild black hole}",
    eprint = "1511.04533",
    archivePrefix = "arXiv",
    primaryClass = "gr-qc",
    doi = "10.1103/PhysRevD.93.064023",
    journal = "Phys. Rev. D",
    volume = "93",
    number = "6",
    pages = "064023",
    year = "2016"
}

@article{Bini:2016dvs,
    author = "Bini, Donato and Damour, Thibault and Geralico, Andrea",
    title = "{High post-Newtonian order gravitational self-force analytical results for eccentric equatorial orbits around a Kerr black hole}",
    eprint = "1602.08282",
    archivePrefix = "arXiv",
    primaryClass = "gr-qc",
    doi = "10.1103/PhysRevD.93.124058",
    journal = "Phys. Rev. D",
    volume = "93",
    number = "12",
    pages = "124058",
    year = "2016"
}

@article{vandeMeent:2015lxa,
    author = "van de Meent, Maarten and Shah, Abhay G.",
    title = "{Metric perturbations produced by eccentric equatorial orbits around a Kerr black hole}",
    eprint = "1506.04755",
    archivePrefix = "arXiv",
    primaryClass = "gr-qc",
    doi = "10.1103/PhysRevD.92.064025",
    journal = "Phys. Rev. D",
    volume = "92",
    number = "6",
    pages = "064025",
    year = "2015"
}

@article{Darwin:1961,
    title = {The gravity field of a particle. II},
    author = {C. Darwin},
    journal = {Proc. Roy. Soc. Lond. A},
    volume = {263}, 
    year = {1961}, 
    pages = {39--50},
    doi = {10.1098/rspa.1961.0142}
}

@article{Dolan:2012jg,
    author = "Dolan, Sam R. and Barack, Leor",
    title = "{Self-force via $m$-mode regularization and 2+1D evolution: III. Gravitational field on Schwarzschild spacetime}",
    eprint = "1211.4586",
    archivePrefix = "arXiv",
    primaryClass = "gr-qc",
    doi = "10.1103/PhysRevD.87.084066",
    journal = "Phys. Rev. D",
    volume = "87",
    pages = "084066",
    year = "2013"
}

@article{Merlin:2016boc,
    author = "Merlin, Cesar and Ori, Amos and Barack, Leor and Pound, Adam and van de Meent, Maarten",
    title = "{Completion of metric reconstruction for a particle orbiting a Kerr black hole}",
    eprint = "1609.01227",
    archivePrefix = "arXiv",
    primaryClass = "gr-qc",
    doi = "10.1103/PhysRevD.94.104066",
    journal = "Phys. Rev. D",
    volume = "94",
    number = "10",
    pages = "104066",
    year = "2016"
}

@article{vanDeMeent:2017oet,
    author = "van De Meent, Maarten",
    title = "{The mass and angular momentum of reconstructed metric perturbations}",
    eprint = "1702.00969",
    archivePrefix = "arXiv",
    primaryClass = "gr-qc",
    doi = "10.1088/1361-6382/aa71c3",
    journal = "Class. Quant. Grav.",
    volume = "34",
    number = "12",
    pages = "124003",
    year = "2017"
}

@article{Toomani:2021jlo,
    author = "Toomani, Vahid and Zimmerman, Peter and Spiers, Andrew and Hollands, Stefan and Pound, Adam and Green, Stephen R.",
    title = "{New metric reconstruction scheme for gravitational self-force calculations}",
    eprint = "2108.04273",
    archivePrefix = "arXiv",
    primaryClass = "gr-qc",
    doi = "10.1088/1361-6382/ac37a5",
    journal = "Class. Quant. Grav.",
    volume = "39",
    number = "1",
    pages = "015019",
    year = "2022"
}

@article{Munna:2023wce,
    author = "Munna, Christopher",
    title = "{High-order post-Newtonian expansion of the generalized redshift invariant for eccentric-orbit, equatorial extreme-mass-ratio inspirals with a spinning primary}",
    eprint = "2307.11158",
    archivePrefix = "arXiv",
    primaryClass = "gr-qc",
    doi = "10.1103/PhysRevD.108.084012",
    journal = "Phys. Rev. D",
    volume = "108",
    number = "8",
    pages = "084012",
    year = "2023"
}

@article{Paul:2024ujx,
    author = "Paul, Kaushik and Maurya, Akash and Henry, Quentin and Sharma, Kartikey and Satheesh, Pranav and Divyajyoti and Kumar, Prayush and Mishra, Chandra Kant",
    title = "{Eccentric, spinning, inspiral-merger-ringdown waveform model with higher modes for the detection and characterization of binary black holes}",
    eprint = "2409.13866",
    archivePrefix = "arXiv",
    primaryClass = "gr-qc",
    doi = "10.1103/PhysRevD.111.084074",
    journal = "Phys. Rev. D",
    volume = "111",
    number = "8",
    pages = "084074",
    year = "2025"
}

@article{Ramond:2022ctc,
    author = "Ramond, Paul and Le Tiec, Alexandre",
    title = "{First law of mechanics for spinning compact binaries: Dipolar order}",
    eprint = "2202.09345",
    archivePrefix = "arXiv",
    primaryClass = "gr-qc",
    doi = "10.1103/PhysRevD.106.044057",
    journal = "Phys. Rev. D",
    volume = "106",
    number = "4",
    pages = "044057",
    year = "2022"
}

@phdthesis{Ramond:2021mip,
    author = "Ramond, Paul",
    title = "{The First Law of Mechanics in General Relativity {\&} Isochrone Orbits in Newtonian Gravity}",
    doi = "10.1007/978-3-031-17964-8",
    school = "Observatoire de Paris",
    year = "2023"
}

@article{LeTiec:2013iux,
    author = "Le Tiec, Alexandre",
    title = "{A Note on Celestial Mechanics in Kerr Spacetime}",
    eprint = "1311.3836",
    archivePrefix = "arXiv",
    primaryClass = "gr-qc",
    doi = "10.1088/0264-9381/31/9/097001",
    journal = "Class. Quant. Grav.",
    volume = "31",
    pages = "097001",
    year = "2014"
}

@article{Gralla:2012dm,
    author = "Gralla, Samuel E. and Le Tiec, Alexandre",
    title = "{Thermodynamics of a Black Hole with Moon}",
    eprint = "1210.8444",
    archivePrefix = "arXiv",
    primaryClass = "gr-qc",
    doi = "10.1103/PhysRevD.88.044021",
    journal = "Phys. Rev. D",
    volume = "88",
    pages = "044021",
    year = "2013"
}

@software{pybhpt,
    author = "Nasipak, Zachary",
    year = 2025,
    url = "https://pybhpt.readthedocs.io/en/latest/"

}

@article{Skoupy:2023lih,
    author = "Skoupy, Viktor and Lukes-Gerakopoulos, Georgios and Drummond, Lisa V. and Hughes, Scott A.",
    title = "{Asymptotic gravitational-wave fluxes from a spinning test body on generic orbits around a Kerr black hole}",
    eprint = "2303.16798",
    archivePrefix = "arXiv",
    primaryClass = "gr-qc",
    doi = "10.1103/PhysRevD.108.044041",
    journal = "Phys. Rev. D",
    volume = "108",
    number = "4",
    pages = "044041",
    year = "2023"
}

@article{Piovano:2024yks,
    author = "Piovano, Gabriel Andres and Pantelidou, Christiana and Mac Uilliam, Jake and Witzany, Vojt{\v{e}}ch",
    title = "{Spinning particles near Kerr black holes: Orbits and gravitational-wave fluxes through the Hamilton-Jacobi formalism}",
    eprint = "2410.05769",
    archivePrefix = "arXiv",
    primaryClass = "gr-qc",
    doi = "10.1103/PhysRevD.111.044009",
    journal = "Phys. Rev. D",
    volume = "111",
    number = "4",
    pages = "044009",
    year = "2025"
}

@article{Ramond:2024sfp,
    author = "Ramond, Paul and Isoyama, Soichiro",
    title = "{Symplectic mechanics of relativistic spinning compact bodies: Canonical formalism in the Schwarzschild spacetime}",
    eprint = "2402.05049",
    archivePrefix = "arXiv",
    primaryClass = "gr-qc",
    doi = "10.1103/PhysRevD.111.064027",
    journal = "Phys. Rev. D",
    volume = "111",
    number = "6",
    pages = "064027",
    year = "2025"
}

@article{Miller:2020bft,
    author = "Miller, Jeremy and Pound, Adam",
    title = "{Two-timescale evolution of extreme-mass-ratio inspirals: waveform generation scheme for quasicircular orbits in Schwarzschild spacetime}",
    eprint = "2006.11263",
    archivePrefix = "arXiv",
    primaryClass = "gr-qc",
    doi = "10.1103/PhysRevD.103.064048",
    journal = "Phys. Rev. D",
    volume = "103",
    number = "6",
    pages = "064048",
    year = "2021"
}

@article{Wei:2025lva,
    author = "Wei, Yi-Xiang and Zhu, Xian-Long and Zhang, Jian-dong and Mei, Jianwei",
    title = "{Toward second-order self-force for eccentric extreme-mass ratio inspirals in Schwarzschild spacetimes}",
    eprint = "2504.09640",
    archivePrefix = "arXiv",
    primaryClass = "gr-qc",
    doi = "10.1103/42fh-sw5h",
    journal = "Phys. Rev. D",
    volume = "112",
    number = "6",
    pages = "064048",
    year = "2025"
}

@article{Miller:2023ers,
    author = "Miller, Jeremy and Leather, Benjamin and Pound, Adam and Warburton, Niels",
    title = "{Worldtube puncture scheme for first- and second-order self-force calculations in the Fourier domain}",
    eprint = "2401.00455",
    archivePrefix = "arXiv",
    primaryClass = "gr-qc",
    doi = "10.1103/PhysRevD.109.104010",
    journal = "Phys. Rev. D",
    volume = "109",
    number = "10",
    pages = "104010",
    year = "2024"
}

@article{Wittek:2024pis,
    author = "Wittek, Nikolas A. and Barack, Leor and Pfeiffer, Harald P. and Pound, Adam and Deppe, Nils and Kidder, Lawrence E. and Macedo, Alexandra and Nelli, Kyle C. and Throwe, William and Vu, Nils L.",
    title = "{Relieving Scale Disparity in Binary Black Hole Simulations}",
    eprint = "2410.22290",
    archivePrefix = "arXiv",
    primaryClass = "gr-qc",
    doi = "10.1103/kskl-8dcj",
    journal = "Phys. Rev. Lett.",
    volume = "134",
    number = "25",
    pages = "251402",
    year = "2025"
}

@article{Pound:2015fma,
    author = "Pound, Adam",
    title = "{Gauge and motion in perturbation theory}",
    eprint = "1506.02894",
    archivePrefix = "arXiv",
    primaryClass = "gr-qc",
    doi = "10.1103/PhysRevD.92.044021",
    journal = "Phys. Rev. D",
    volume = "92",
    number = "4",
    pages = "044021",
    year = "2015"
}

@book{AbramowitzStegun64,
  author    = {Abramowitz, Milton and Stegun, Irene A.},
  title     = {Handbook of Mathematical Functions with Formulas, Graphs, and Mathematical Tables},
  publisher = {Dover Publications},
  year      = {1964}, 
}

@article{Barack:2011ed,
    author = "Barack, Leor and Sago, Norichika",
    title = "{Beyond the geodesic approximation: conservative effects of the gravitational self-force in eccentric orbits around a Schwarzschild black hole}",
    eprint = "1101.3331",
    archivePrefix = "arXiv",
    primaryClass = "gr-qc",
    doi = "10.1103/PhysRevD.83.084023",
    journal = "Phys. Rev. D",
    volume = "83",
    pages = "084023",
    year = "2011"
}

@article{Nasipak:2025tby,
    author = "Nasipak, Zachary",
    title = "{Metric reconstruction and the Hamiltonian for eccentric, precessing binaries in the small-mass-ratio limit}",
    eprint = "2507.07746",
    archivePrefix = "arXiv",
    primaryClass = "gr-qc",
    month = "7",
    year = "2025", 
    journal = ""
}

@article{Detweiler:2002mi,
    author = "Detweiler, Steven L. and Whiting, Bernard F.",
    title = "{Selfforce via a Green's function decomposition}",
    eprint = "gr-qc/0202086",
    archivePrefix = "arXiv",
    doi = "10.1103/PhysRevD.67.024025",
    journal = "Phys. Rev. D",
    volume = "67",
    pages = "024025",
    year = "2003"
}

@article{Detweiler:2011tt,
    author = "Detweiler, Steven",
    title = "{Gravitational radiation reaction and second order perturbation theory}",
    eprint = "1107.2098",
    archivePrefix = "arXiv",
    primaryClass = "gr-qc",
    doi = "10.1103/PhysRevD.85.044048",
    journal = "Phys. Rev. D",
    volume = "85",
    pages = "044048",
    year = "2012"
}

@article{Harte:2011ku,
    author = "Harte, Abraham I.",
    title = "{Mechanics of extended masses in general relativity}",
    eprint = "1103.0543",
    archivePrefix = "arXiv",
    primaryClass = "gr-qc",
    doi = "10.1088/0264-9381/29/5/055012",
    journal = "Class. Quant. Grav.",
    volume = "29",
    pages = "055012",
    year = "2012"
}

@article{Pound:2012nt,
    author = "Pound, Adam",
    title = "{Second-order gravitational self-force}",
    eprint = "1201.5089",
    archivePrefix = "arXiv",
    primaryClass = "gr-qc",
    doi = "10.1103/PhysRevLett.109.051101",
    journal = "Phys. Rev. Lett.",
    volume = "109",
    pages = "051101",
    year = "2012"
}

@article{Pound:2015tma,
    author = "Pound, Adam",
    editor = {Puetzfeld, Dirk and L{\"a}mmerzahl, Claus and Schutz, Bernard},
    title = "{Motion of small objects in curved spacetimes: An introduction to gravitational self-force}",
    eprint = "1506.06245",
    archivePrefix = "arXiv",
    primaryClass = "gr-qc",
    doi = "10.1007/978-3-319-18335-0_13",
    journal = "Fund. Theor. Phys.",
    volume = "179",
    pages = "399--486",
    year = "2015"
}

@article{Upton:2025bja,
    author = "Upton, Samuel D. and Wardell, Barry and Pound, Adam and Warburton, Niels and Barack, Leor",
    title = "{Effective source for second-order self-force calculations: quasicircular orbits in Schwarzschild spacetime}",
    eprint = "2508.00087",
    archivePrefix = "arXiv",
    primaryClass = "gr-qc",
    month = "7",
    year = "2025", 
    journal = ""
}

@article{Upton:2021oxf,
    author = "Upton, Samuel D. and Pound, Adam",
    title = "{Second-order gravitational self-force in a highly regular gauge}",
    eprint = "2101.11409",
    archivePrefix = "arXiv",
    primaryClass = "gr-qc",
    doi = "10.1103/PhysRevD.103.124016",
    journal = "Phys. Rev. D",
    volume = "103",
    number = "12",
    pages = "124016",
    year = "2021"
}

@article{Spiers:2023mor,
    author = "Spiers, Andrew and Pound, Adam and Wardell, Barry",
    title = "{Second-order perturbations of the Schwarzschild spacetime: Practical, covariant, and gauge-invariant formalisms}",
    eprint = "2306.17847",
    archivePrefix = "arXiv",
    primaryClass = "gr-qc",
    doi = "10.1103/PhysRevD.110.064030",
    journal = "Phys. Rev. D",
    volume = "110",
    number = "6",
    pages = "064030",
    year = "2024"
}

@article{Warburton:2021kwk,
    author = "Warburton, Niels and Pound, Adam and Wardell, Barry and Miller, Jeremy and Durkan, Leanne",
    title = "{Gravitational-Wave Energy Flux for Compact Binaries through Second Order in the Mass Ratio}",
    eprint = "2107.01298",
    archivePrefix = "arXiv",
    primaryClass = "gr-qc",
    doi = "10.1103/PhysRevLett.127.151102",
    journal = "Phys. Rev. Lett.",
    volume = "127",
    number = "15",
    pages = "151102",
    year = "2021"
}

@article{Damour:2016abl,
    author = {Damour, Thibault and Jaranowski, Piotr and Sch{\"a}fer, Gerhard},
    title = "{Conservative dynamics of two-body systems at the fourth post-Newtonian approximation of general relativity}",
    eprint = "1601.01283",
    archivePrefix = "arXiv",
    primaryClass = "gr-qc",
    doi = "10.1103/PhysRevD.93.084014",
    journal = "Phys. Rev. D",
    volume = "93",
    number = "8",
    pages = "084014",
    year = "2016"
}

@article{Pound:2019lzj,
    author = "Pound, Adam and Wardell, Barry and Warburton, Niels and Miller, Jeremy",
    title = "{Second-Order Self-Force Calculation of Gravitational Binding Energy in Compact Binaries}",
    eprint = "1908.07419",
    archivePrefix = "arXiv",
    primaryClass = "gr-qc",
    doi = "10.1103/PhysRevLett.124.021101",
    journal = "Phys. Rev. Lett.",
    volume = "124",
    number = "2",
    pages = "021101",
    year = "2020"
}

@article{Warburton:2024xnr,
    author = "Warburton, Niels and Wardell, Barry and Trestini, David and Henry, Quentin and Pound, Adam and Blanchet, Luc and Durkan, Leanne and Faye, Guillaume and Miller, Jeremy",
    title = "{Comparison of 4.5PN and 2SF gravitational energy fluxes from quasicircular compact binaries}",
    eprint = "2407.00366",
    archivePrefix = "arXiv",
    primaryClass = "gr-qc",
    doi = "10.1103/tzsw-kcyt",
    journal = "Phys. Rev. D",
    volume = "113",
    number = "8",
    pages = "084050",
    year = "2026"
}

@article{LeTiec:2011dp,
    author = "Le Tiec, Alexandre and Barausse, Enrico and Buonanno, Alessandra",
    title = "{Gravitational Self-Force Correction to the Binding Energy of Compact Binary Systems}",
    eprint = "1111.5609",
    archivePrefix = "arXiv",
    primaryClass = "gr-qc",
    doi = "10.1103/PhysRevLett.108.131103",
    journal = "Phys. Rev. Lett.",
    volume = "108",
    pages = "131103",
    year = "2012"
}

@article{Drummond:2023wqc,
    author = "Drummond, Lisa V. and Lynch, Philip and Hanselman, Alexandra G. and Becker, Devin R. and Hughes, Scott A.",
    title = "{Extreme mass-ratio inspiral and waveforms for a spinning body into a Kerr black hole via osculating geodesics and near-identity transformations}",
    eprint = "2310.08438",
    archivePrefix = "arXiv",
    primaryClass = "gr-qc",
    doi = "10.1103/PhysRevD.109.064030",
    journal = "Phys. Rev. D",
    volume = "109",
    number = "6",
    pages = "064030",
    year = "2024"
}

@article{Lynch:2021ogr,
    author = "Lynch, Philip and van de Meent, Maarten and Warburton, Niels",
    title = "{Eccentric self-forced inspirals into a rotating black hole}",
    eprint = "2112.05651",
    archivePrefix = "arXiv",
    primaryClass = "gr-qc",
    doi = "10.1088/1361-6382/ac7507",
    journal = "Class. Quant. Grav.",
    volume = "39",
    number = "14",
    pages = "145004",
    year = "2022"
}

@article{Trestini:2025nzr,
    author = "Trestini, David",
    title = "{Schott term in the binding energy for compact binaries on circular orbits at fourth post-Newtonian order}",
    eprint = "2504.13245",
    archivePrefix = "arXiv",
    primaryClass = "gr-qc",
    doi = "10.1103/lsbb-sv45",
    journal = "Phys. Rev. D",
    volume = "112",
    number = "2",
    pages = "024076",
    year = "2025"
}

@article{Leather:2023dzj,
    author = "Leather, Benjamin and Warburton, Niels",
    title = "{Applying the effective-source approach to frequency-domain self-force calculations for eccentric orbits}",
    eprint = "2306.17221",
    archivePrefix = "arXiv",
    primaryClass = "gr-qc",
    doi = "10.1103/PhysRevD.108.084045",
    journal = "Phys. Rev. D",
    volume = "108",
    number = "8",
    pages = "084045",
    year = "2023"
}

@article{Spiers:2023cip,
    author = "Spiers, Andrew and Pound, Adam and Moxon, Jordan",
    title = "{Second-order Teukolsky formalism in Kerr spacetime: Formulation and nonlinear source}",
    eprint = "2305.19332",
    archivePrefix = "arXiv",
    primaryClass = "gr-qc",
    doi = "10.1103/PhysRevD.108.064002",
    journal = "Phys. Rev. D",
    volume = "108",
    number = "6",
    pages = "064002",
    year = "2023"
}

@article{Cunningham:2024dog,
    author = "Cunningham, Kevin and Kavanagh, Chris and Pound, Adam and Trestini, David and Warburton, Niels and Neef, Jakob",
    title = "{Gravitational memory: new results from post-Newtonian and self-force theory}",
    eprint = "2410.23950",
    archivePrefix = "arXiv",
    primaryClass = "gr-qc",
    doi = "10.1088/1361-6382/adbc3d",
    journal = "Class. Quant. Grav.",
    volume = "42",
    number = "13",
    pages = "135009",
    year = "2025",
    note = "[Addendum: Class.Quant.Grav. 42, 199401 (2025)]"
}

@article{Osburn:2022bby,
    author = "Osburn, Thomas and Nishimura, Nami",
    title = "{New self-force method via elliptic partial differential equations for Kerr inspiral models}",
    eprint = "2206.07031",
    archivePrefix = "arXiv",
    primaryClass = "gr-qc",
    doi = "10.1103/PhysRevD.106.044056",
    journal = "Phys. Rev. D",
    volume = "106",
    number = "4",
    pages = "044056",
    year = "2022"
}

@article{PanossoMacedo:2024pox,
    author = "Panosso Macedo, Rodrigo and Bourg, Patrick and Pound, Adam and Upton, Samuel D.",
    title = "{Multidomain spectral method for self-force calculations}",
    eprint = "2404.10083",
    archivePrefix = "arXiv",
    primaryClass = "gr-qc",
    doi = "10.1103/PhysRevD.110.084008",
    journal = "Phys. Rev. D",
    volume = "110",
    number = "8",
    pages = "084008",
    year = "2024"
}

@article{Bourg:2024cgh,
    author = "Bourg, Patrick and Pound, Adam and Upton, Samuel D. and Panosso Macedo, Rodrigo",
    title = "{Simple, efficient method of calculating the Detweiler-Whiting singular field to very high order}",
    eprint = "2404.10082",
    archivePrefix = "arXiv",
    primaryClass = "gr-qc",
    doi = "10.1103/PhysRevD.110.084007",
    journal = "Phys. Rev. D",
    volume = "110",
    number = "8",
    pages = "084007",
    year = "2024"
}

@article{Wardell:2024yoi,
    author = "Wardell, Barry and Kavanagh, Chris and Dolan, Sam R.",
    title = "{Sourced metric perturbations of Kerr spacetime in Lorenz gauge}",
    eprint = "2406.12510",
    archivePrefix = "arXiv",
    primaryClass = "gr-qc",
    doi = "10.1088/1361-6382/ae0918",
    journal = "Class. Quant. Grav.",
    volume = "42",
    number = "20",
    pages = "205007",
    year = "2025"
}

@article{Dolan:2021ijg,
    author = "Dolan, Sam R. and Kavanagh, Chris and Wardell, Barry",
    title = "{Gravitational Perturbations of Rotating Black Holes in Lorenz Gauge}",
    eprint = "2108.06344",
    archivePrefix = "arXiv",
    primaryClass = "gr-qc",
    doi = "10.1103/PhysRevLett.128.151101",
    journal = "Phys. Rev. Lett.",
    volume = "128",
    number = "15",
    pages = "151101",
    year = "2022"
}

@article{Dolan:2023enf,
    author = "Dolan, Sam R. and Durkan, Leanne and Kavanagh, Chris and Wardell, Barry",
    title = "{Metric perturbations of Kerr spacetime in Lorenz gauge: circular equatorial orbits}",
    eprint = "2306.16459",
    archivePrefix = "arXiv",
    primaryClass = "gr-qc",
    doi = "10.1088/1361-6382/ad52e3",
    journal = "Class. Quant. Grav.",
    volume = "41",
    number = "15",
    pages = "155011",
    year = "2024"
}

@article{Chua:2020stf,
    author = "Chua, Alvin J. K. and Katz, Michael L. and Warburton, Niels and Hughes, Scott A.",
    title = "{Rapid generation of fully relativistic extreme-mass-ratio-inspiral waveform templates for LISA data analysis}",
    eprint = "2008.06071",
    archivePrefix = "arXiv",
    primaryClass = "gr-qc",
    doi = "10.1103/PhysRevLett.126.051102",
    journal = "Phys. Rev. Lett.",
    volume = "126",
    number = "5",
    pages = "051102",
    year = "2021"
}

@article{Pound:2017psq,
    author = "Pound, Adam",
    title = "{Nonlinear gravitational self-force: second-order equation of motion}",
    eprint = "1703.02836",
    archivePrefix = "arXiv",
    primaryClass = "gr-qc",
    doi = "10.1103/PhysRevD.95.104056",
    journal = "Phys. Rev. D",
    volume = "95",
    number = "10",
    pages = "104056",
    year = "2017"
}

@article{Compere:2019gft,
    author = "Comp{\`e}re, Geoffrey and Oliveri, Roberto and Seraj, Ali",
    title = "{The Poincar{\'e} and BMS flux-balance laws with application to binary systems}",
    eprint = "1912.03164",
    archivePrefix = "arXiv",
    primaryClass = "gr-qc",
    doi = "10.1007/JHEP10(2020)116",
    journal = "JHEP",
    volume = "10",
    pages = "116",
    year = "2020",
    note = "[Erratum: JHEP 06, 045 (2024)]"
}

@article{Ashtekar:2004cn,
    author = "Ashtekar, Abhay and Krishnan, Badri",
    title = "{Isolated and dynamical horizons and their applications}",
    eprint = "gr-qc/0407042",
    archivePrefix = "arXiv",
    doi = "10.12942/lrr-2004-10",
    journal = "Living Rev. Rel.",
    volume = "7",
    pages = "10",
    year = "2004"
}

@article{Chandrasekaran:2018aop,
    author = "Chandrasekaran, Venkatesa and Flanagan, {\'E}anna {\'E}. and Prabhu, Kartik",
    title = "{Symmetries and charges of general relativity at null boundaries}",
    eprint = "1807.11499",
    archivePrefix = "arXiv",
    primaryClass = "hep-th",
    doi = "10.1007/JHEP11(2018)125",
    journal = "JHEP",
    volume = "11",
    pages = "125",
    year = "2018",
    note = "[Erratum: JHEP 07, 224 (2023)]"
}

@unpublished{Grant:InPrep,
    author = {Alexander M. Grant and Alexandre Le Tiec and Adam Pound},
    title = {Energy balance and the first law of binary black hole mechanics},
    note = {in preparation}
}

@article{Detweiler:2008ft,
    author = "Detweiler, Steven L.",
    title = "{A Consequence of the gravitational self-force for circular orbits of the Schwarzschild geometry}",
    eprint = "0804.3529",
    archivePrefix = "arXiv",
    primaryClass = "gr-qc",
    doi = "10.1103/PhysRevD.77.124026",
    journal = "Phys. Rev. D",
    volume = "77",
    pages = "124026",
    year = "2008"
}

@book{Arnold,
  title={Mathematical Methods of Classical Mechanics},
  author={Arnold, Vladimir I.},
  volume={60},
  series={Graduate Texts in Mathematics},
  year={1989},
  publisher={Springer-Verlag},
  edition={2nd ed.},
  address={New York},
  isbn={978-0-387-96890-2},
  translator={Weinstein, Alan D. and Vogtmann, Karen}
}

@article{Hughes:2005qb,
    author = "Hughes, Scott A. and Drasco, Steve and Flanagan, Eanna E. and Franklin, Joel",
    title = "{Gravitational radiation reaction and inspiral waveforms in the adiabatic limit}",
    eprint = "gr-qc/0504015",
    archivePrefix = "arXiv",
    doi = "10.1103/PhysRevLett.94.221101",
    journal = "Phys. Rev. Lett.",
    volume = "94",
    pages = "221101",
    year = "2005"
}

@article{Sago:2005fn,
    author = "Sago, Norichika and Tanaka, Takahiro and Hikida, Wataru and Ganz, Katsuhiko and Nakano, Hiroyuki",
    title = "{The Adiabatic evolution of orbital parameters in the Kerr spacetime}",
    eprint = "gr-qc/0511151",
    archivePrefix = "arXiv",
    doi = "10.1143/PTP.115.873",
    journal = "Prog. Theor. Phys.",
    volume = "115",
    pages = "873--907",
    year = "2006"
}

@article{Albertini:2022rfe,
    author = "Albertini, Angelica and Nagar, Alessandro and Pound, Adam and Warburton, Niels and Wardell, Barry and Durkan, Leanne and Miller, Jeremy",
    title = "{Comparing second-order gravitational self-force, numerical relativity, and effective one body waveforms from inspiralling, quasicircular, and nonspinning black hole binaries}",
    eprint = "2208.01049",
    archivePrefix = "arXiv",
    primaryClass = "gr-qc",
    doi = "10.1103/PhysRevD.106.084061",
    journal = "Phys. Rev. D",
    volume = "106",
    number = "8",
    pages = "084061",
    year = "2022"
}

@article{Harte:2025tmd,
    author = "Harte, Abraham I. and Blanco, Francisco M. and Flanagan, Eanna E.",
    title = "{Nonlinearly Self-Interacting Extended Bodies Move as Test Bodies in Effective External Fields}",
    eprint = "2504.11912",
    archivePrefix = "arXiv",
    primaryClass = "gr-qc",
    doi = "10.1103/k74d-dj7y",
    journal = "Phys. Rev. Lett.",
    volume = "135",
    number = "15",
    pages = "151401",
    year = "2025"
}

@misc{supp,
  note = {The Supplemental Material consists of 4 files. High-order post-Newtonian expressions are provided in machine-readable form in \texttt{PN\_expressions.wl}. Data for generic eccentric orbits is provided in \texttt{data\_generic.csv}. Data for circular orbits is provided in \texttt{data\_circ.csv}. Data on the separatrix is provided in \texttt{data\_sep.csv}.}
}

@ARTICLE{2020SciPy-NMeth,
  author  = {Virtanen, Pauli and Gommers, Ralf and Oliphant, Travis E. and
            Haberland, Matt and Reddy, Tyler and Cournapeau, David and
            Burovski, Evgeni and Peterson, Pearu and Weckesser, Warren and
            Bright, Jonathan and {van der Walt}, St{\'e}fan J. and
            Brett, Matthew and Wilson, Joshua and Millman, K. Jarrod and
            Mayorov, Nikolay and Nelson, Andrew R. J. and Jones, Eric and
            Kern, Robert and Larson, Eric and Carey, C J and
            Polat, {\.I}lhan and Feng, Yu and Moore, Eric W. and
            {VanderPlas}, Jake and Laxalde, Denis and Perktold, Josef and
            Cimrman, Robert and Henriksen, Ian and Quintero, E. A. and
            Harris, Charles R. and Archibald, Anne M. and
            Ribeiro, Ant{\^o}nio H. and Pedregosa, Fabian and
            {van Mulbregt}, Paul and {SciPy 1.0 Contributors}},
  title   = {{{SciPy} 1.0: Fundamental Algorithms for Scientific
            Computing in Python}},
  journal = {Nature Methods},
  year    = {2020},
  volume  = {17},
  pages   = {261--272},
  adsurl  = {https://rdcu.be/b08Wh},
  doi     = {10.1038/s41592-019-0686-2},
}

@article{Isoyama:2014mja,
    author = "Isoyama, Soichiro and Barack, Leor and Dolan, Sam R. and Le Tiec, Alexandre and Nakano, Hiroyuki and Shah, Abhay G. and Tanaka, Takahiro and Warburton, Niels",
    title = "{Gravitational Self-Force Correction to the Innermost Stable Circular Equatorial Orbit of a Kerr Black Hole}",
    eprint = "1404.6133",
    archivePrefix = "arXiv",
    primaryClass = "gr-qc",
    reportNumber = "KUNS-2490, YITP-14-33",
    doi = "10.1103/PhysRevLett.113.161101",
    journal = "Phys. Rev. Lett.",
    volume = "113",
    number = "16",
    pages = "161101",
    year = "2014"
}

@software{pybhpt-0.9.8,
  author       = {Zachary Nasipak},
  title        = {znasipak/pybhpt: v0.9.8},
  month        = oct,
  year         = 2025,
  publisher    = {Zenodo},
  version      = {v0.9.8},
  doi          = {10.5281/zenodo.17408385},
  url          = {https://doi.org/10.5281/zenodo.17408385},
  swhid        = {swh:1:dir:2566ac37fcf1d7ca84af6a1e33835f832d247a3d
                   ;origin=https://doi.org/10.5281/zenodo.15627817;vi
                   sit=swh:1:snp:678ef7af1b148178a152b05f0535e321340b
                   0936;anchor=swh:1:rel:0b576a4e5364a3463f4b9ed3c9e8
                   09524fbe8273;path=znasipak-pybhpt-07a8ea3
                  },
}

\end{document}